\documentclass[onecolumn,journal,romanappendices]{IEEEtran}
\usepackage{amsmath}
\usepackage{amssymb}
\usepackage{amsthm}
\usepackage{algorithmic}
\usepackage{algorithm}
\usepackage{array}
\usepackage{mathrsfs}
\usepackage[caption=false,font=normalsize,labelfont=sf,textfont=sf]{subfig}
\usepackage{textcomp}
\usepackage{stfloats}
\usepackage{multirow}
\usepackage{url}
\usepackage{mathtools}
\usepackage{verbatim}
\usepackage{graphicx}
\usepackage{cite} 
\usepackage[bb=boondox]{mathalfa}
\usepackage{braket}
\usepackage{enumitem}
\usepackage{cuted} 
\newtheorem{theorem}{Theorem}[section]
\newtheorem{lemma}[theorem]{Lemma}
\newtheorem{proposition}[theorem]{Proposition}
\newtheorem{corollary}[theorem]{Corollary}
\newtheorem{definition}[theorem]{Definition}
\newtheorem{example}[theorem]{Example}

\makeatletter
\providecommand*{\boxast}{%
	\mathbin{
		\mathpalette\@boxit{*}%
	}%
}
\newcommand*{\@boxit}[2]{%
	\sbox0{$\m@th#1\Box$}%
	\ifx#1\displaystyle \ht0=\dimexpr\ht0+.05ex\relax \fi
	\ifx#1\textstyle \ht0=\dimexpr\ht0+.05ex\relax \fi
	\ifx#1\scriptstyle \ht0=\dimexpr\ht0+.04ex\relax \fi
	\ifx#1\scriptscriptstyle \ht0=\dimexpr\ht0+.065ex\relax \fi
	\sbox2{$#1\vcenter{}$}
	\rlap{%
		\hbox to \wd0{%
			\hfill
			\raisebox{%
				\dimexpr.5\dimexpr\ht0+\dp0\relax-\ht2\relax
			}{$\m@th#1#2$}%
			\hfill
		}%
	}%
	\Box
}
\makeatother
\begin{document}
	
	\title{SC-LDPC Codes Over $\mathbb{F}_q$: Minimum Distance, Decoding Analysis and Threshold Saturation}
	\author{Jiaxin Lyu, Guanghui He, ~\IEEEmembership{Member,~IEEE,}}
	
	\markboth{Lyu et al. SC-LDPC Codes Over $\mathbb{F}_q$: Minimum Distance, Decoding Analysis and Threshold Saturation}%
	{Shell \MakeLowercase{\textit{et al.}}: A Sample Article Using IEEEtran.cls for IEEE Journals}
	
	
	\maketitle
	
	\begin{abstract}
		We study random spatially coupled low-density parity-check (SC-LDPC) code ensembles over arbitrary finite fields, constructed through variable-node edge spreading and independent random monomial maps. We consider two spreading rules: independent uniform edge offsets (the standard ensemble) and uniformly sampled distinct offsets (the improved ensemble). For fixed coupling parameters and degrees \(d_r\ge d_l\ge3\), we establish linear lower bounds on the minimum distance and minimum stopping-set size, together with exact polynomial orders of the corresponding failure probabilities. For belief-propagation (BP) decoding over \(q\)-ary memoryless symmetric channels (QMSCs), we develop a unified density-evolution and potential-functional framework based on symmetric probability measures and degradation, allowing general measurable output spaces. We prove that, for complete QMSC families ordered by degradation, the BP thresholds of both coupled ensembles converge to a well-defined fixed-point potential threshold \(h^{\mathrm{FP}}\) of the uncoupled system as the coupling length and subsequently the coupling width tend to infinity. More\,\,generally, under an appropriate stability condition, this saturation result extends to exchangeable edge-spreading laws satisfying a mild shift-regularity criterion. Together, these results provide theoretical tools for evaluating and designing edge-spreading rules. 
	\end{abstract}


\section{Introduction}
\IEEEPARstart{L}{OW-DENSITY} parity-check (LDPC) convolutional codes, originally introduced by Felström and Zigangirov\cite{LDPCC1}, have shown remarkable performance under low-complexity iterative decoding \cite{LDPCC2,LDPCC3,LDPCC4,LDPCC5}. Compared to their block code counterparts, terminated LDPC convolutional code ensembles typically exhibit better asymptotic belief propagation (BP) thresholds, which can be further improved by increasing the connectivity in their Tanner graphs. The underlying mechanism, referred to as \textit{threshold saturation via spatial coupling}, was first introduced by Kudekar \textit{et al.} and proven for regular LDPC ensembles over the binary erasure channel (BEC) \cite{SCLDPC1}, and the corresponding results were subsequently extended to general binary-input memoryless output-symmetric (BMS) channels and irregular LDPC ensembles \cite{SCLDPC,SCLDPC3,SCLDPC4}. In a typical spatially coupled (SC) ensemble derived from a random LDPC block code ensemble, multiple copies of variable and check nodes from the underlying ensemble are sequentially arranged, locally and randomly coupled, and terminated at both boundaries, forming a chain structure. Termination can be represented by fixing virtual boundary
symbols to known values. This reduces the effective check
degrees near the boundaries, facilitating local decoding
and the inward propagation of reliable messages under BP. For the ensembles and channel families covered by existing
binary results, threshold saturation is established in terms
of an intrinsic threshold of the underlying
ensemble\cite{SCLDPC1,SCLDPC,SCLDPC3,SCLDPC4}.
Taking the chain-length limit before the coupling-width
limit also eliminates the termination-induced rate loss. Moreover, carefully designed SC-LDPC codes can inherit asymptotically good minimum distance performance from their underlying ensembles, e.g., protograph-based regular SC-LDPC codes due to Mitchell \textit{et al.} \cite{SCLDPC_minimum_distance1,SCLDPC_minimum_distance2,SCLDPC_minimum_distance3,SCLDPC_minimum_distance4}. Hence, spatial coupling exhibits great potential in designing good LDPC codes with both capacity-approaching thresholds and low error floors under low-complexity BP decoding over general BMS channels. For more details of SC-LDPC codes, see \cite{SC_magzine} for a comprehensive review, \cite{SCLDPC1}, \cite{SCLDPC} for rigorous construction and asymptotic analysis of random ensembles, and \cite{SCLDPC_minimum_distance3}, \cite{SCLDPC_finite_length1,SCLDPC_finite_length2,SCLDPC_finite_length3} for finite-length scaling.

While extensive theoretical results have been established for binary SC-LDPC ensembles transmitted over BMS channels, the theoretical results for their nonbinary extensions, e.g., SC-LDPC ensembles over finite fields, remain relatively underdeveloped. Compared to their binary counterparts, nonbinary LDPC codes can offer improved performance at short-to-moderate block lengths and can be advantageous for bandwidth-efficient modulation schemes. For example, Davey and MacKay showed that LDPC codes defined over binary extension fields can significantly improve performance at short-to-moderate code lengths when transmitted over binary-input channels\cite{Davey1998}. Many works have already considered the analysis, construction, or optimization of SC-LDPC codes over nonbinary finite fields. Piemontese \textit{et al.}\cite{NBSC1} studied nonbinary ensembles with general-linear-group edge labels on the BEC. They derived upper bounds on the MAP thresholds of the underlying uncoupled ensembles and observed threshold saturation in the coupled ensembles.  In\cite{NBSC2}, Wei \textit{et al.} studied the asymptotic performance of nonbinary coupled ensembles with windowed decoding, under both the  BEC  and the binary-input additive white Gaussian noise channel (BIAWGNC). In\cite{NBSC3}, Andriyanova \textit{et al.} proved   the   saturation  of the BP threshold of coupled ensembles   over binary extension fields to the potential threshold, defined by Kumar \textit{et al.}\cite{SCLDPC4}, under the BEC. In\cite{NBSC4}, Huang \textit{et al.} compared the finite-length performance of nonbinary protograph-based block-LDPC   and SC-LDPC codes with their binary counterparts, highlighting their trade-offs in performance, complexity and latency. In\cite{Zhang2018}, Zhang \textit{et al.} investigated random and protograph-based coupled ensembles over binary extension fields, and by optimizing non-uniform  variable-to-check node connections, they achieved better BP thresholds and reduced rate loss. In\cite{Hareedy2019}, Hareedy \textit{et al.} optimized nonbinary SC-LDPC codes using the weight consistency matrix  framework,  and significantly improved their FER performance. To the best of the authors' knowledge, the threshold saturation property of coupled ensembles over nonbinary finite fields has not yet been rigorously proven for general nonbinary-input channels. In our view, this might stem from the absence of comprehensive theoretical tools and results, similar to those available for BMS channels\cite[Sec. IV]{ModernCode}, when performing   iterative decoding analysis in the nonbinary setting. Although asymptotic distance properties of nonbinary protograph-based SC-LDPC ensembles have been studied\cite{Huang2015Distance}, the effects of edge-spreading rules on minimum-distance and stopping-set guarantees merit further investigation.

In this paper, we investigate random SC-LDPC code
ensembles over arbitrary finite fields $\mathbb F_q$,
where $q\ge2$ is a prime\,\,power.
We establish their asymptotic distance properties
and develop an analytical framework for BP decoding
and threshold saturation over general $q$-ary
memoryless symmetric channels (QMSCs).
Our main contributions are summarized as follows.

\noindent\,\,\,\textit{i) Coupled ensemble construction
	and distance analysis.}
We construct random coupled ensembles through
variable-node edge spreading and independent
uniformly random monomial maps, which specify
the edge connections and nonzero edge labels.
We introduce a general edge-spreading profile
and its equivalent representation by an exchangeable
random offset vector.
This framework accommodates different spreading
rules and provides a common foundation for  distance
analysis and coupled density evolution. We focus on two coupled ensembles.
The \emph{standard coupled ensemble} uses independent
uniform edge offsets, extending the binary construction
in \cite[Sec. II-B]{SCLDPC1}.
The \emph{improved coupled ensemble} uses uniformly
sampled distinct offsets, ensuring that the edges
of each variable node reach distinct check-node
positions. By analyzing\,their\,average\,weight\,and\,stopping-set
distributions, we establish asymptotic distance
bounds and precise polynomial estimates for the
probabilities of bad minimum distance and
stopping-set size.
In particular, the following result holds.
\begin{theorem}[Informal version of Theorem III.16]
Fix a prime power $q\ge2$ and integers
$d_r\ge d_l\ge3$, $w\ge d_l$, and $L\ge1$.
Consider either the standard or improved
$(d_l,d_r,w,L)$ coupled ensemble over $\mathbb F_q$.
Let $\mathcal C_n$ denote its random code of block
length $2Ln$, and $\mathcal G_n$ its associated
Tanner graph. Then there exist constants
$\alpha,\widetilde\alpha,c>0$ such that
\[
\Pr\left\{
d_{\min}(\mathcal C_n)\ge 2L\alpha n,\,
d_{\mathrm{ss}}(\mathcal G_n)\ge
2L\widetilde\alpha n
\right\}
=1-O(n^{-c})
\]
as $n\to\infty$ through admissible block lengths.
Here, $d_{\min}$ and $d_{\mathrm{ss}}$ denote the
minimum distance and minimum  
stopping-set size, respectively.	 
\end{theorem}
Beyond these probabilistic results, our
growth-rate formulas provide analytical criteria
for comparing different edge-spreading profiles.
In particular, the improved ensemble has better  explicit distance
and stopping-set lower bounds before capping
them\,\,at 1. These results suggest that controlling the
concentration of a variable node's edges at
individual check positions may improve the
  distance performance.
Numerical comparisons further illustrate
the possible benefits of improved edge spreading.

 \textit{ii) A unified framework for
	$q$-ary density evolution.}
We develop analytical tools for iterative BP
decoding over $\mathbb F_q$ on\,\,QMSCs,
whose symmetry is described by the doubly
transitive affine group of $\mathbb F_q$.
Using posterior probability vectors, we
characterize symmetric probability measures
and their reference measures on the probability
simplex $\mathcal S_q$, including for channels
with general measurable output spaces. We establish   basic properties of BP updates and
derive the fundamental properties of several useful information functionals. We also equip the spaces of symmetric distributions with compact metric topologies and develop their degradation theory. These results yield the continuity and
monotonicity of density evolution (DE),
the convergence of monotone DE sequences,
and a stability condition
for   perfect BP decoding. Unlike the binary setting, where message
distributions admit a scalar 
representation, the nonbinary analysis involves
  measures on a higher-dimensional
simplex. Our framework handles\,\,this structure
directly and provides the mathematical
foundation for the potential-functional
analysis for threshold saturation.
It also supplies reusable tools for analyzing
other sparse graph codes over \(\mathbb{F}_q\) under iterative decoding.

 \textit{iii) Threshold saturation over general QMSCs.}
We establish threshold saturation for coupled
ensembles over   $\mathbb F_q$
on complete QMSC families ordered by degradation.
For such a family
$\{\mathsf c_h:0\le h\le\log q\}$,
parameterized by its conditional entropy $h$,
we define a fixed-point threshold
$h^{\mathrm{FP}}$ through the potential values
of nontrivial fixed points of the uncoupled
DE recursion.\,\,This threshold depends only on the underlying
  degree pair and the channel family,
independently of the coupling parameters. Let $h_c^{\mathrm{BP}}(w,L)$ denote the
coupled BP threshold under DE.
Our main result is as follows.
\begin{theorem}[Informal consequence of Theorem V.4
	and Lemmas V.5, V.6]
For every prime power $q\ge2$ and
$d_r\ge d_l\ge3$, both the standard and improved
$(d_l,d_r,w,L)$ coupled ensembles satisfy
\[
\lim_{w\to\infty}\lim_{L\to\infty}
h_c^{\mathrm{BP}}(w,L)
=
h^{\mathrm{FP}}.
\]
\end{theorem}
Our proof first establishes the concentration
and DE characterization of BP decoding for
every fixed number of iterations.
We then extend the potential-functional
approach of\cite{SCLDPC4} to symmetric
message measures over $\mathbb F_q$.
The resulting analysis establishes matching
achievability and converse bounds at
$h^{\mathrm{FP}}$. More generally, our   analysis applies
to exchangeable edge-spreading laws \(\pi_w\) satisfying
a mild shift-regularity condition. Specifically, let $\eta_w$ denote the minimum
total-variation distance between $\pi_w$ and
its positive-integer shifts in one coordinate,
as defined in Section V-B.
Under the stability condition of the uncoupled
system, $\eta_w\to0$ as $w\to\infty$ is sufficient
for threshold saturation. This includes the standard and improved ensembles,
for which
\[
\eta_w=
\begin{cases}
	1/w,&\text{standard ensemble},\\[2pt]
	d_l/w,&\text{improved ensemble}.
\end{cases}
\]
The stability condition is automatic when
$d_l\ge3$. These results also provide theoretical guidance for designing and evaluating edge-spreading rules beyond the two ensembles studied here. Specifically, the explicit lower bounds on the minimum-distance and stopping-set thresholds depend on the maximum number of edges directed to a single check position, whereas a sufficient condition for threshold saturation depends on the variation of the joint edge-offset distribution under a spatial shift. The\,\,distinct criteria suggest that improving the distance bounds does not necessarily reduce the coupling width required by the saturation guarantee. This observation may  help guide the design of new spreading profiles that balance distance performance, decoding thresholds, and termination-induced rate loss.  Finally, our threshold saturation theorem establishes
that the BP threshold of the coupled ensemble
converges to $h^{\mathrm{FP}}$ as the coupling length
$L\to\infty$, followed by the coupling width
$w\to\infty$. Our theorem establishes threshold saturation to \(h^{\mathrm{FP}}\) for fixed degrees \(d_l,d_r\), but it does not address the large-degree behavior of \(h^{\mathrm{FP}}\). For the binary case (\(q=2\)),  it is known that \(h^{\mathrm{FP}}\) approaches the Shannon threshold \(r\log q\) as \(d_l,d_r\to\infty\) with \(d_l/d_r\to r\in(0,1)\) \cite{SCLDPC,SCLDPC4}. We plan to present a corresponding rigorous result for general nonbinary QMSC families in future work.

The remainder of this paper is organized as follows.
Section II introduces the notation and conventions.
Section III defines\,the uncoupled and coupled
ensembles over finite fields and analyzes their
minimum-distance and stopping-set properties.
Section IV develops the theoretical framework for
QMSCs, symmetric probability measures, degradation,
and uncoupled density evolution.
Section V establishes coupled density evolution
and proves threshold saturation using potential
functionals. Section VI concludes the paper.
Proofs of all new theoretical results are provided,
with the proof of the threshold saturation theorem (Theorem V.4) presented in Section V and the remaining
proofs collected in the appendices.



\section{Notation and Conventions}
We take $\mathbb{N}=\{0,1,\ldots\}$ and write
$\mathbb{S}_n$ for the symmetric group on $n$ letters.
For a prime power $q\geq2$, let $\mathbb{F}_q$ denote
the finite field of order $q$, with multiplicative group
$\mathbb{F}_q^\times=\mathbb{F}_q\setminus\{0\}$.
Vectors are denoted by boldface or underlined symbols,
depending on context. We write $\|\cdot\|$ for the
Euclidean norm and $\|\cdot\|_p$ for the usual
$\ell_p$ norm. For real-valued sequences \(f(n)\) and \(g(n)\), with \(g(n)>0\) for all sufficiently large \(n\), we write \(f(n)=O(g(n))\) if \(|f(n)|\le Cg(n)\) for some constant \(C>0\) and all sufficiently large \(n\), and \(f(n)=o(g(n))\) if \(|f(n)|/g(n)\to0\) as \(n\to\infty\). Similarly, \(f(n)=\Omega(g(n))\) if \(f(n)\ge cg(n)\) for some constant \(c>0\) and all sufficiently large \(n\), and \(f(n)=\Theta(g(n))\) if both \(f(n)=O(g(n))\) and \(f(n)=\Omega(g(n))\). Other limiting variables or parameter dependencies are indicated by subscripts when needed. For example, \(o_n(1)\) denotes a quantity tending to zero as \(n\to\infty\), while \(O_n(\varepsilon^2)\) denotes a bound as \(\varepsilon\to0\), with an implied constant that may depend on \(n\). Throughout, all logarithms are natural, with the conventions
$\log0=-\infty$ and $0\log0=0$. The $q$-ary entropy function
$H_q:[0,1]\to[0,\log q]$ is defined by \(H_q(\alpha)
\coloneqq
-\alpha\log\alpha
-(1-\alpha)\log(1-\alpha)
+\alpha\log(q-1)\). It is easy to check that \(H_q\) is strictly concave on \([0,1]\).
For any integer sequence $0\leq k_n\leq n$
satisfying $k_n/n\to\alpha\in[0,1]$, Stirling's formula gives
\[
\lim_{n\to\infty}
\frac1n\log\left[
\binom{n}{k_n}(q-1)^{k_n}
\right]
=H_q(\alpha).
\]

For a metric space \((X,d)\), recall that \(X\) is complete if every Cauchy sequence converges in \(X\), and totally bounded if\,\,it\,\,admits a finite \(\varepsilon\)-net for every \(\varepsilon>0\). Compactness is equivalent to completeness and total boundedness, or equivalently, to sequential compactness: every sequence in \(X\) admits a subsequence converging to a  point in \(X\). A metric space is called Polish if it is separable and completely metrizable. In particular, every compact metric space is Polish. Closed subsets and finite products of compact metric spaces are compact, and continuous images of compact spaces are also compact. Every continuous real-valued function on a nonempty compact space attains its minimum and maximum. Moreover, a continuous bijection from a compact space onto a Hausdorff space is a homeomorphism.

We work on an underlying probability space
$(\Omega,\mathcal F,\operatorname{Pr})$ and write
$\mathbb E$ for expectation.
A random element $Z$ taking values in a measurable space
$(X,\mathcal A)$ is an
$(\mathcal F,\mathcal A)$-measurable map
$Z:\Omega\to X$.
Its distribution, denoted by $\operatorname{Law}(Z)$,
is the probability measure defined by
$\operatorname{Law}(Z)(B)=\operatorname{Pr}(Z\in B)$
for every $B\in\mathcal A$. We write \(Z\sim\mu\) if \(\operatorname{Law}(Z)=\mu\), and \(Z\overset{\mathrm d}{=}Z'\) if\,\,\(\operatorname{Law}(Z)=\operatorname{Law}(Z')\). An event \(A\in\mathcal F\) is said to occur almost surely (a.s.) if \(\operatorname{Pr}(A)=1\). For a measurable map $T:X\to Y$
and a\,\,measure $\mu$ on $X$, the pushforward of \(\mu\) under \(T\) is the measure \(T_\#\mu\) defined by  \((T_\#\mu)(B)=\mu(T^{-1}(B))\) for every measurable $B\subseteq Y$.
In particular,
$\operatorname{Law}(Z)=Z_\#\operatorname{Pr}$. For a metric space $(X,d)$, let $\mathcal B(X)$ denote
its Borel $\sigma$-algebra and $\mathcal P(X)$
the space of Borel probability measures on $X$.
A sequence $\mu_n\in\mathcal P(X)$ converges weakly
to $\mu\in\mathcal P(X)$ if
$\int_X f\,d\mu_n\to\int_X f\,d\mu$
for every bounded continuous function
$f:X\to\mathbb R$.
All signed measures considered in this paper are finite. For $1\le p<\infty$, the $p$-Wasserstein space is
\[
\mathcal P_p(X)
\coloneqq
\left\{
\mu\in\mathcal P(X):
\int_X d(x,x_0)^p\,\mu(dx)<\infty
\right\},
\]
where $x_0\in X$ is arbitrary.
For $\mu,\nu\in\mathcal P_p(X)$,
the $p$-Wasserstein distance is defined by
\[
W_p(\mu,\nu)
\coloneqq
\left(
\inf_{\gamma\in\Pi(\mu,\nu)}
\int_{X\times X}
d(x,y)^p\,\gamma(dx,dy)
\right)^{1/p},
\]
where $\Pi(\mu,\nu)$ denotes the set of probability
measures on $X\times X$ with marginals $\mu$ and $\nu$
\cite[Def.~6.1]{OT}.
If $(X,d)$ is complete and separable, then
$(\mathcal P_p(X),W_p)$ is also complete and separable
\cite[Thm.~6.18]{OT}.
Moreover, by\cite[Thm.~6.9]{OT}, $W_p(\mu_n,\mu)\to0$ if and only if
$\mu_n$ converges weakly to $\mu$ and
\[
\int_X d(x,x_0)^p\,\mu_n(dx)
\longrightarrow
\int_X d(x,x_0)^p\,\mu(dx).
\] 
If $X$ is compact, then
$\mathcal P_p(X)=\mathcal P(X)$,
$W_p$ metrizes weak convergence, and
$(\mathcal P(X),W_p)$ is compact.
These properties apply, in particular, to the
probability simplex $\mathcal S_q$ introduced
in Section IV, equipped with the Euclidean metric.

\section{Coupled Ensembles and Distance Analysis}
\!This section introduces the LDPC code ensembles over finite fields considered in this paper and studies their distance\,properties. We first recall the uncoupled ensembles, and describe the spatially coupled construction through edge spreading, including\,\,the standard and improved   ensembles induced by two distinct spreading rules. Then, we establish asymptotic bounds and probability estimates for the minimum distance and minimum stopping-set size of the coupled ensembles, and highlight the impact of\,\,edge spreading on these properties.

\subsection{Uncoupled LDPC Code Ensembles over $\mathbb{F}_q$}
Let $q\geq 2$ be a prime power. LDPC codes over $\mathbb{F}_q$, similar to their binary counterparts, can be defined by   Tanner graphs, where the variable nodes represent codeword symbols in $\mathbb{F}_q$, and the check nodes represent parity checks. Each edge is assigned a nonzero coefficient from \(\mathbb F_q^\times\), and each check node imposes a homogeneous linear parity-check constraint. We refer to such   code graphs as $\mathbb{F}^{\times}_q$-labelled. An LDPC code is called \textit{regular} if its Tanner graph is regular, i.e., all the variable nodes  have the same degree and all the check nodes have the same degree. Otherwise, the code is referred to as \textit{irregular}. The first ensemble of regular LDPC codes was proposed and studied by Gallager\cite{Gallager}. Subsequently, many (uncoupled) LDPC codes or code ensembles were defined\cite{GoodCodes,Luby,LDPC1,margulis1982explicit,vasic2004combinatorial}. Though most of these definitions are over $\mathbb{F}_2$, their extension to an arbitrary finite field is straightforward. In this work, we mainly focus on graph-based constructions, where graph connections can be defined by random permutations, due to Luby \textit{et al.}\cite{Luby} and Richardson \textit{et al.}\cite{LDPC1}. When  $\mathbb{F}_q^{\times}$-labelled Tanner graphs are considered, extending a random permutation to a \textit{random monomial map} admits a more compact definition.
\begin{definition}[Monomial map]
For   $\boldsymbol{c}\in (\mathbb{F}_q^{\times})^n$  and  permutation  $\pi_n\in\mathbb{S}_n$, define the  monomial  map  $\xi_{\boldsymbol{c},\pi_n}:\mathbb{F}_q^n\rightarrow \mathbb{F}_q^n$  by 
$$\boldsymbol{x}\mapsto (c_1x_{\pi_n^{-1}(1)},c_2x_{\pi_n^{-1}(2)},\ldots,c_nx_{\pi_n^{-1}(n)})\quad \forall \boldsymbol{x}\in\mathbb{F}_q^n.$$
Moreover, let $\Xi_n$ denote a \textit{uniformly random monomial map} on $n$ letters in $\mathbb{F}_q$, i.e., the random map $\Xi_n$ is uniformly distributed over the set  $\{\xi_{\boldsymbol{c},\pi_n}:\boldsymbol{c}\in (\mathbb{F}_q^{\times})^n, \pi_n\in\mathbb{S}_n\}$.
\end{definition}

We will briefly review the definition of regular LDPC code ensembles over $\mathbb{F}_q$ and some of their properties. A $(d_l,d_r)$-regular Tanner graph has variable-node degree $d_l$ and check-node degree $d_r$, and we   use $n$ and $m$ to denote   the numbers of its variable nodes and check nodes, respectively. Clearly, $m= {d_ln}/{d_r}$. A block length \(n\) is called admissible if \(m\) is an integer.
\begin{definition}[Repetition and check maps]
A single repetition map $f_{d_l}^{\mathrm{rep}}:\mathbb{F}_q\rightarrow\mathbb{F}_q^{d_l}$ with degree parameter $d_l$ is given by $v\mapsto (v,v,\ldots,v)$, and its $n$-fold Cartesian product is denoted by $f_{d_l,n}^{\mathrm{rep}}:\mathbb{F}_q^n\rightarrow\mathbb{F}_q^{d_l n}$. A single parity-check map $f_{d_r}^{\mathrm{chk}}:\mathbb{F}_q^{d_r}\rightarrow\mathbb{F}_q$ with degree parameter $d_r$ is given by $\boldsymbol{v}\mapsto \sum_{i=1}^{d_r} v_i$, and its $m$-fold Cartesian product is denoted by $f_{d_r,m}^{\mathrm{chk}}:\mathbb{F}_q^{d_rm}\rightarrow\mathbb{F}_q^{m}$. 
\end{definition} 
\begin{definition}[\((d_l,d_r)\)-regular ensemble over \(\mathbb F_q\)]
For \(nd_l=md_r\), the \((d_l,d_r)\)-regular LDPC code ensemble over \(\mathbb F_q\) of\,\,block length \(n\) is the random linear code
\[
\mathcal C_{d_l,d_r,n}\coloneqq \ker\left(
f_{d_r,m}^{\mathrm{chk}}
\circ\Xi_{nd_l}
\circ f_{d_l,n}^{\mathrm{rep}}
\right).
\]
\end{definition} 

Readers familiar with the random bipartite graph model in\cite{Luby, LDPC1} may note that Definition III.3 is a direct extension of the random Tanner graph definition for a binary regular ensemble, where the connections between variable nodes and check nodes (along with $d_ln$ edge labels) are determined by the uniformly random monomial map $\Xi_{d_l n}$, and there might exist \textit{multi-edge connections} in this random graph. The associated random Tanner graph is denoted by \(\mathcal G_{d_l,d_r,n}\).

\begin{definition}[Weight and stopping-set distributions]
Let \(C_n\subseteq\mathbb F_q^n\) be a linear code and \(G_n\) a Tanner graph representing\,\,it. For each \(0\le\ell\le n\), let \(A_\ell(C_n)\) and \(\widetilde A_\ell(G_n)\) denote the numbers of weight-\(\ell\) codewords and size-\(\ell\) stopping sets, respectively.\footnote{A stopping set is a set of variable nodes such that every check node has either zero or at least two incident edges from the set, counting multiplicities.} The minimum distance of \(C_n\) and the minimum size of a nonempty stopping set in \(G_n\) are denoted by \(d_{\min}(C_n)\) and \(d_{\mathrm{ss}}(G_n)\), respectively, with the conventions \(d_{\min}(C_n)=+\infty\) if \(C_n=\{\boldsymbol 0\}\) and \(d_{\mathrm{ss}}(G_n)=+\infty\) if \(G_n\) has no nonempty stopping set.
\end{definition}

Note that the stopping-set distribution depends only on the Tanner graph and is independent of the edge labels and field\,\,size \(q\). The ensemble-average weight and stopping-set distributions are recalled below.
\begin{theorem}[Average weight and stopping-set distributions, cf.\cite{Weight_Distribution1,Weight_Distribution2,Weight_Finite_Field,Di2002,Orlitsky2005}]
For \(0\le\ell\le n\),
\[
\begin{aligned}
	\mathbb E[A_\ell(\mathcal C_{d_l,d_r,n})] =
	\frac{
		\binom n\ell
		[z^{d_l\ell}]W_{q,d_r}(z)^m
	}{
		\binom{d_ln}{d_l\ell}(q-1)^{(d_l-1)\ell}
	},\qquad
	\mathbb E[\widetilde A_\ell(\mathcal G_{d_l,d_r,n})] =
	\frac{
		\binom n\ell
		[z^{d_l\ell}]\widetilde W_{d_r}(z)^m
	}{
		\binom{d_ln}{d_l\ell}
	},
\end{aligned}
\] 
where \([z^k]P(z)\) denotes the coefficient of \(z^k\) in \(P(z)\), and
\begin{equation}
W_{q,d_r}(z)\coloneqq 
\frac{[1+(q-1)z]^{d_r}+(q-1)(1-z)^{d_r}}q,\qquad
\widetilde W_{d_r}(z) \coloneqq(1+z)^{d_r}-d_rz.\label{eq:weight_enumerator}
\end{equation}
Here, \(W_{q,d_r}\) is the weight enumerator of the length-\(d_r\) single parity-check code over \(\mathbb F_q\), while \(\widetilde W_{d_r}\) counts subsets of check-node sockets of sizes other than one.
\end{theorem}

The asymptotic growth rate functions of the above average weight and stopping set distributions have also been well studied. For \(0\le\alpha\le1\), define the asymptotic growth-rate functions of the two average distributions by
\[
\begin{aligned}
	g_{q,d_l,d_r}(\alpha)\coloneqq\limsup_{n\to\infty}
	\frac1n\log\mathbb E
	\left[A_{\lfloor\alpha n\rfloor}(\mathcal C_{d_l,d_r,n})\right],\qquad
	\widetilde g_{d_l,d_r}(\alpha)\coloneqq\limsup_{n\to\infty}
	\frac1n\log\mathbb E
	\left[\widetilde A_{\lfloor\alpha n\rfloor}
	(\mathcal G_{d_l,d_r,n})\right],
\end{aligned}
\] 
where \(n\) ranges over admissible block lengths.  Some related properties are reviewed as follows.
\begin{theorem}[Asymptotic growth rates, cf.\cite{Weight_Finite_Field,Orlitsky2005}]
For \(0\le\alpha\le1\),
\[
\begin{aligned}
	g_{q,d_l,d_r}(\alpha)
	&=
	\frac{d_l}{d_r}
	\log\inf_{z>0}
	\frac{W_{q,d_r}(z)}{z^{d_r\alpha}}
	-(d_l-1)H_q(\alpha),\\
	\widetilde g_{d_l,d_r}(\alpha)
	&=
	\frac{d_l}{d_r}
	\log\inf_{z>0}
	\frac{\widetilde W_{d_r}(z)}{z^{d_r\alpha}}
	-(d_l-1)H_2(\alpha).
\end{aligned}
\] 
Both \(g_{q,d_l,d_r}\) and \(\widetilde g_{d_l,d_r}\) are continuous on their respective finite domains.
For \(d_r\ge d_l\ge3\), \(g_{q,d_l,d_r}\) has a unique zero \(\alpha_{q,d_l,d_r}\) in \((0,1-1/q]\). For \(d_l,d_r\ge3\), \(\widetilde g_{d_l,d_r}\) has a unique zero \(\widetilde\alpha_{d_l,d_r}\) in \((0,1)\).
\end{theorem} 
The zeros \(\alpha_{q,d_l,d_r}\) and \(\widetilde{\alpha}_{d_l,d_r}\) provide asymptotic lower bounds on the normalized minimum distance and minimum stopping-set size, respectively. As \(d_l,d_r\to\infty\) with the design rate \(R=1-d_l/d_r\) fixed, \(\alpha_{q,d_l,d_r}\) approaches the \(q\)-ary asymptotic\,Gilbert-Varshamov (GV) bound, whereas \(\widetilde{\alpha}_{d_l,d_r}\to0\) (see\cite[Rem. 5.7]{Weight_Finite_Field}  and\cite[p. 934]{Orlitsky2005}). The   probability estimates are given below.
\begin{theorem}[cf.\cite{Bound_binary,Weight_Finite_Field,Orlitsky2005}]
Let \(n\to\infty\) through admissible block lengths. If \(d_r\ge d_l\ge3\),   for any \(0<\alpha<\alpha_{q,d_l,d_r}\),
\[
\Pr\!\left\{
d_{\min}(\mathcal C_{d_l,d_r,n})\le\alpha n
\right\}
=
\begin{cases}
	\Theta(n^{2-d_l}),
	& q=2,\ d_l\text{ odd},\\[2mm]
	\Theta\!\left(n^{1-\lceil d_l/2\rceil}\right),
	& \text{otherwise}.
\end{cases}
\]
If \(d_l,d_r\ge3\), then for any \(0<\widetilde\alpha<\widetilde\alpha_{d_l,d_r}\), \(\Pr  \{
d_{\mathrm{ss}}(\mathcal G_{d_l,d_r,n})
\le\widetilde\alpha n
 \}
=
\Theta (n^{1-\lceil d_l/2\rceil} )\).
\end{theorem}
We also consider irregular LDPC ensembles over \(\mathbb F_q\), characterized by their degree profiles. We assume that all variable and check nodes have degrees at least \(2\). A node-perspective degree profile \((\Lambda,P)\) is a pair of polynomials
\[
\Lambda(x)=\sum_{l=2}^{l_{\max}}\Lambda_l x^l,
\qquad
P(x)=\sum_{k=2}^{k_{\max}}P_k x^k, 
\]
with nonnegative coefficients satisfying \(\Lambda(1)=P(1)=1\), where \(\Lambda_i\) (\(P_i\)) is the fraction of variable (check) nodes of degree \(i\), and \(\Lambda'(1)\) and \(P'(1)\) are the respective average node degrees. The associated edge-perspective degree profile \((\lambda,\rho)\) is given\,\,by
\[
\lambda(x)
=\sum_l\lambda_l x^{l-1}
=\frac{\Lambda'(x)}{\Lambda'(1)},
\qquad
\rho(x)
=\sum_k\rho_k x^{k-1}
=\frac{P'(x)}{P'(1)},
\]
where \(\lambda_i\) (\(\rho_i\)) is the fraction of edges incident to variable (check) nodes of degree \(i\). Conversely,
\[
\Lambda(x)
=\frac{\int_0^x\lambda(z)\,\mathrm dz}
{\int_0^1\lambda(z)\,\mathrm dz},
\qquad
P(x)
=\frac{\int_0^x\rho(z)\,\mathrm dz}
{\int_0^1\rho(z)\,\mathrm dz}.
\]
\begin{definition}[Irregular ensemble over \(\mathbb F_q\)]
Given a node-perspective degree profile \((\Lambda,P)\), let \(n\) and \(m\) be   positive integers satisfying \(M\coloneqq n\Lambda'(1)=mP'(1)\), with \(n\Lambda_l\) and \(mP_k\) being nonnegative integers for all \(l,k\). Define the irregular repetition map $f^{\mathrm{rep}}_{\Lambda,n}:\mathbb{F}_q^{n}\rightarrow \mathbb{F}_q^{\Lambda^{\prime}(1)n}$ and the irregular parity-check map $f^{\mathrm{chk}}_{P,m}:\mathbb{F}_q^{P^{\prime}(1)m}\rightarrow \mathbb{F}_q^{m}$ by the Cartesian products
$$f^{\mathrm{rep}}_{\Lambda,n}\coloneqq f^{\mathrm{rep}}_{2,\Lambda_2 n}\times \cdots\times f^{\mathrm{rep}}_{l_{\max},\Lambda_{l_{\max}} n},\quad f^{\mathrm{chk}}_{P,m}\coloneqq f^{\mathrm{chk}}_{2,P_2 m}\times \cdots\times f^{\mathrm{chk}}_{k_{\max},P_{k_{\max}} m}.$$
The irregular LDPC code ensemble over \(\mathbb F_q\) with degree profile \((\Lambda,P)\) (equivalently, \((\lambda,\rho)\)) and block length \(n\) is defined by
\[
\mathcal C_{\Lambda,P,n}
\coloneqq
\ker\!\left(
f_{P,m}^{\mathrm{chk}}
\circ\Xi_M
\circ f_{\Lambda,n}^{\mathrm{rep}}
\right).
\]
\end{definition}
\subsection{Coupled Ensembles over \(\mathbb{F}_q\)}
We now construct spatially coupled LDPC ensembles based on the regular ensembles of Section III-A. We first place variable and check nodes along a finite chain. The construction then proceeds in two steps: spreading each variable node's sockets over nearby check-node positions and then randomly matching the incoming variable sockets to check-node sockets at each\,\,position. We describe the spreading by edge types, which are convenient for distance enumeration, and equivalently by an exchangeable offset vector, which will be used for density evolution analysis in Section V.

Fix integers $d_l,d_r\ge2$ and $w,L\ge1$. For a positive integer $n$, put
\[
N\coloneqq2L,\qquad K\coloneqq N+w-1,\qquad
m\coloneqq\frac{d_l}{d_r}n.
\]
Here $w$ is the \emph{coupling width}: a variable node at position $k$ can connect only to check-node positions $k,k+1,\ldots,k+w-1$. The parameter $L$ measures the length of the terminated coupling chain, with $N=2L$ variable-node positions. Place $n$ physical variable nodes at each position $1,\ldots,N$, and $m$ degree-$d_r$ check nodes at each position $1,\ldots,K$. To implement termination, also place $n$ \emph{virtual} variable nodes at each position in $\{2-w,\ldots,0\}\cup\{N+1,\ldots,K\}$. All virtual symbols are fixed to zero; only the $Nn$ physical variables are transmitted codeword symbols.

\begin{figure}[t]
	\centering
	\includegraphics[width=0.6\textwidth]{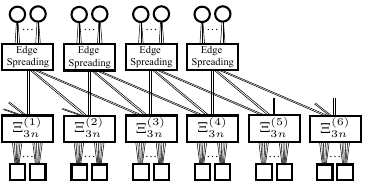} 
	\caption{An example of a coupled Tanner graph with $(d_l,d_r,w,L)=(3,6,3,2)$. Circles and squares represent variable and check nodes, respectively. Virtual variable nodes at the boundaries are omitted.}
	\label{fig2}
\end{figure}

\begin{definition}[Edge type and edge-spreading profile]
	\label{def:sec3_edge_profile}
	An edge type is a vector $\underline t=(t_0,\ldots,t_{w-1})$ in
	\[
	\mathcal T_{w,d_l}\coloneqq
	\left\{\underline t\in\mathbb Z_{\ge0}^w:
	\sum_{b=0}^{w-1}t_b=d_l\right\}.
	\]
	A variable node at position $k$ has type $\underline t$ if exactly $t_b$ of its $d_l$ sockets are directed to check position $k+b$. An \textbf{edge-spreading profile} $(\mathcal T,p)$ consists of a probability distribution $p$ on $\mathcal T_{w,d_l}$ and its support $\mathcal T$. At every physical and virtual variable position, exactly $np(\underline t)$ nodes are assigned type $\underline t\in\mathcal T$, in any fixed order. We assume that $p$ has rational masses and call $n$ admissible if $m$ and all type counts $np(\underline t)$ are integers.
\end{definition}
The same edge-spreading  profile \((\mathcal{T},p)\) is used at all variable positions, including the virtual ones. Then at each check\,position $i\in\{1,\ldots,K\}$, the incoming variable-side socket count is
\begin{equation}\label{eq:sec3_socket_balance}
	n\sum_{b=0}^{w-1}\sum_{\underline t\in\mathcal T}
	p(\underline t)t_b
	=nd_l=md_r.\notag
\end{equation}
Therefore, the incoming variable sockets can be matched uniformly to the $md_r$ check sockets at each position. We implement the random matching together with edge labeling  by independent uniformly random monomial maps $\Xi_{d_ln}^{(i)}$, $1\le i\le K$.
\begin{definition}[Coupled ensemble over \(\mathbb{F}_q\)]\label{def:sec3_coupled_ensemble}
	For every check position $i=1,\ldots,K$, choose an independent uniformly\,random monomial map $\Xi_{nd_l}^{(i)}$ to match and label its incoming variable sockets with its check-node sockets. Thus both the matching and the nonzero edge labels are uniformly random. For $\boldsymbol v\in\mathbb F_q^{Nn}$, let $\boldsymbol u_i(\boldsymbol v)\in\mathbb F_q^{nd_l}$ list the symbols on the variable sockets\,\,entering check position $i$, with zero assigned to every virtual socket. The coupled LDPC code ensemble is 
	\begin{equation}\notag
		\mathcal C_{d_l,d_r,w,L,n}
		\coloneqq\left\{\boldsymbol v\in\mathbb F_q^{Nn}:
		f_{d_r,m}^{\mathrm{chk}}\!\left(
		\Xi_{nd_l}^{(i)}(\boldsymbol u_i(\boldsymbol v))
		\right)=\boldsymbol0,\ \ 1\le i\le K\right\}.
	\end{equation}
	Its associated $\mathbb F_q^\times$-labelled Tanner graph ensemble is denoted by $\mathcal G_{d_l,d_r,w,L,n}$. See Fig.\ref{fig2} for an example.
\end{definition}
The edge type specifies how many sockets of a variable node are directed to each check position, but does not distinguish\,\,the individual sockets. For the subsequent density evolution analysis, we also need to describe where the other sockets of a variable node lead when one of its edges is selected. To this end, fix any physical variable position $k$, choose a variable node uniformly, and order its $d_l$ sockets uniformly at random. Let $\underline T\sim p$ denote its edge type. Define the corresponding random offset vector\,\,by
\[
\boldsymbol J=(J_1,\ldots,J_{d_l})\sim\pi_w,
\qquad J_a\in\{0,\ldots,w-1\},
\]
where $J_a=b$ means that the $a$th socket of this node is directed to check position $k+b$. Conditioned on $\underline T=\underline t$, the vector $\boldsymbol J$ is uniformly distributed over all permutations of the multiset containing $t_b$ copies of $b$  for $b=0,1,\ldots,w-1$. Putting
\[\underline{t}(\boldsymbol{j})=\left(\sum_{a=1}^{d_l}\mathbf1\{j_a=b\}\right)_{0\le b\le w-1},\qquad
\boldsymbol j\in\{0,\ldots,w-1\}^{d_l},
\]
we can describe the offset distribution \(\pi_w\) through the edge-spreading profile by 
\begin{equation}\notag
	\pi_w(\boldsymbol j)
	=p(\underline t(\boldsymbol j))/\binom{d_l}{\underline t(\boldsymbol j)}. 
\end{equation}
By uniformly random ordering, the random offset vector $\boldsymbol J$ is \emph{exchangeable}, meaning that, for every permutation $\sigma\in\mathbb S_{d_l}$,
\[
(J_{\sigma(1)},\ldots,J_{\sigma(d_l)})
\overset{\mathrm d}=\boldsymbol J.
\]
Conversely, every exchangeable law $\pi_w$ with rational
masses on $\{0,1,\ldots,w-1\}^{d_l}$ uniquely determines
a  profile through
\[
p(\underline t)
=\sum_{\boldsymbol j:\,\underline t(\boldsymbol j)=\underline t}
\pi_w(\boldsymbol j).
\]
Hence, $p$ and $\pi_w$ provide equivalent descriptions of the same edge-spreading rule. The random ordering is only an auxiliary representation of the fixed edge types and does not alter the finite ensemble.  Now consider an incoming edge selected uniformly at any check position $i$, and designate it as the first socket of its variable endpoint, with the remaining sockets ordered uniformly. Since exactly $np(\underline t)t_b$ incoming sockets originate from variable nodes of edge type $\underline t$ at position $i-b$, the resulting offset vector again has law $\pi_w$, with
\begin{equation}\label{eq:sec3_rooted_offset}
	\Pr\{\underline T=\underline t,J_1=b\}
	=\frac{t_b}{d_l}p(\underline t),
	\qquad
	\Pr\{J_1=b\}
	=\frac1{d_l}\sum_{\underline t\in\mathcal T}
	p(\underline t)t_b.
\end{equation}
The variable endpoint of the selected edge lies at position $i-J_1$, while its remaining sockets are directed to positions $i-J_1+J_a$, $2\le a\le d_l$. This  will be used to derive the density evolution equations for the coupled system in Section V-A.

Of particular interest are edge-spreading profiles \((\mathcal{T},p)\) satisfying
\begin{equation}\label{Tpp}
	\sum_{\underline t\in\mathcal T}p(\underline t)t_b
	= {d_l/w},
	\qquad b=0,1,\ldots,w-1.
\end{equation}
By \eqref{eq:sec3_rooted_offset} and exchangeability, this   is equivalent to \(\Pr\{J_a=b\}=\frac1w\) for each \(1\le a\le d_l\) and \(0\le b\le w-1\). Consequently,\,\,each check position receives exactly $nd_l/w$ variable sockets from each of its $w$ neighboring variable positions.  Below, we give three examples of $(\mathcal{T},p)$   satisfying (\ref{Tpp}), each of which specifies a class of coupled ensembles over $\mathbb{F}_q$.
\begin{example}[$(d_l,d_r,L)$ spreading]
 Let \(w=d_l\) and \(\mathcal{T}=\{(1,\ldots,1)\}\) consist of a single all-ones edge type. Equivalently, \(\boldsymbol J\) is uniformly distributed over all permutations of \((0,\ldots,d_l-1)\). This gives the $(d_l,d_r,L)$ ensemble, following the edge-spreading rule
 of \cite[Sec. II-A]{SCLDPC1}.
\end{example}
\begin{example}[Standard spreading]\label{ex:sec3_standard_spreading}
	Let $J_1,\ldots,J_{d_l}$ be independent and uniform on $\{0,\ldots,w-1\}$. Equivalently, $\mathcal T=\mathcal T_{w,d_l}$, and for each \(\underline{t}\in \mathcal T_{w,d_l}\),
	\begin{equation}\label{eq:sec3_standard_profile} 
		p(\underline t)=\frac{1}{w^{d_l}}\binom{d_l}{\underline t}.
	\end{equation}
	This defines the standard $(d_l,d_r,w,L)$ ensemble, which extends
	the binary construction in \cite[Sec. II-B]{SCLDPC1}.
	Multiple sockets of\,\,a variable node may reach the same check position,
	so parallel edges are possible after random matching.
\end{example}
\begin{example}[Improved spreading]\label{ex:sec3_improved_spreading}
	Assume $w\ge d_l$ and choose $\boldsymbol J$ uniformly from the ordered $d_l$-tuples of  \textbf{distinct} elements in $\{0,\ldots,w-1\}$. Equivalently,
	\begin{equation}\label{eq:sec3_improved_profile}
		\mathcal T=\mathcal T_{w,d_l}\cap\{0,1\}^w,\qquad
		p(\underline t)= 1/{\binom{w}{d_l}}
		\qquad \forall \underline t\in\mathcal T,
	\end{equation}
	and $\pi_w(\boldsymbol j)=1/(w)_{d_l}$ for tuples \(\boldsymbol{j}\) with distinct coordinates, and zero otherwise, where $(w)_{d_l}\coloneqq w(w-1)\cdots(w-d_l+1)$. This defines the improved $(d_l,d_r,w,L)$ ensemble.
	 Its variable-node sockets always reach distinct check positions,  excluding\,\,parallel edges. When $w=d_l$, it reduces to the $(d_l,d_r,L)$ ensemble above. 
\end{example}

Related distinct-offset spreading has also been used in protograph-based
SC-LDPC constructions \cite{SCLDPC_minimum_distance2}. As shown in Section III-C, different edge-spreading rules can lead to different distance properties   for fixed \(d_l,d_r,w,L\). In particular, the improved ensemble exhibits stronger lower bounds on the minimum distance and minimum stopping-set size, motivating its name.
\subsection{Distance Analysis of Coupled Ensembles}
\label{subsec:sec3_distance_analysis}
We derive probabilistic results on the minimum distance and minimum stopping-set size of the \((d_l,d_r,w,L)\) coupled\,ensembles  in Definition III.10 by analyzing the  average weight and stopping-set distributions. In the binary case, see\cite{Sridharan2007,Truhachev2010,SCLDPC_minimum_distance1,SCLDPC_minimum_distance2,SCLDPC_minimum_distance3,SCLDPC_minimum_distance4} for related distance analyses of various ensembles of LDPC convolutional codes. Throughout, $q,d_l,d_r,w,L$ and the rational edge-spreading profile $(\mathcal T,p)$ are fixed, $N=2L$, $K=N+w-1$,   $n\to\infty$ through admissible values, and \(m=d_ln/d_r\).  All weights and\,\,stopping-set sizes count only physical variables at positions \(1,2,\ldots,N\).

Both the weight and stopping-set distributions can be enumerated by classifying subsets of physical variable nodes according to their \textit{support types}. For a subset \(S\) of physical variables, its support type is the array\[
\underline\ell=(\ell_{k,\underline t})_{
	1\le k\le N,\;\underline t\in\mathcal T},
\]
where \(\ell_{k,\underline t}\) is the number of variables in \(S\) at position \(k\) with edge type \(\underline t\). For a codeword, \(S\) is the set of its nonzero coordinates; for a stopping set, \(S\) is the stopping set itself. A support type \(\underline\ell\) is feasible if its entries are integers satisfying \(0\le \ell_{k,\underline t}\le np(\underline t)\). For such a type, let \(\ell\) denote the total number of selected variables and \(e_i\) the number of their sockets directed to check position \(i\), counting multiplicities. Then
\begin{equation}
\ell=\sum_{k=1}^{N}\sum_{\underline t\in\mathcal T}
\ell_{k,\underline t},\qquad
e_i=\sum_{b=0}^{w-1}\sum_{\underline t\in\mathcal T}
t_b\ell_{i-b,\underline t},
\quad 1\le i\le K,\qquad
\sum_{i=1}^{K}e_i=d_l\ell,\label{eq:ell_e_i}
\end{equation}
where we set \(\ell_{k,\underline t}=0\) for \(k\notin\{1,\ldots,N\}\). The last identity holds since each selected variable node has exactly \(d_l\) sockets.\,\,Let \(A_{\underline\ell}(\mathcal C_{d_l,d_r,w,L,n})\) and \(\widetilde A_{\underline\ell}(\mathcal G_{d_l,d_r,w,L,n})\)  denote the numbers of codewords and stopping sets, respectively, with support type \(\underline\ell\). The zero type includes the all-zero codeword and the empty stopping set. Recall the single-check enumerators \(W_{q,d_r}\) and \(\widetilde W_{d_r}\) in \eqref{eq:weight_enumerator}, and let \([z^e]F(z)\) denote the coefficient of \(z^e\) in \(F(z)\).
\begin{proposition} 
	\label{prop:sec3_enumerators}
	For every feasible support type \(\underline\ell\), with \(\ell\) and \(e_i\) defined in \eqref{eq:ell_e_i}, the following identities hold:
	\begin{align}
		\mathbb E  A_{\underline\ell}(\mathcal C_{d_l,d_r,w,L,n}) 
		&=(q-1)^{-(d_l-1)\ell}
		\left[\prod_{k=1}^{N}\prod_{\underline t\in\mathcal T}
		\binom{np(\underline t)}{\ell_{k,\underline t}}\right]
		\prod_{i=1}^{K}
		\frac{[z^{e_i}]W_{q,d_r}(z)^m}{\binom{nd_l}{e_i}},
		\label{AVE_WTD1}\\
		\mathbb E \widetilde A_{\underline\ell}(\mathcal G_{d_l,d_r,w,L,n})
		&=\left[\prod_{k=1}^{N}\prod_{\underline t\in\mathcal T}
		\binom{np(\underline t)}{\ell_{k,\underline t}}\right]
		\prod_{i=1}^{K}
		\frac{[z^{e_i}]\widetilde W_{d_r}(z)^m}{\binom{nd_l}{e_i}}.
		\label{AVE_STD1}
	\end{align}
	The average weight-$\ell$ and size-$\ell$ enumerators   are obtained by summing these expressions over the feasible support-type set
	\[\left\{\underline{\ell}=(\ell_{k,\underline t})_{
		1\le k\le N,\;\underline t\in\mathcal T}:0\le \ell_{k,\underline t}\le np(\underline t) \ \forall k,\underline{t},\ \sum_{k=1}^{N}\sum_{\underline t\in\mathcal T}
		\ell_{k,\underline t}=\ell\right\}.\]
\end{proposition}
For $0\le\alpha\le1$, define the feasible set of normalized support types with total normalized weight \(\alpha\) by
\begin{align} 
\mathcal A(\alpha) \coloneqq\left\{\underline a=(a_{k,\underline t})_{
	1\le k\le N,\;\underline t\in\mathcal T}:
0\le a_{k,\underline t}\le p(\underline t)\ \forall k,\underline{t},\
\sum_{k=1}^{N}\sum_{\underline t\in\mathcal T}a_{k,\underline t}=N\alpha\right\}.\notag 
\end{align}
Again,   extend  $a_{k,\underline t}$ by zero for \(k\notin\{1,2,\ldots,N\}\). Given any \(\underline{a}\in\mathcal{A}(\alpha)\), define
\[\beta_i \coloneqq\frac1{d_l}\sum_{b=0}^{w-1}\sum_{\underline t\in\mathcal T}
t_ba_{i-b,\underline t},\qquad 1\le i\le K.\]
 Thus $a_{k,\underline t}=\ell_{k,\underline t}/n$ and $\beta_i=e_i/(nd_l)$ for a feasible support type \(\underline{\ell}\). Define the asymptotic growth-rate functions  
\begin{align}
	g_{q,d_l,d_r,w,L}(\alpha)
	&\coloneqq\frac1N\max_{\underline a\in\mathcal A(\alpha)}
	\left\{\sum_{k,\underline t}p(\underline t)
	H_q\!\left(\frac{a_{k,\underline t}}{p(\underline t)}\right)
	+d_l\sum_{i=1}^{K}\left[
	\frac1{d_r}\log\inf_{z>0}\frac{W_{q,d_r}(z)}{z^{d_r\beta_i}}
	-H_q(\beta_i)\right]\right\},\label{gggggg}\\
	\widetilde g_{d_l,d_r,w,L}(\alpha)
	&\coloneqq\frac1N\max_{\underline a\in\mathcal A(\alpha)}
	\left\{\sum_{k,\underline t}p(\underline t)
	H_2\!\left(\frac{a_{k,\underline t}}{p(\underline t)}\right)
	+d_l\sum_{i=1}^{K}\left[
	\frac1{d_r}\log\inf_{z>0}\frac{\widetilde W_{d_r}(z)}{z^{d_r\beta_i}}
	-H_2(\beta_i)\right]\right\}.
	\label{ggggggg}
\end{align} 
Define the spectral thresholds by
\begin{align}
	\alpha_{q,d_l,d_r,w,L}&\coloneqq
	\inf\bigl(\{\alpha\in(0,1]:g_{q,d_l,d_r,w,L}(\alpha)\ge0\}\cup\{1\}\bigr),\notag\\
	\widetilde\alpha_{d_l,d_r,w,L}&\coloneqq
	\inf\bigl(\{\alpha\in(0,1]:\widetilde g_{d_l,d_r,w,L}(\alpha)\ge0\}\cup\{1\}\bigr).\notag
\end{align}
Roughly speaking, $\alpha_{q,d_l,d_r,w,L}$ can be treated as the smallest positive  zero of the function $g_{q,d_l,d_r,w,L}$ on $(0,1)$, and if no zero exists, then $\alpha_{q,d_l,d_r,w,L}=1$. A similar observation holds for  \(\widetilde\alpha_{d_l,d_r,w,L}\).
\begin{proposition} 
	\label{prop:sec3_spectral_bounds}
	The growth-rate functions \(g_{q,d_l,d_r,w,L}\) and \(\widetilde g_{d_l,d_r,w,L}\) are continuous on their  respective finite domains, with \(g_{q,d_l,d_r,w,L}(0)=\widetilde g_{d_l,d_r,w,L}(0)=0\). Uniformly for $0\le\ell\le Nn$,
	\begin{equation}\label{eq:sec3_uniform_spectral_bound}
	\begin{aligned}
		\mathbb E A_\ell(\mathcal C_{d_l,d_r,w,L,n})
		&\le\exp\!\left\{Nn\,g_{q,d_l,d_r,w,L}\!\left(\frac\ell{Nn}\right)+O(\log n)\right\}, \\
		\mathbb E\widetilde A_\ell(\mathcal G_{d_l,d_r,w,L,n})
		&\le\exp\!\left\{Nn\,\widetilde g_{d_l,d_r,w,L}\!\left(\frac\ell{Nn}\right)+O(\log n)\right\}. 
	\end{aligned}
	\end{equation}
	If $d_l\ge3$, then both $\alpha_{q,d_l,d_r,w,L}$ and $\widetilde\alpha_{d_l,d_r,w,L}$ are strictly positive. Put
	\[
	t_{\max}\coloneqq\max_{\underline t\in\mathcal T}\max_{0\le b\le w-1}t_b.
	\]
	Using the zeros \(\alpha_{q,d_l,d_r}\) and \(\widetilde{\alpha}_{d_l,d_r}\) in Theorem III.6, the following explicit bounds hold:
	\begin{align}
		\alpha_{q,d_l,d_r,w,L}
		&\ge\alpha_{\mathrm{lb}}\coloneqq
		\min\!\left\{1,\frac{d_l}{Nt_{\max}}\alpha_{q,d_l,d_r}\right\}\quad (d_r\ge d_l\ge3),\notag\\
		\widetilde\alpha_{d_l,d_r,w,L}
		&\ge\widetilde\alpha_{\mathrm{lb}}\coloneqq
		\min\!\left\{1,\frac{d_l}{Nt_{\max}}\widetilde\alpha_{d_l,d_r}\right\}  \quad (d_l,d_r\ge3).\notag
	\end{align}
\end{proposition}
For standard spreading in Example III.12, \(t_{\max}=d_l\), whereas for improved spreading in Example III.13  with \(w\ge d_l\),\,\(t_{\max}=1\). Thus, for fixed \(d_l,d_r,w,L\), the improved ensemble gives a factor-\(d_l\) gain in both explicit lower bounds before the cap at \(1\). This comparison concerns the proven lower bounds \(\alpha_{\mathrm{lb}}\) and \(\widetilde{\alpha}_{\mathrm{lb}}\), and does not imply an ordering of the exact   thresholds.
\begin{theorem}
Let \(q\ge2\), \(d_r\ge d_l\ge3\),   \(w\ge d_l\). Consider either the standard or improved \((d_l,d_r,w,L)\) coupled\,ensemble\,over \(\mathbb F_q\), with \(d_l,d_r,w,L\) fixed. For any  \(0<\alpha<\alpha_{q,d_l,d_r,w,L}\) and \(0<\widetilde\alpha<\widetilde\alpha_{d_l,d_r,w,L}\),
as \(n\to\infty\) through admissible block lengths,
\[
\Pr\!\left\{
d_{\min}(\mathcal C_{d_l,d_r,w,L,n})
\le N\alpha n
\right\}
=
\begin{cases}
	\Theta(n^{2-d_l}),
	&\text{improved ensemble, or standard ensemble with }q=2,\ d_l\text{ odd},\\[1mm]
	\Theta\!\left(n^{1-\lceil d_l/2\rceil}\right),
	&\text{otherwise},
\end{cases}
\] 
and
\[
\Pr\!\left\{
d_{\mathrm{ss}}(\mathcal G_{d_l,d_r,w,L,n})
\le N\widetilde\alpha n
\right\}
=
\begin{cases}
	\Theta(n^{2-d_l}),
	&\text{improved ensemble},\\[1mm]
	\Theta\!\left(n^{1-\lceil d_l/2\rceil}\right),
	&\text{standard ensemble}.
\end{cases}
\] 
Thus, both ensembles have asymptotically good minimum distance and minimum stopping-set size with probability tending to\,\(1\).
\end{theorem}
By Theorem III.16 and the Borel-Cantelli lemma, the following almost-sure achievability  result is immediate.
\begin{corollary}[Almost-sure distance and stopping-set bounds]
	\label{cor:sec3_distance_stopping_as}
	Consider the ensembles in the preceding theorem, realized for all admissible
	\(n\). If the ensemble is standard with
	\(d_l\ge5\), or improved with \(d_l\ge4\), then
	\begin{align}
		\liminf_{n\to\infty}
		\frac{d_{\min}(\mathcal C_{d_l,d_r,w,L,n})}{Nn} \ge\alpha_{q,d_l,d_r,w,L} ,\qquad
		\liminf_{n\to\infty}
		\frac{d_{\mathrm{ss}}(\mathcal G_{d_l,d_r,w,L,n})}{Nn}
	 \ge\widetilde\alpha_{d_l,d_r,w,L}
		\qquad\mathrm{a.s.}\notag
	\end{align}
\end{corollary}

For \(d_l\ge3\), Theorem III.16 already implies the achievability result in probability. The stronger degree conditions in Corollary III.17 ensure summability of the failure probabilities and thus the almost-sure assertion.
\begin{figure}[t]
	\centering
	\includegraphics[width=0.5\textwidth]{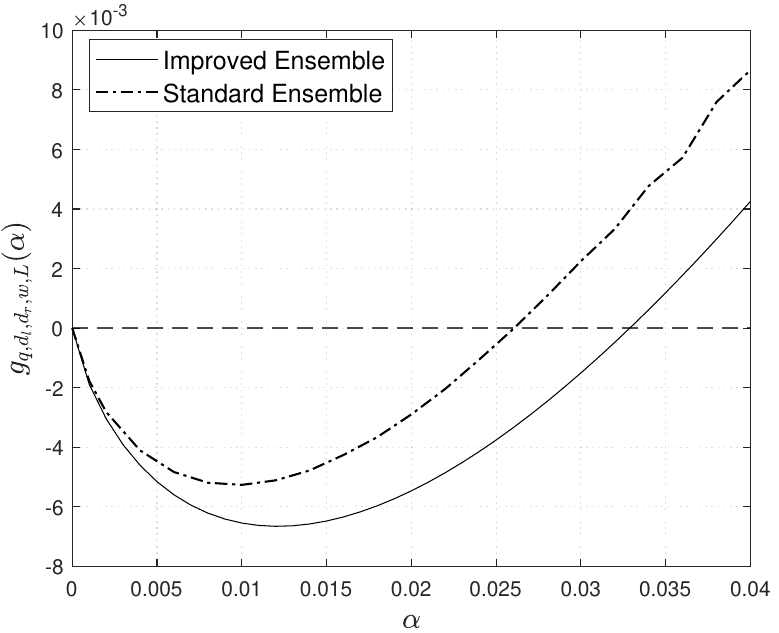} 
	\caption{The growth rate function $g_{q,d_l,d_r,w,L}(\alpha)$ of the average weight distribution of the two coupled $(d_l,d_r,w,L)$ ensembles over $\mathbb{F}_4$ for small normalized weight $\alpha$, where $(d_l,d_r,w,L)=(3,6,3,6)$. The smallest positive zero $\alpha_{q,d_l,d_r,w,L}$ is approximately $0.02607$  for the standard coupled ensemble and $0.03289$ for the improved  coupled ensemble.}
	\label{fig3}
\end{figure} 
\begin{figure}[t]
	\centering
	\includegraphics[width=0.5\textwidth]{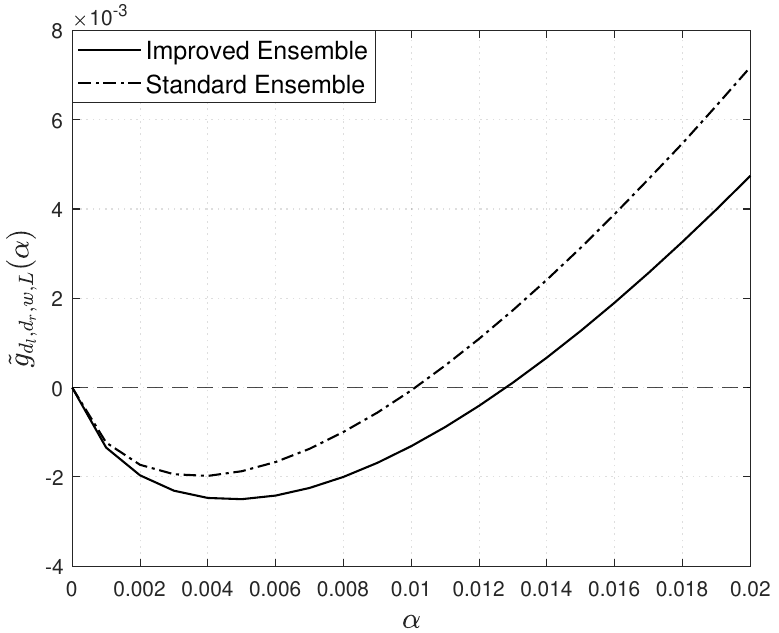} 
	\caption{The growth rate function $\tilde{g}_{d_l,d_r,w,L}(\alpha)$ of the average stopping set distribution of the two coupled $(d_l,d_r,w,L)$ ensembles for small normalized size $\alpha$, where $(d_l,d_r,w,L)=(3,6,3,6)$. The smallest positive zero $\tilde{\alpha}_{d_l,d_r,w,L}$ is  approximately $0.01011$ for the standard coupled ensemble and $0.01278$ for the improved coupled ensemble.}
	\label{fig4}
\end{figure} 

We close this subsection with numerical evaluations of the
minimum-distance and stopping-set growth-rate functions and bounds. Numerical asymptotic distance analyses have also been reported
for several related protograph-based ensembles.
For binary terminated $\mathcal C(3,6,L)$ SC-LDPC ensembles,
Mitchell \textit{et al.} obtained minimum-distance growth rates
of $0.0449$ and $0.0258$ for $L=6$ and $L=10$, respectively
\cite[Table I]{SCLDPC_minimum_distance2}.
For periodically time-varying binary $(3,6)$-regular
LDPC convolutional ensembles, Mitchell \textit{et al.}
reported a free-distance growth rate of approximately $0.086$
at period $T=12$ \cite[Sec. IV-C]{Mitchell2011FreeDistance}.
Huang \textit{et al.} further investigated free-distance
growth rates for protograph-based SC-LDPC ensembles over
$\mathbb F_q$ \cite{Huang2015Distance}.
The\,coupling\,length\\ in \cite{SCLDPC_minimum_distance2}
counts variable-node positions (corresponding to $N=2L$ in
our notation), whereas $T$ in \cite{Mitchell2011FreeDistance}
denotes the convolutional period.
Moreover, the reported minimum- and free-distance growth rates
are normalized by block length and decoding constraint length,
respectively. Our numerical evaluations illustrate how distinct edge-spreading
rules can affect the achievable distance bounds under identical parameters. Fig.~\ref{fig3} compares the minimum-distance growth-rate functions
$g_{q,d_l,d_r,w,L}(\alpha)$ of the standard and improved ensembles
over $\mathbb F_4$, with $(d_l,d_r,w,L)=(3,6,3,6)$.
The numerically estimated spectral thresholds
$\alpha_{q,d_l,d_r,w,L}$ are $0.02607$ and $0.03289$,
respectively.
Fig.~\ref{fig4} presents the corresponding stopping-set
growth-rate functions $\widetilde g_{d_l,d_r,w,L}(\alpha)$
for the same coupling parameters.
The estimated thresholds $\widetilde\alpha_{d_l,d_r,w,L}$
are $0.01011$ and $0.01278$ for the standard and improved
ensembles, respectively.
Thus, in both cases, the improved ensemble has a larger
estimated achievable bound. The distinction arises from the edge-spreading rule:
the improved ensemble directs the $d_l$ sockets of each
variable node to distinct check positions, whereas the
standard ensemble permits repeated offsets.
These examples illustrate that the choice of edge-spreading
rule can improve the achievable distance bounds even when
$d_l,d_r,w$, and $L$ are fixed. Tables~I and II provide further numerical results for the
improved $(d_l,d_r,L)$ ensemble, corresponding to $w=d_l$. In Tables~I and II, we write
$\alpha_{d_l,d_r,L}
\coloneqq\alpha_{4,d_l,d_r,d_l,L}$
and
$\widetilde\alpha_{d_l,d_r,L}
\coloneqq\widetilde\alpha_{d_l,d_r,d_l,L}$ for brevity.
Table~I reports the estimated achievable normalized minimum
distance over $\mathbb F_4$, together with the explicit
lower bound $\alpha_{\mathrm{lb}}$ and the asymptotic
design rate $r_{\mathrm d}$.
Under our convention $N=2L$ and with \(2L\ge d_l-1\), 
\begin{equation}
	r_{\mathrm d}
	=1-\frac{d_l}{d_r}
	-\frac{d_l}{d_r}
	\frac{d_l-1-
		2\sum_{i=1}^{d_l-1}(i/d_l)^{d_r}}{2L},\notag
\end{equation}
which follows by counting check nodes incident to physical
variables, as in \cite[Lem. 3]{SCLDPC1}.
Table~II reports the associated estimates of the
achievable normalized minimum stopping-set size and its
explicit lower bound $\widetilde\alpha_{\mathrm{lb}}$.
Since the two tables use identical ensemble parameters,
the design rates are omitted from Table~II.
For these improved ensembles, the explicit bounds are
\[
\alpha_{\mathrm{lb}}
=\frac{d_l}{2L}\alpha_{q,d_l,d_r},
\qquad
\widetilde\alpha_{\mathrm{lb}}
=\frac{d_l}{2L}\widetilde\alpha_{d_l,d_r},
\]
where the cap at $1$ in
Proposition~\ref{prop:sec3_spectral_bounds}
is inactive for the displayed parameters. The results show that the numerical spectral thresholds
are generally larger than their explicit lower bounds.
Nevertheless, the latter are substantially easier to
evaluate because they depend only on the growth-rate
functions of the underlying uncoupled ensembles.
The tables also illustrate the rate--distance tradeoff
associated with termination.
For example, when $(d_l,d_r)=(6,12)$, increasing $L$
from $8$ to $14$ increases the design rate from
$0.35126$ to $0.41500$, while the estimated
minimum-distance and stopping-set thresholds decrease
from $0.14994$ to $0.08572$ and from $0.04652$
to $0.02658$, respectively.
These   thresholds are estimated by numerically solving the nonconvex optimization problems in \(\eqref{gggggg}\) and \(\eqref{ggggggg}\). 
\begin{table}[htbp]
	\centering
	\caption{Numerical estimates of the minimum-distance   threshold and explicit lower bounds for the improved \((d_l,d_r,L)\) ensemble over \(\mathbb F_4\)}
	\resizebox{\linewidth}{!}{
		\begin{tabular}{cccccc|cccccc|cccccc}
			$d_l$ & $d_r$ & $L$ & $r_{\mathrm d}$
			& $\alpha_{d_l,d_r,L}$ & $\alpha_{\mathrm{lb}}$
			& $d_l$ & $d_r$ & $L$ & $r_{\mathrm d}$
			& $\alpha_{d_l,d_r,L}$ & $\alpha_{\mathrm{lb}}$
			& $d_l$ & $d_r$ & $L$ & $r_{\mathrm d}$
			& $\alpha_{d_l,d_r,L}$ & $\alpha_{\mathrm{lb}}$ \\
			\hline
			$3$ & $6$ & $6$ & $0.42410$ & $0.03289$ & $0.01163$
			& $4$ & $6$ & $10$ & $0.24626$ & $0.12558$ & $0.04539$
			& $6$ & $12$ & $8$ & $0.35126$ & $0.14994$ & $0.06364$ \\
			
			$4$ & $8$ & $6$ & $0.38367$ & $0.09450$ & $0.03895$
			& $6$ & $9$ & $10$ & $0.18145$ & $0.25337$ & $0.08371$
			& $6$ & $12$ & $10$ & $0.38101$ & $0.12105$ & $0.05091$ \\
			
			$5$ & $10$ & $6$ & $0.34279$ & $0.14847$ & $0.06342$
			& $8$ & $12$ & $10$ & $0.11579$ & $0.38449$ & $0.11558$
			& $6$ & $12$ & $12$ & $0.40084$ & $0.10004$ & $0.04242$ \\
			
			$6$ & $12$ & $6$ & $0.30168$ & $0.19992$ & $0.08485$
			& $10$ & $15$ & $10$ & $0.04975$ & $0.51818$ & $0.14555$
			& $6$ & $12$ & $14$ & $0.41500$ & $0.08572$ & $0.03636$
	\end{tabular}}
\end{table}
\begin{table}[htbp]
	\centering
	\caption{Numerical estimates of the stopping-set   threshold and explicit lower bounds for the improved \((d_l,d_r,L)\) ensemble}
	\begin{tabular}{ccccc|ccccc|ccccc} 
		$d_l$  & $d_r$ & $L$  &  $\tilde{\alpha}_{d_l,d_r,L}$ &$\tilde{\alpha}_{\mathrm{lb}}$  & $d_l$  & $d_r$ & $L$  &  $\tilde{\alpha}_{d_l,d_r,L}$ &$\tilde{\alpha}_{\mathrm{lb}}$ &$d_l$  & $d_r$ & $L$  &  $\tilde{\alpha}_{d_l,d_r,L}$ &$\tilde{\alpha}_{\mathrm{lb}}$\\
		\hline
		$3$ & $6$ & $6$ & $0.01278$ & $0.00450$ & $4$ & $6$ & $10$ & $0.04173$ & $0.01756$ & $6$ & $12$ & $8$ & $0.04652$ & $0.02363$\\ 
		$4$ & $8$ & $6$ & $0.03498$ & $0.01511$ & $6$ & $9$ & $10$ & $0.06179$ & $0.03041$ & $6$ & $12$ & $10$ & $0.03721$ & $0.01891$\\
		$5$ & $10$ & $6$ & $0.05061$ & $0.02415$ & $8$ & $12$ & $10$ & $0.07336$ & $0.03896$ & $6$ & $12$ & $12$ & $0.03104$ & $0.01575$\\
		$6$ & $12$ & $6$ & $0.06202$ & $0.03151$ & $10$ & $15$ & $10$ & $0.08152$ & $0.04540$ & $6$ & $12$ & $14$ & $0.02658$ & $0.01350$ 
	\end{tabular}
\end{table}

\section{Symmetric Channels and Density Evolution}
This section develops an analytical framework for density evolution (DE) of LDPC code ensembles over \(\mathbb F_q\) transmitted\,\,over \(q\)-ary memoryless symmetric channels (QMSCs). Our main objective is to establish the fundamental properties of the DE\,\,recursion
\[
\mathsf{x}_{\ell+1}
=
\mathsf{c}\circledast
\lambda^{\circledast}\bigl(\rho^{\boxast}(\mathsf{x}_{\ell})\bigr), 
\] 
where \(\mathsf c\) and \(\mathsf x_\ell\) represent the channel-message distribution and the variable-to-check message distribution, respectively. We\,\,first characterize symmetric message distributions through posterior probability vectors and establish their closure under BP updates. Then, we develop the relevant information functionals, metric topology, and degradation ordering, and investigate the continuity and monotonicity properties of DE. These theoretical tools allow us to establish the convergence of forward DE, characterize the BP threshold, and derive a stability condition for perfect iterative decoding. They also provide the theoretical foundation\,\,for the potential-function analysis and threshold saturation results in Section V.
\subsection{Channel Symmetry and Posterior Representation} 
Let \(q\ge 2\) be a prime power. Consider the affine group
\[
G\coloneqq 
\{g_{a,b}:i\mapsto a+bi
\mid a\in\mathbb F_q,\ b\in\mathbb F_q^\times\} ,
\] 
which is generated by the additive and multiplicative transformations of \(\mathbb{F}_q\). This group  acts doubly transitively on \(\mathbb F_q\): for any \(i,j,u,v\in\mathbb{F}_q\) with \(i\ne j\) and \(u\ne v\), the affine map \(g_{a,b}\) with \(b= (v-u)/(j-i)\) and \(a=u-bi\)  maps \((i,j)\) to \((u,v)\).
\begin{definition}[QMSC]
A memoryless channel (MC) \(W\) with input alphabet \(\mathbb F_q\) and output measurable space \((\mathcal Y,\mathcal A)\) is\,\,called a \(q\)-ary memoryless symmetric channel (QMSC) if \(G\) admits a measurable action on \(\mathcal Y\) such that
\[
W(gA\,|\, gi)=W(A\,|\, i)\qquad \forall g\in G, i\in\mathbb{F}_q, A\in\mathcal{A}.
\] 
Here, the action is assumed to be measurable in the sense that each map \(y\mapsto gy\) and its inverse are measurable.
\end{definition}

The following two channels, whose transition probabilities are invariant under the corresponding permutations of their input and output alphabets, are standard examples of QMSCs.

\begin{example}[\!\!\cite{Gallager_information}]
A \(q\)-ary symmetric channel (QSC) has input and output alphabets \(\mathbb F_q\) and transition probabilities
\[
W(y\,|\, x)=
\begin{cases}
	1-\epsilon, & y=x,\\[2pt]
	\dfrac{\epsilon}{q-1}, & y\ne x,
\end{cases}
\qquad x,y\in\mathbb F_q,
\] 
where \(0\le\epsilon\le(q-1)/q\) is the crossover probability.
\end{example}
\begin{example}[\!\!\cite{QPEC}]
A \(q\)-ary partial erasure channel (QPEC) with erasure size \(M\in\{2,\ldots,q\}\) and erasure probability\,\,\(\epsilon\in[0,1]\) has input alphabet \(\mathbb F_q\) and output alphabet
\[
\mathcal Y
=
\mathbb F_q\sqcup
\binom{\mathbb F_q}{M},
\]
where \(\binom{\mathbb F_q}{M}\) denotes the collection of all \(M\)-element subsets of \(\mathbb F_q\). Its transition probabilities are
\[
W(y\mid x)=
\begin{cases}
	1-\epsilon,
	& y=x\in\mathbb F_q,\\[2pt]
	\dfrac{\epsilon}{\binom{q-1}{M-1}},
	& y\in\binom{\mathbb F_q}{M},\ x\in y,\\[6pt]
	0,
	& \text{otherwise}.
\end{cases}
\] 
With probability \(\epsilon\), the channel output specifies a uniformly random \(M\)-element subset of \(\mathbb{F}_q\) containing the transmitted symbol. The case \(M=q\) reduces to the \(q\)-ary erasure channel (QEC).
\end{example}
For both channels, every permutation of \(\mathbb F_q\) induces a corresponding permutation of the output alphabet, acting on set-valued outputs by \(gS=\{gi:i\in S\}\). Their transition probabilities satisfy \(W(gy\,|\, gx)=W(y\,|\, x)\), so both channels are QMSCs.

We represent channel-output and decoding messages by a posteriori probability (APP) vectors. Let
\[
\mathcal S_q
\coloneqq
\left\{
\underline y\in[0,1]^{\mathbb F_q}:
\sum_{i\in\mathbb F_q}y_i=1
\right\}
\] 
denote the probability simplex. For a channel \(W\) with input alphabet \(\mathbb{F}_q\) and output space \(\mathcal{Y}\), let
\[
m\coloneqq\frac1q\sum_{i\in\mathbb F_q}W(\cdot\mid i)
\]
be its output distribution under uniform input. Since \(W(\cdot\,|\, i)\ll m\) for every \(i\), the APP operator \(\psi:\mathcal Y\to\mathcal S_q\) can be defined\,\,by
\[
[\psi(y)]_i
\coloneqq
\frac1q\frac{dW(\cdot\mid i)}{dm}(y),
\qquad i\in\mathbb F_q.
\]
The above expression is defined \(m\)-a.e.; on a common \(m\)-null set, we may assign \(\psi(y)\) the uniform probability vector. Thus,\,\,\(\psi\) is well-defined as an \(\mathcal S_q\)-valued measurable map. If \(X\) is uniform on \(\mathbb F_q\) and \(Y|X\sim W(\cdot\,|\,X)\), then
\[
[\psi(Y)]_i=\Pr\{X=i\mid Y\}
\qquad\text{a.s.}
\]
The affine group \(G\) induces a natural left action on \(\mathcal S_q\), given by \[(g\cdot\underline y)_i
\coloneqq y_{g^{-1}i}\qquad g\in G,  \ \underline{y}\in\mathcal{S}_q,\ i\in\mathbb{F}_q. \]   For a QMSC, channel symmetry implies that \(m\) is \(G\)-invariant, so by the uniqueness of Radon–Nikodym derivatives,  \([\psi(gy)]_i=[\psi(y)]_{g^{-1}i}\) for every \(g\in G\), \(i\in\mathbb{F}_q\) and \(m\)-a.e. \(y\). Equivalently, for every \(g\in G\),\[
\psi(gY)=g\cdot\psi(Y)
\qquad\text{a.s.}
\] 

We now characterize the   message distributions. Let \(\mathcal P(\mathcal S_q)\) denote the space of Borel probability measures on \(\mathcal S_q\), and define
\[
\overline{\mathcal X}_q
\coloneqq
\left\{
\mu\in\mathcal P(\mathcal S_q):
g_\#\mu=\mu\ \ \forall g\in G
\right\},
\]
where \(g_\#\mu\) denotes the pushforward measure of \(\mu\) under the above action. For every \(\mu\in\overline{\mathcal X}_q\), define
\[
(\Phi\mu)(B)
\coloneqq q\int_B y_0\,\mu(d\underline y),
\qquad B\in\mathcal B(\mathcal S_q).
\]
Since \(G\) acts transitively on \(\mathbb F_q\) and \(\sum_i y_i=1\), \(\int y_i\,\mu(d\underline y)=1/q\) for all \(i\in\mathbb{F}_q\), so \(\Phi\mu\) is a probability measure. Define\[
\mathcal X_q\coloneqq\Phi(\overline{\mathcal X}_q).
\]
For every \(\underline y\in\mathcal S_q\), define the coordinate reindexings
\[
(\underline y^{+a})_i\coloneqq y_{i+a},
\qquad
(\underline y^{\times b})_i\coloneqq y_{bi},
\qquad
a\in\mathbb F_q,\ b\in\mathbb F_q^\times.
\]
For every Borel set \(B\subseteq\mathcal S_q\), let \(B^{+a}\) and \(B^{\times b}\) denote its images under the above reindexings. For \(\mathsf x\in\mathcal X_q\), define
\[
(\Psi\mathsf x)(B)
\coloneqq
\frac1q\sum_{a\in\mathbb F_q}\mathsf x(B^{+a}).
\]
The following proposition shows that \(\Phi\) and \(\Psi\) are mutually inverse bijections between \(\overline{\mathcal X}_q\) and \(\mathcal X_q\). 
\begin{proposition}
Let \((X,Y)\) denote the input-output pair of a QMSC \(W\) with \(X\) uniformly distributed over \(\mathbb F_q\). Denote by
\[
\bar{\mathsf x}
\coloneqq\operatorname{Law}(\psi(Y)),
\qquad
\mathsf x_i
\coloneqq\operatorname{Law}(\psi(Y)\mid X=i),\qquad i\in\mathbb{F}_q
\]
the marginal and conditional distributions of the APP vector, respectively. Then \(\bar{\mathsf x}\in\overline{\mathcal X}_q\) and
\[
\mathsf x_i(B)
=
q\int_B y_i\,\bar{\mathsf x}(d\underline y)
\qquad\forall i\in\mathbb F_q.
\]
In particular, \(\mathsf x\coloneqq\mathsf x_0\in\mathcal X_q\), and
\[
\mathsf x=\Phi\bar{\mathsf x},
\qquad
\bar{\mathsf x}=\Psi\mathsf x.
\]
Conversely, every \(\mu\in\overline{\mathcal X}_q\) is the marginal APP distribution of a QMSC. Thus, \(\Phi\) and \(\Psi\) are mutually inverse bijections between \(\overline{\mathcal X}_q\) and \(\mathcal X_q\).
\end{proposition}
Proposition IV.4 shows that \(\mathcal X_q\) consists precisely of the APP distributions of QMSCs conditioned on transmission of \(0\in\mathbb F_q\), while \(\overline{\mathcal X}_q\) consists of their associated marginal APP distributions under uniform input. We call \(\mathsf x\in\mathcal{X}_q\) a symmetric distribution and \(\bar{\mathsf x}=\Psi\mathsf x\) the reference measure of \(\mathsf x\). Thus, \(\mathsf x\in\mathcal{X}_q\) and \(\bar{\mathsf x}=\Psi\mathsf x\in\overline{\mathcal{X}}_q\) uniquely determine each other: 
\[ 
	\mathsf x(B)
	=
	q\int_B y_0\,\bar{\mathsf x}(d\underline y),
	\qquad
	\bar{\mathsf x}(B)
	=
	\frac1q\sum_{a\in\mathbb F_q}\mathsf x(B^{+a}),\qquad \forall B\in\mathcal{B}(\mathcal{S}_q). 
\] 
The first identity implies \(y_0>0\) \(\mathsf x\)-a.e., whereas \(\bar{\mathsf x}\) may assign positive mass to \(\{y_0=0\}\). Moreover, since \(\bar{\mathsf x}\) is \(G\)-invariant,  
\[
\mathsf x(B^{+a})
=
q\int_B y_a\,\bar{\mathsf x}(d\underline y),
\qquad
\mathsf x(B^{\times b})=\mathsf x(B),\qquad \forall B\in\mathcal{B}(\mathcal{S}_q),\ a\in\mathbb{F}_q,\ b\in\mathbb{F}_q^\times.
\]
These identities will be used throughout the subsequent analysis.
\begin{definition}[Perfect and Uninformative Message Distributions]
Let \(e_i\in\mathcal S_q\) denote the \(i\)-th standard basis vector, and let \(u=(1/q,\ldots,1/q)\in\mathcal{S}_q\). The perfect and completely uninformative message distributions  are defined, respectively, by
\[ 
	\Delta_\infty\coloneqq\delta_{e_0},
	\qquad
	\Delta_0\coloneqq\delta_u, 
\]
where \(\delta_y\) denotes the Dirac measure at \(y\in\mathcal S_q\). Then, \(\Delta_\infty,\Delta_0\in \mathcal X_q\). Their corresponding reference probability measures are
\[
\Psi\Delta_\infty
=
\frac1q\sum_{i\in\mathbb F_q}\delta_{e_i},
\qquad
\Psi\Delta_0=\delta_u.
\]  
\end{definition}
A QMSC can be constructed from any \(\mathbb F_q\)-input MC \(W\) by augmenting its output with a uniformly random affine transformation. Specifically, let \(W\) have output space \((\mathcal Y,\mathcal A)\). Given an input \(i\in\mathbb F_q\), choose \(g\) uniformly from \(G\), transmit \(gi\) through\,\,\(W\), and reveal both \(g\) and the resulting output \(y\in\mathcal{Y}\) to the receiver. The resulting channel \(\widehat W\) has output space \((\mathcal Y\times G,\mathcal A\otimes 2^G)\) and transition kernel
\[
\widehat W(A\times\{g\}\,|\, i)
=
\frac1{|G|}  W(A\,|\, gi),\qquad A\in\mathcal{A},\ g\in G,\ i\in\mathbb{F}_q.
\]
Under the output action \(h\cdot(y,g)=(y,gh^{-1})\), the channel \(\widehat W\) is a QMSC. Since \(g\) is revealed to the receiver, this construction preserves the mutual information of \(W\) under uniform input. However, it does not imply identical   decoding performance over the original channel \(W\).  Related   ideas have been used in coding theory, e.g., the random-coset constructions in\cite{Coset1,Coset2,Coset3}.

\subsection{BP Updates on Message Distributions}
We   describe BP updates directly in terms of message distributions. On a computation tree, incoming messages from distinct branches are independent under the all-zero codeword assumption. The following operations allow us to track their distributions within \(\mathcal X_q\).
For \(\underline y,\underline z\in\mathcal S_q\), let
\[
D(\underline y,\underline z)
\coloneqq
\sum_{i\in\mathbb F_q}y_i z_i.
\]
The two-input variable-node and check-node APP message updates are defined, respectively, by
\[ 
	[\mathcal V(\underline y,\underline z)]_i \coloneqq
	\frac{y_i z_i}{D(\underline y,\underline z)} \quad \forall D(\underline y,\underline z)>0,\qquad
	[\mathcal C(\underline y,\underline z)]_i \coloneqq
	\sum_{ u,v\in\mathbb F_q:u+v=i} y_u z_v,\qquad i\in\mathbb F_q. 
\]
When \(D(\underline y,\underline z)=0\), we assign \(\mathcal V(\underline y,\underline z)\) the uniform probability vector. This convention is immaterial for the message distributions considered here, since \(y_0>0\) \(\mathsf x\)-a.e. for every \(\mathsf x\in\mathcal X_q\), and thus \(D(\underline y,\underline z)\ge y_0z_0>0\) \((\mathsf x_1\otimes\mathsf x_2)\)-a.e. for all \(\mathsf x_1,\mathsf x_2\in\mathcal X_q\). Under\,\,a uniform input prior, \(\mathcal V\) combines observations that are conditionally independent given the input symbol, while \(\mathcal C\) computes\,\,the APP of the sum of two independent input symbols observed through independent channels. For a check equation \(au+bv+ci=0\), with edge labels \(a,b,c\in\mathbb F_q^\times\),  the corresponding check-node update is given by
\[
[\mathcal C_{a,b,c}(\underline y,\underline z)]_i
\coloneqq
\sum_{ {u,v\in\mathbb F_q:au+bv=-ci}}
y_u z_v. 
\]
In particular, \(\mathcal C=\mathcal C_{1,1,-1}\).
\begin{definition}[BP convolutions]
For \(\mathsf x_1,\mathsf x_2\in\mathcal X_q\), define
\[ 
	\mathsf x_1\circledast\mathsf x_2 \coloneqq
	\mathcal V_\#(\mathsf x_1\otimes\mathsf x_2),\qquad 
	\mathsf x_1\boxast\mathsf x_2 \coloneqq
	\mathcal C_\#(\mathsf x_1\otimes\mathsf x_2). 
\] 
Thus, \(\circledast\) and \(\boxast\) describe the variable-node and check-node updates, respectively, for independent incoming messages.
\end{definition}
\begin{lemma}[Basic properties of \(\circledast,\boxast\)]\label{Closure,  Edge-Label Invariance, and Reference Distributions of BP Updates}Let \(\mathsf x_1,\mathsf x_2\in\mathcal X_q\), and write \(\bar{\mathsf x}_j=\Psi\mathsf x_j\) for \(j=1,2\). Then, 
\[\mathsf x_1\circledast\mathsf x_2\in\mathcal{X}_q,\qquad\mathsf x_1\boxast\mathsf x_2\in\mathcal X_q,\]
and, for all \(a,b,c\in\mathbb F_q^\times\),
\[
(\mathcal C_{a,b,c})_\#
(\mathsf x_1\otimes\mathsf x_2)
=
\mathsf x_1\boxast\mathsf x_2. 
\]
Moreover, for every bounded Borel function \(f:\mathcal S_q\to\mathbb R\),
\begin{align}
	\int f\,d\Psi(\mathsf x_1\circledast\mathsf x_2)&=
	q\int_{\mathcal S_q\times \mathcal{S}_q}
	D(\underline y,\underline z)
	f\bigl(\mathcal V(\underline y,\underline z)\bigr)
	\,\bar{\mathsf x}_1(d\underline y)
	\bar{\mathsf x}_2(d\underline z), \label{eq:f_dPsi_circledast} \\
	\int f\,d\Psi(\mathsf x_1\boxast\mathsf x_2)&=
	\int_{\mathcal S_q \times \mathcal S_q}
	f\bigl(\mathcal C(\underline y,\underline z)\bigr)
	\,\bar{\mathsf x}_1(d\underline y)
	\bar{\mathsf x}_2(d\underline z),\label{eq:f_dPsi_boxast}
\end{align}
where the   integrand in \eqref{eq:f_dPsi_circledast} is defined to be zero when \(D(\underline y,\underline z)=0\).
\end{lemma}
Lemma \ref{Closure,  Edge-Label Invariance, and Reference Distributions of BP Updates} shows that \(\mathcal X_q\) is closed under \(\circledast\) and \(\boxast\), and that   check-edge labels in \(\mathbb{F}_q^\times\) do not affect the resulting message distributions.  Both \(\circledast\) and \(\boxast\) are commutative and associative. Moreover, for every \(\mathsf x\in\mathcal X_q\),
\begin{equation}
\mathsf x\circledast\Delta_0
=\mathsf x, \qquad
\mathsf x\circledast\Delta_\infty
=\Delta_\infty,\qquad
\mathsf x\boxast\Delta_\infty =\mathsf x,\qquad
\mathsf x\boxast\Delta_0 =\Delta_0.  \label{eq:algebra}
\end{equation}
Hence, \((\mathcal X_q,\circledast)\) and \((\mathcal X_q,\boxast)\) are commutative monoids with identity elements \(\Delta_0\) and \(\Delta_\infty\), respectively. These properties are verified in Appendix II-C. For a degree-\(r\) variable node with channel-message distribution \(\mathsf c\) and independent input distributions \(\mathsf x_1,\ldots,\mathsf x_{r-1}\), the outgoing distribution is \(\mathsf c\circledast
\mathsf x_1\circledast\cdots\circledast\mathsf x_{r-1}\). The outgoing distribution of a degree-\(r\) check node is \(\mathsf x_1\boxast\cdots\boxast\mathsf x_{r-1}\),  independently of its nonzero edge labels. For later use, we adopt the conventions
\[
\mathsf x^{\circledast0}\coloneqq\Delta_0,
\qquad
\mathsf x^{\boxast0}\coloneqq\Delta_\infty,
\]
and let \(\mathsf x^{*k}\) denote the \(k\)-fold convolution for \(*\in\{\circledast,\boxast\}\). For a polynomial \(p(t)=\sum_{k=0}^d p_kt^k\) with \(p_k\geq0\) and \(\sum_kp_k=1\),\,put
\[
p^*(\mathsf x)
\coloneqq
\sum_{k=0}^{d}p_k\mathsf x^{*k}. 
\] 

\subsection{Linear Functionals and Metric Topology}
Let \(\mathcal M_q
\coloneqq
\operatorname{span}_{\mathbb R}\mathcal X_q\) be the space of finite signed measures generated by \(\mathcal X_q\). The map \(\Psi\) extends linearly to \(\mathcal M_q\) through
\[
(\Psi\alpha)(B)
=
\frac1q\sum_{a\in\mathbb F_q}\alpha(B^{+a}),
\qquad \alpha\in\mathcal M_q,\ B\in\mathcal{B}(\mathcal{S}_q).
\]
Likewise, both \(\circledast\) and \(\boxast\) extend bilinearly to \(\mathcal M_q\) through
\[
\alpha\circledast\beta
=
\mathcal V_\#(\alpha\otimes\beta),
\qquad
\alpha\boxast\beta
=
\mathcal C_\#(\alpha\otimes\beta)\qquad \alpha,\beta\in\mathcal{M}_q.
\]
These extensions are justified in Appendix II-D. For every bounded Borel function \(f:\mathcal S_q\to\mathbb R\), define the linear functional
\[
F_f(\alpha)
\coloneqq
\int_{\mathcal S_q}f(\underline y)
\,(\Psi\alpha)(d\underline y),
\qquad \alpha\in\mathcal M_q.
\]
The function \(f\) is called a kernel of \(F_f\). A functional may admit distinct kernels. Consider the following functionals.
\begin{definition}[Information functionals]
For \(\mathsf x\in\mathcal X_q\), let \(\bar{\mathsf x}=\Psi\mathsf x\). Define the Bhattacharyya functional \(\mathfrak B\), entropy functional\,\,\(\mathrm H\), pairwise MAP  error functional   \(\mathfrak P\), and squared-norm functional \(\mathrm Q\) by 
\begin{align*}
	\mathfrak B(\mathsf x)
	 \coloneqq
	q\int\sqrt{y_0y_1}\,d\bar{\mathsf x},\qquad \mathrm H(\mathsf x) \coloneqq
	-\int\sum_{i\in\mathbb F_q}y_i\log y_i
	\,d\bar{\mathsf x},\qquad
	\mathfrak P(\mathsf x) \coloneqq
	\frac q2\int\min\{y_0,y_1\}\,d\bar{\mathsf x},\qquad
	\mathrm Q(\mathsf x) \coloneqq
	\int\|\underline y\|^2\,d\bar{\mathsf x},
\end{align*}  
where \(0\log0=0\). These definitions extend linearly to \(\mathcal M_q\).
\end{definition}
For the QMSC described by \(\mathsf x\), the functionals \(\mathfrak B\) and \(\mathrm H\) give its   Bhattacharyya parameter and  conditional entropy, respectively. The functional \(\mathfrak P\) equals the minimum error rate for distinguishing any two distinct input symbols under equal priors; for \(q=2\), it coincides with the MAP bit error rate. Since \(\bar{\mathsf x}\) is \(G\)-invariant, \(\mathfrak B\) and \(\mathfrak P\) also admit permutation-invariant kernels:
\[ 
	\mathfrak B(\mathsf x) =
	\frac1{q-1}\int
	\sum_{ {i,j\in\mathbb F_q:i\ne j}}
	\sqrt{y_iy_j}\,d\bar{\mathsf x},\qquad
	\mathfrak P(\mathsf x) =
	\frac1{2(q-1)}\int
	\sum_{ {i,j\in\mathbb F_q: i\ne j}}
	\min\{y_i,y_j\}\,d\bar{\mathsf x}. 
\]
The kernels of \(\mathfrak B\), \(\mathrm H\), and \(\mathfrak P\) can be chosen concave, strictly so for \(\mathfrak B\) and \(\mathrm H\), while \(\mathrm Q\) has a strictly convex kernel.
\begin{lemma}[Extremal values]
For every \(\mathsf x\in\mathcal X_q\),
\[ 
	0\le\mathfrak B(\mathsf x)\le1,\qquad
	0\le\mathrm H(\mathsf x)\le\log q,\qquad
	0\le\mathfrak P(\mathsf x)\le\frac12,\qquad 
	\frac1q\le\mathrm Q(\mathsf x)\le1. 
\] 
For \(\mathfrak{B},\mathrm{H}\) and \(\mathfrak{P}\), the lower and upper bounds are attained if and only if \(\mathsf x=\Delta_\infty\) and \(\mathsf x=\Delta_0\). The reverse holds for \(\mathrm Q\). 
\end{lemma}
Further properties, bounds, and large-deviation results for \(\mathfrak B\) and \(\mathfrak P\) are given in Appendix II-F.
\begin{lemma}[Entropy duality]
For all \(\mathsf x_1,\mathsf x_2\in\mathcal X_q\),
\[ 
	\mathrm H(\mathsf x_1\circledast\mathsf x_2)
	+
	\mathrm H(\mathsf x_1\boxast\mathsf x_2)
	=
	\mathrm H(\mathsf x_1)+\mathrm H(\mathsf x_2). 
\] 
More generally, \(\mathrm H(\alpha\circledast\beta)
+\mathrm H(\alpha\boxast\beta) =
\alpha(\mathcal S_q)\mathrm H(\beta)
+\beta(\mathcal S_q)\mathrm H(\alpha)\) for all \(\alpha,\beta\in\mathcal M_q\). If \(\alpha(\mathcal{S}_q)=\beta(\mathcal{S}_q)=0\), then 
\[\mathrm H(\alpha\circledast\beta)
+
\mathrm H(\alpha\boxast\beta)=0.\]
\end{lemma}
To study convergence of message distributions, let \(W_2\) denote the \(2\)-Wasserstein metric induced by the Euclidean metric on \(\mathcal S_q\), and define the induced metric 
\[
d_W(\mathsf x_1,\mathsf x_2)
\coloneqq
W_2(\Psi\mathsf x_1,\Psi\mathsf x_2),
\qquad
\mathsf x_1,\mathsf x_2\in\mathcal X_q.
\]
\begin{proposition}[Compactness and continuity]We have the following topological results.
\begin{enumerate}
\item Both \((\overline{\mathcal X}_q,W_2)\) and \((\mathcal X_q,d_W)\) are compact metric spaces. Moreover, \(d_W\) metrizes weak convergence on \(\mathcal X_q\).
\item  For every continuous \(f:\mathcal S_q\to\mathbb R\), the functional \(F_f\) is continuous on \((\mathcal X_q,d_W)\)  and attains its extrema.
\item Both \(\circledast\) and \(\boxast\) are jointly continuous: if \(\mathsf x_{j,n}\xrightarrow{d_W}\mathsf x_j\) for \(j=1,2\), then
\[
\mathsf x_{1,n}*\mathsf x_{2,n}
\xrightarrow{d_W}
\mathsf x_1*\mathsf x_2,
\qquad
*\in\{\circledast,\boxast\}.
\] 
Thus, finite compositions and convex combinations of BP convolutions are continuous.
\end{enumerate}
\end{proposition}
  
 \subsection{Degradation   of Symmetric Probability Measures}
 Degradation provides a partial order on message distributions according to their informativeness. For \(\mathsf x\in\mathcal X_q\), let \(\bar{\mathsf x}=\Psi\mathsf{x}\)\,\,and \(W_{\mathsf x}\) denote the canonical QMSC with output space \((\mathcal S_q,\mathcal{B}(\mathcal{S}_q))\) defined by
 \[
 W_{\mathsf x}(B\,|\, i)
 \coloneqq
 q\int_B y_i\,\bar{\mathsf x}(d\underline y),
 \qquad
 i\in\mathbb F_q,\quad
 B\in\mathcal B(\mathcal S_q).
 \] 
 Under uniform input, its output is an APP vector with distribution \(\bar{\mathsf x}\) and conditional distribution \(\mathsf x\) when \(0\in\mathbb{F}_q\) is transmitted.
 \begin{definition}[Degradation] 
 For \(\mathsf x_1,\mathsf x_2\in\mathcal X_q\), we write \(\mathsf x_1\preceq\mathsf x_2\) if there exists a Markov kernel \(K\) on \(\mathcal S_q\) such that
 	\[
 	W_{\mathsf x_2}(B\,|\, i)
 	=
 	\int_{\mathcal S_q}
 	K(B\,|\,\underline y)\,
 	W_{\mathsf x_1}(d\underline y\,|\, i)
 	\]
for all \(i\in\mathbb F_q\) and  \(B\in\mathcal B(\mathcal S_q)\). In this case, \(\mathsf x_2\) is degraded with respect to \(\mathsf x_1\). The kernel \(K\) can be chosen \(G\)-equivariant,\,\,i.e., \(K(gB\,|\, g\cdot\underline y)
=
K(B\,|\,\underline y)\) for all \(g\in G\), \(\underline y\in\mathcal S_q\), and \(B\in\mathcal B(\mathcal S_q)\). We write \(\mathsf x_1\prec\mathsf x_2\) when \(\mathsf x_1\preceq\mathsf x_2\) and \(\mathsf x_1\ne\mathsf x_2\). 
 \end{definition}
 \begin{theorem}[Characterizations of degradation]
 Let \(\mathsf x_1,\mathsf x_2\in\mathcal X_q\) and \(\bar{\mathsf x}_j=\Psi\mathsf x_j\), \(j=1,2\). The following are equivalent:
 \begin{enumerate}[label= \roman*)]
\item \(\mathsf x_1\preceq\mathsf x_2\).
\item   There exists a coupling \((\underline Y,\underline Z)\) such that \(\underline Y\sim\bar{\mathsf x}_1\), \(\underline Z\sim\bar{\mathsf x}_2\) and \(\mathbb E[\underline Y\,|\,\underline Z]
=\underline Z\) a.s.
\item   For every bounded concave \(f:\mathcal S_q\to\mathbb R\),
\[
\int f\,d\bar{\mathsf x}_1
\le
\int f\,d\bar{\mathsf x}_2.
\]
\item  For every  \(\eta\in\mathcal P_2(\mathbb R^q)\),
\[
W_2(\bar{\mathsf x}_1,\eta)^2
-
W_2(\bar{\mathsf x}_2,\eta)^2
\le
\mathrm Q(\mathsf x_1)-\mathrm Q(\mathsf x_2).
\]
 \end{enumerate}
 \end{theorem}
By Theorem IV.13, \(\preceq\) is a partial order on \(\mathcal X_q\), with \( \Delta_\infty\preceq\mathsf x\preceq\Delta_0\) for \(\mathsf x\in\mathcal X_q\). It is preserved under convex combinations: if \(\mathsf x_j\preceq\mathsf y_j\), \(j=1,2\), then \(\lambda\mathsf x_1+(1-\lambda)\mathsf x_2
\preceq
\lambda\mathsf y_1+(1-\lambda)\mathsf y_2\) for every \(0\le\lambda\le1\).
\begin{proposition}[Strict monotonicity and metric inequalities]
Let \(\mathsf{x}_1,\mathsf{x}_2\in\mathcal{X}_q\) satisfy \(\mathsf{x}_1\preceq\mathsf{x}_2\). Then,
\begin{equation}
d_W(\mathsf x_1,\mathsf x_2)^2
\le
\mathrm Q(\mathsf x_1)-\mathrm Q(\mathsf x_2)
\le
\mathrm H(\mathsf x_2)-\mathrm H(\mathsf x_1).\label{eq:dQH}
\end{equation}
Moreover, if \(\mathsf x_1\prec\mathsf x_2\), then  \(F_f(\mathsf x_1)<F_f(\mathsf x_2)\) for every bounded, strictly concave \(f:\mathcal S_q\to\mathbb R\).
\end{proposition}
The next decomposition lemma relates an arbitrary  distribution in \(\mathcal{X}_q\) to a QPEC, which  will be useful in the stability\,analysis.
 \begin{lemma}[Partial-erasure decomposition]
 Let \(\mathsf e_\epsilon\in\mathcal X_q\) denote the conditional APP distribution of a QPEC with erasure size \(2\) and erasure probability \(\epsilon\). Then, for every \(\mathsf x\in\mathcal X_q\),
 \[
\mathsf e_{2\mathfrak P(\mathsf x)}
\preceq
\mathsf x.
 \]
 \end{lemma}
The next proposition shows that degradation is preserved under  \(\circledast,\boxast\); moreover, strict degradation can be preserved under\,\,\(\circledast\).
\begin{proposition}[BP monotonicity]
For \(\mathsf x_1,\mathsf x_2,\mathsf y_1,\mathsf y_2\in\mathcal X_q\), if \(\mathsf x_j\preceq\mathsf y_j\) for \(j=1,2\), then
\[
\mathsf x_1*\mathsf x_2
\preceq
\mathsf y_1*\mathsf y_2,
\qquad
*\in\{\circledast,\boxast\}.
\]
Moreover, if \(\mathsf x_1\prec\mathsf x_2\) and \(\mathsf z\in\mathcal{X}_q\setminus\{\Delta_\infty\}\), then \(\mathsf x_1\circledast\mathsf z
\prec
\mathsf x_2\circledast\mathsf z\).
\end{proposition}
Combining BP monotonicity with entropy duality, we obtain the following entropy bounds for signed measures.
\begin{proposition}[Signed entropy inequalities]
Let \(\mathsf x,\mathsf x'\in\mathcal X_q\) satisfy \(\mathsf x\preceq\mathsf x'\). For every \(\mathsf z,\mathsf z_1,\mathsf z_2,\mathsf z_3\in\mathcal X_q\) and \(*\in\{\circledast,\boxast\}\),
\[
0
\le
\mathrm H\bigl((\mathsf x'-\mathsf x)*\mathsf z\bigr)
\le
\mathrm H(\mathsf x'-\mathsf x),\quad\left|
\mathrm H\bigl((\mathsf x'-\mathsf x)*
(\mathsf z_1-\mathsf z_2)\bigr)
\right|
\le
\mathrm H(\mathsf x'-\mathsf x),\quad\left|
\mathrm H\bigl((\mathsf x'-\mathsf x)*
(\mathsf z_1-\mathsf z_2)*\mathsf z_3\bigr)
\right|
\le
\mathrm H(\mathsf x'-\mathsf x).
\] 
\end{proposition}
The following topological properties induced by degradation will be used in analyzing DE sequences.
\begin{proposition} 
The partial order \(\preceq\) is closed and satisfies the monotone convergence property:
	\begin{enumerate}[label=\roman*)]
	\item  If \(\mathsf x_n\preceq\mathsf y_n\ \forall n\), \(\mathsf x_n\xrightarrow{d_W}\mathsf x\) and \(\mathsf y_n\xrightarrow{d_W}\mathsf y\), then \(\mathsf x\preceq\mathsf y\).
	\item If \(\mathsf x_n\preceq(\succeq)\mathsf x_{n+1}\ \forall n\), then \(\mathsf x_n\xrightarrow{d_W}\mathsf x\) for some \(\mathsf x\in\mathcal X_q\), with \(\mathsf x_n\preceq(\succeq)\mathsf x \ \forall n\). 
	\end{enumerate} 
\end{proposition}
\begin{definition}
A QMSC family \(\{\mathsf c_h:0\le h\le\log q\}
\subset\mathcal X_q\) is said to be complete and ordered by degradation if
\[
\mathrm H(\mathsf c_h)=h,
\qquad
h_1\le h_2
\ \Longrightarrow\
\mathsf c_{h_1}\preceq\mathsf c_{h_2}.
\]
Then, \(\mathsf c_0=\Delta_\infty\),  \(\mathsf c_{\log q}=\Delta_0\), and the degradation is strict whenever \(h_1<h_2\).
\end{definition}
 We will consider such a  QMSC family in the subsequent threshold analysis. By \eqref{eq:dQH}, the following   is immediate.
\begin{corollary}
Every complete ordered QMSC family \(\{\mathsf c_h:0\le h\le\log q\}\) satisfies \(d_W(\mathsf c_{h_1},\mathsf c_{h_2})
\le
\sqrt{|h_1-h_2|}\).
Therefore, \(h\mapsto\mathsf c_h\) is continuous, its image is compact, and \(h\mapsto F(\mathsf c_h)\) is continuous for every continuous functional \(F:\mathcal X_q\to\mathbb{R}\).
\end{corollary}
\subsection{Density Evolution and Stability of the Uncoupled System}
Consider an uncoupled LDPC code ensemble over \(\mathbb F_q\) with edge-perspective degree profile \((\lambda,\rho)\), as defined in Section III-A. Under all-zero transmission over a QMSC with message distribution \(\mathsf c\in\mathcal X_q\), density evolution (DE) can be described by
\[
\mathsf x_{\ell+1}
=
\mathsf T_s(\mathsf x_\ell;\mathsf c),
\qquad
\mathsf T_s(\mathsf x;\mathsf c)
\coloneqq
\mathsf c\circledast
\lambda^{\circledast}
\bigl(\rho^{\boxast}(\mathsf x)\bigr),
\] 
where
\[
\lambda^{\circledast}(\mathsf x) =
\sum_{i\ge2}\lambda_i
\mathsf x^{\circledast(i-1)},\qquad
\rho^{\boxast}(\mathsf x) =
\sum_{j\ge2}\rho_j
\mathsf x^{\boxast(j-1)}.
\] 
Let \(\mathsf T_s^{(\ell)}\) denote the \(\ell\)-fold iteration of \(\mathsf T_s\), with \(\mathsf T_s^{(0)}(\mathsf x;\mathsf c)=\mathsf x\). The edge-label invariance of the check-node update ensures\,\,that the DE operator is independent of the edge labels. A distribution \(\mathsf x\in\mathcal X_q\) is called a \textit{DE fixed point} if
\[
\mathsf T_s(\mathsf x;\mathsf c)=\mathsf x.
\]
In particular, \(\Delta_\infty\) is always a fixed point.
\begin{proposition} [Monotonicity and convergence of uncoupled DE]
For any \(\mathsf c,\mathsf c',\mathsf x,\mathsf x'\in\mathcal X_q\), the following hold.
\begin{enumerate}[label=\roman*)]
\item If \(\mathsf x\preceq\mathsf x'\) and \(\mathsf c\preceq\mathsf c'\), then \(\mathsf T_s^{(\ell)}(\mathsf x;\mathsf c)
\preceq
\mathsf T_s^{(\ell)}(\mathsf x';\mathsf c')\ \forall \ell\ge0\).
\item If \(\mathsf T_s(\mathsf x;\mathsf c)\preceq\mathsf x\) or \(\mathsf T_s(\mathsf x;\mathsf c)\succeq\mathsf x\), then the sequence \( \{\mathsf T_s^{(\ell)}(\mathsf x;\mathsf c) \}_{\ell\ge0}\) is monotone under degradation and converges in \(d_W\) to a DE fixed point.
\item The forward DE sequence, initialized at \(\mathsf x_0=\Delta_0\), satisfies \(\Delta_\infty
\preceq
\mathsf x_{\ell+1}
\preceq
\mathsf x_\ell
\preceq
\Delta_0\)  and converges in \(d_W\) to a fixed point \(\mathsf x_\infty(\mathsf c)
\coloneqq
\lim_{\ell\to\infty}
\mathsf T_s^{(\ell)}(\Delta_0;\mathsf c)\). Moreover, \(\mathsf x_\infty(\mathsf c)\) is the greatest DE fixed point under degradation. Thus, \(\mathsf x_\infty(\mathsf c)=\Delta_\infty\) if and only if \(\Delta_\infty\) is  the unique DE fixed point.
\end{enumerate} 
\end{proposition}
Let \(\{\mathsf c_h:0\le h\le\log q\}\) be a complete QMSC family ordered by degradation. Proposition IV.21 implies that \(h\mapsto\mathsf x_\infty(\mathsf c_h)\)\,\,is nondecreasing under \(\preceq\). Thus, the BP threshold of the uncoupled ensemble is defined by
\[
h_s^{\mathrm{BP}}
\coloneqq
\sup\left\{
h\in[0,\log q]:
\mathsf x_\infty(\mathsf c_h)=\Delta_\infty
\right\}.
\]
Forward DE converges to \(\Delta_\infty\) for every \(h<h_s^{\mathrm{BP}}\), whereas it does not for \(h>h_s^{\mathrm{BP}}\).  The following theorem gives stability\,\,and instability criteria for the perfect-decoding fixed point, generalizing\cite[Thm. 4.127]{ModernCode}.
\begin{theorem} 
Let \(\mathsf c\in\mathcal X_q\), and define \(\alpha
\coloneqq
\mathfrak B(\mathsf c)\lambda'(0)\rho'(1)\). For any \(\mathsf x_0\in\mathcal X_q\), consider the DE sequence \(\mathsf x_{\ell+1}
=
\mathsf T_s(\mathsf x_\ell;\mathsf c)\).
\begin{enumerate}
\item \textbf{Stability}.  If \(\alpha<1\), there exist \(\kappa>0\) and \(\theta\in(0,1)\), depending only on \((\lambda,\rho,\mathsf c)\), such that, if \(\mathfrak B(\mathsf x_{\ell_0})\le\kappa\)
for some\,\,\(\ell_0\ge 0\), then \(\mathfrak B(\mathsf x_\ell)
\le
\theta^{\ell-\ell_0}
\mathfrak B(\mathsf x_{\ell_0})\) for all \(\ell\ge\ell_0\), and     \(\mathsf x_\ell\xrightarrow{d_W}\Delta_\infty\).
\item \textbf{Instability}. If \(\alpha>1\), there exists a constant \(\xi>0\), depending only on \((\lambda,\rho,\mathsf c)\), such that for every \(\mathsf x_0\in\mathcal X_q\setminus\{\Delta_\infty\}\),
\[\liminf_{\ell\to\infty}
\mathfrak P(\mathsf x_\ell)\ge\xi.\] 
\end{enumerate}    
\end{theorem}
We call the uncoupled system \((\lambda,\rho,\mathsf c)\) stable if \(\mathfrak B(\mathsf c)\lambda'(0)\rho'(1)<1\). In particular, if the minimum variable-node degree\,\,is\,\,at least \(3\), then \(\lambda'(0)=0\), and   the system \((\lambda,\rho,\mathsf c)\) is stable for every \(\mathsf c\in\mathcal{X}_q\).  For a complete QMSC family \(\{\mathsf c_h:0\le h\le\log q\}\) ordered by degradation, since \(h\mapsto\mathsf c_h\) and \(\mathfrak B\) are continuous (see Corollary IV.20), stability at a given channel entropy\,\,\(h\in (0,\log q)\) also holds in a neighborhood of \(h\).
\begin{corollary} [Compactness of nontrivial fixed points]
For \(\mathsf c\in\mathcal X_q\), define the set of nontrivial DE fixed points by
\[
\mathcal F(\mathsf c)
\coloneqq
\left\{
\mathsf x\in\mathcal X_q\setminus\{\Delta_\infty\}:
\mathsf T_s(\mathsf x;\mathsf c)=\mathsf x
\right\}.
\]
If \(\mathfrak B(\mathsf c)\lambda'(0)\rho'(1)<1\), then \(\mathcal F(\mathsf c)\) is compact (possibly empty) and separated from \(\Delta_\infty\). More generally, if \(\mathcal C\subseteq\mathcal X_q\) is compact and \(\sup_{\mathsf c\in\mathcal C}
\mathfrak B(\mathsf c)\lambda'(0)\rho'(1)<1\), then there exists \(\kappa_{\mathcal C}>0\) such that \(\mathfrak B(\mathsf x)\ge\kappa_{\mathcal C}\) for all \(\mathsf c\in\mathcal C\) and \(\mathsf x\in\mathcal F(\mathsf c)\), and moreover, the set \(\left\{
(\mathsf c,\mathsf x):
\mathsf c\in\mathcal C,\ 
\mathsf x\in\mathcal F(\mathsf c)
\right\}\) is compact. 
\end{corollary}
\section{Threshold Saturation of Coupled Ensembles}
In this section, we establish threshold saturation for spatially coupled LDPC ensembles over \(\mathbb F_q\) on general complete QMSC families ordered by degradation. We first formulate the coupled DE update using the random offset vector \(\boldsymbol J\sim\pi_w\), and\,\,establish its connection to finite-length BP decoding via concentration.  Then, we introduce the potential functional and the fixed-point threshold of the underlying uncoupled system, and prove threshold saturation under an edge-spreading regularity condition and the stability condition on the uncoupled system. The former holds for both the standard and improved coupled  ensembles,\,while the latter is automatic when \(d_l\ge3\). 
\subsection{Coupled Density Evolution and Concentration}
Consider a \((d_l,d_r,w,L)\) coupled ensemble over \(\mathbb F_q\)  as defined in Section III-B, with \(N\coloneqq2L\) physical variable-node\,positions and \(K\coloneqq N+w-1\) check-node positions. Recall that the exchangeable random vector 
\[\boldsymbol J=(J_1,\ldots,J_{d_l})\sim\pi_w,\qquad J_a\in \{0,1,\ldots,w-1\}\]
 records the check-node offsets of a variable node’s edges in a uniformly random order.   
 Moreover, by \eqref{eq:sec3_rooted_offset}, if an incoming edge is
 selected uniformly at any check-node position, designated as the
 first edge of its variable endpoint, and the remaining edges are
 ordered uniformly at random, then the resulting offset vector also has
 law $\pi_w$, with virtual variable nodes included.
 Thus, $\pi_w$ determines the spatial averaging in the coupled
 DE operator below. 

Given a conditional APP distribution \(\mathsf c\in\mathcal X_q\),   define the effective channel-message distribution at variable-node position \(i\)\,\,by
\[
\mathsf c_i=
\begin{cases}
	\mathsf c,&1\le i\le N,\\
	\Delta_\infty,&\text{otherwise}.
\end{cases}
\] 
Let \(\underline{\mathsf x}=(\mathsf x_1,\ldots,\mathsf x_K)\in\mathcal X_q^K\) denote the vector of check-node-input   distributions, where \(\mathsf x_i\in\mathcal{X}_q\) denotes the conditional  APP distribution of a uniformly selected variable-to-check message entering a check node at position \(i\). We also set \(\mathsf x_i=\Delta_\infty\)\,outside \(1,\ldots,K\). Under the ideal cycle-free assumption, the coupled DE operator \(\mathsf T_c:\mathcal X_q^K\times\mathcal X_q
\longrightarrow\mathcal X_q^K\) is given by
\begin{equation} 
		[\mathsf T_c(\underline{\mathsf x};\mathsf c)]_i\coloneqq
		\mathbb E_{\boldsymbol{J}} \left[
		\mathsf c_{i-J_1}
		\circledast\left(\mathop{\circledast}\limits_{a=2}^{d_l}
		\mathsf x_{i-J_1+J_a}^{\boxast(d_r-1)}\right)
		\right],\qquad 1\le i\le K.\label{eq:coupled_DE}
\end{equation}
where the expectation is taken with respect to the fixed distribution \(\pi_w\). To see this, consider a uniformly selected check-node-input edge at position \(i\). Its variable-node endpoint is at position \(i-J_1\), and the other \(d_l-1\) edges of that variable node\,\,connect\\ to check-node positions \(i-J_1+J_a\) for \(2\le a\le d_l\). Conditioned on \(\boldsymbol J\),  the incoming check-node messages on the computation tree are independent under the all-zero-codeword assumption, with distributions \(\mathsf x_{i-J_1+J_a}^{\boxast(d_r-1)}\), \(2\le a\le d_l\). Combining these\,\,messages with the channel message and averaging over \(\boldsymbol J\) gives  \eqref{eq:coupled_DE}. We have the following examples.
\begin{example}
For the standard coupled ensemble, \(J_1,\ldots,J_{d_l}\) are independent and uniform on \(\{0,\ldots,w-1\}\), so \eqref{eq:coupled_DE} becomes
\[
[\mathsf T_c(\underline{\mathsf x};\mathsf c)]_i
 =
\frac1w\sum_{k=0}^{w-1}
\mathsf c_{i-k}\circledast 
\left(
\frac1w\sum_{j=0}^{w-1}
\mathsf x_{i-k+j}^{\boxast(d_r-1)}
\right)^{\circledast(d_l-1)}.
\]
For the improved coupled ensemble,   \(\boldsymbol J\) is uniform over all ordered \(d_l\)-tuples of distinct offsets, so \eqref{eq:coupled_DE} becomes
\[
[\mathsf T_c(\underline{\mathsf x};\mathsf c)]_i =
\frac{1}{w\binom{w-1}{d_l-1}}
\sum_{k=0}^{w-1}
\sum_{\substack{
		S\subseteq\{0,\ldots,w-1\}\setminus\{k\}\\
		|S|=d_l-1}}
\mathsf c_{i-k}
\circledast
\left(\mathop{\circledast}\limits_{j\in S}
\mathsf x_{i-k+j}^{\boxast(d_r-1)}\right).
\]
\end{example}

Throughout, ties are broken uniformly by independent random seeds for distinct hard decisions, with all the seeds independent of the random code and channel outputs. For a linear code over \(\mathbb{F}_q\) transmitted over an \(\mathbb{F}_q\)-input symmetric memoryless channel, it is known that BP decoding has\,\,the uniform error property: its symbol or block conditional error probability is independent of the transmitted codeword (cf.\cite[Lem. 1]{LDPC1}; see also the nonbinary coset analogue in\cite[Lem. 1]{Coset3}).  For a QMSC, translating each channel output by the negative of the corresponding transmitted symbol reduces its distribution to the zero-codeword case, while the linear check constraints make BP equivariant under these translations. Thus,  we may assume that the all-zero codeword is transmitted. 

It is well known that DE can characterize the asymptotic performance of BP decoding on sparse-graph code ensembles for any fixed number of iterations\cite{LDPC1}. To make this connection precise, consider the coupled DE sequence defined by
\begin{equation}
\underline{\mathsf x}^{(0)}=\underline{\Delta_0},
\qquad
\underline{\mathsf x}^{(\ell)}
=
\mathsf T_c\bigl(\underline{\mathsf x}^{(\ell-1)};\mathsf c\bigr),
\quad \ell\ge1,\label{eq:coupled_DE_sequence}
\end{equation}
where \(\mathsf x_i^{(\ell)}\) denotes the conditional distribution of a uniformly selected check-node-input message at position \(i\) after \(\ell\) variable-node updates. For \(\mathsf u\in\mathcal X_q\), let  
\[
P_{\mathrm e}(\mathsf u)
\coloneqq
\int_{\mathcal S_q}
\left(1-\max_{a\in\mathbb F_q}y_a\right)
(\Psi\mathsf u)(d\underline y)
\]
denote the symbol-MAP error probability under uniform tie-breaking. By the linearity of $P_{\mathrm e}$ and \eqref{eq:coupled_DE}, the  variable-to-check message error rate predicted by \eqref{eq:coupled_DE_sequence} can be given by 
\[
p_\ell\coloneqq
\frac1N\sum_{s=1}^{N}
\mathbb E_{\boldsymbol J}
\left[
P_{\mathrm e}
\left(
\mathsf c\circledast
\left(
\mathop{\circledast}\limits_{a=2}^{d_l}
\left(\mathsf x_{s+J_a}^{(\ell-1)}\right)^{\boxast(d_r-1)}
\right)
\right)
\right]=\frac1N\sum_{i=1}^{K}
P_{\mathrm e}\bigl(\mathsf x_i^{(\ell)}\bigr),
\qquad \ell\ge1.
\]
For finite-length BP decoding, let \(P_{\ell,n}\) denote the fraction of erroneous variable-to-check messages after \(\ell\) variable-node updates among the \(Nd_ln\) messages originating from transmitted variable nodes.  Since the computation neighborhood associated with\,\,any fixed number of BP iterations is cycle-free with high probability, and BP is a local algorithm whose messages depend only on this neighborhood, using arguments analogous to those in\cite[Thm. 2]{LDPC1}, we have the following result.  
\begin{theorem} [Concentration and DE characterization]
Consider a \((d_l,d_r,w,L)\) coupled ensemble over \(\mathbb F_q\) with a fixed rational edge-spreading profile \((\mathcal T,p)\), transmitted over a QMSC with conditional APP distribution \(\mathsf c\). Let \(n\to\infty\)  through admissible values. For every fixed \(\ell\ge1\) and \(\delta>0\), there exists \(\beta>0\), independent of \(n\), such that
\begin{equation}
\Pr\left\{
|P_{\ell,n}-\mathbb E P_{\ell,n}|>\delta
\right\}
\le 2e^{-\beta n}.\label{eq:concentration}
\end{equation}
Moreover, \(\mathbb E P_{\ell,n}
=p_\ell+O(n^{-1})\), and   \(P_{\ell,n}\longrightarrow p_\ell\) a.s. 
\end{theorem}

Theorem V.2 holds for every suitable distribution \(\pi_w\), including   the standard and improved coupled ensembles. It shows that  the asymptotic BP performance can be precisely predicted by coupled DE at each fixed iteration. We will take the block-length limit \(n\to\infty\) before the iteration limit  \(\ell\to\infty\).  By the continuity and degradation-preserving properties of \(\circledast,\boxast\) established in Section IV, \(\mathsf T_c\) is continuous and monotone in both arguments. Thus, for the   coupled DE sequence in \eqref{eq:coupled_DE_sequence},  
\[
\underline{\mathsf x}^{(\ell+1)}
\preceq
\underline{\mathsf x}^{(\ell)},
\qquad
\underline{\mathsf x}^{(\ell)} \xrightarrow{d_W}
\underline{\mathsf x}^{(\infty)}, 
\] 
where \(\underline{\mathsf x}^{(\infty)}\in\mathcal{X}_q^K\) is the greatest fixed point of \(\mathsf{T}_c(\cdot;\mathsf{c})\) under degradation. Throughout this section, degradation and convergence of measure vectors in \(\mathcal X_q^K\) are understood \textit{componentwise}, and \(\underline{\Delta_0}=(\Delta_0)_{1\le i\le K}, \underline{\Delta_\infty}=(\Delta_\infty)_{1\le i\le K}\).
 
 For a complete QMSC family \(\{\mathsf c_h:0\le h\le\log q\}\) ordered by degradation, define the coupled BP threshold by
 \[
h_c^{\mathrm{BP}}(w,L)
=
\sup\left\{
h\in[0,\log q]:
\mathsf T_c^{(\infty)}
(\underline{\Delta_0};\mathsf c_h)
=
\underline{\Delta_\infty}
\right\}.
 \]
 The definition above raises a natural question: as \(L\to\infty\) followed by \(w\to\infty\), can \(h_c^{\mathrm{BP}}(w,L)\) be characterized solely by the degree profile and channel family of the underlying uncoupled ensemble, independently of the particular edge-spreading\,\,rule? In the binary case, threshold saturation has been established for the BEC\cite{SCLDPC1} and, more generally, for BMS channels\cite{SCLDPC,SCLDPC4},\\ with the limiting BP threshold characterized by an intrinsic threshold of the uncoupled system. For nonbinary coupled ensembles, rigorous saturation results are also available for the BEC\cite{NBSC3}, but a corresponding general theorem for arbitrary finite fields and complete QMSC families was not available prior to this work. In the remainder of this section, we address this question by introducing a potential functional and a fixed-point threshold for the uncoupled DE, and establishing threshold saturation under the edge-spreading regularity and stability conditions stated below.
 \subsection{Potential Functionals and Threshold Saturation}
We use the \(q\)-ary analogue of the uncoupled potential functional in\cite[Def. 20]{SCLDPC4}.  Unlike the energy gap in\cite[Def. 25]{SCLDPC4},\,which is defined over the complement of the basin of attraction of \(\Delta_\infty\), our energy gap is defined over the nontrivial DE fixed points.
Define the uncoupled potential functional 
 \[
 U_s(\mathsf x;\mathsf c)\coloneqq
 d_l\mathrm H\!\left(\mathsf x^{\boxast(d_r-1)}\right)
 -\frac{d_l(d_r-1)}{d_r}
 \mathrm H\!\left(\mathsf x^{\boxast d_r}\right) 
  -\mathrm H\!\left(
 \mathsf c\circledast
 \left(\mathsf x^{\boxast(d_r-1)}\right)^{\circledast d_l}
 \right),\quad \mathsf{x},\mathsf{c}\in\mathcal{X}_q.
 \]
 Recall the nontrivial fixed-point set \(\mathcal F(\mathsf c)=\{
 \mathsf x\in\mathcal X_q\setminus\{\Delta_\infty\}:
 \mathsf T_s(\mathsf x;\mathsf c)=\mathsf x\}\) in Corollary IV.23. Define the energy gap and\,\,the fixed-point threshold by
 \[
\Delta E(\mathsf c) \coloneqq
\inf_{\mathsf x\in\mathcal F(\mathsf c)}
U_s(\mathsf x;\mathsf c),\qquad
h^{\mathrm{FP}} \coloneqq
\sup\left\{
h\in[0,\log q]:
\Delta E(\mathsf c_h)>0
\right\},
 \]
 with \(\inf\varnothing=+\infty\). When the uncoupled system is stable and \(\mathcal F(\mathsf c)\ne\varnothing\), Corollary IV.23 ensures that \(\mathcal F(\mathsf c)\) is compact, and\,\,thus the infimum is attained by continuity of \(U_s\). Note that the threshold \(h^{\mathrm{FP}}\) depends only on the underlying ensemble and channel family, not on the   coupling parameters \(w,L\). For the coupled system, with \(N=2L\) and \(K=N+w-1\), define
 \[
	U_c(\underline{\mathsf x};\mathsf c)
\coloneqq 
\sum_{i=1}^{K}
\left[
d_l\mathrm H\!\left(\mathsf x_i^{\boxast(d_r-1)}\right)
-\frac{d_l(d_r-1)}{d_r}
\mathrm H\!\left(\mathsf x_i^{\boxast d_r}\right)
\right] -\sum_{j=1}^{N}
\mathbb E_{\boldsymbol J}
\left[
\mathrm H\!\left(
\mathsf c\circledast
\left(
\mathop{\circledast}_{a=1}^{d_l}
\mathsf x_{j+J_a}^{\boxast(d_r-1)}
\right)
\right)
\right],\quad \underline{\mathsf{x}}\in\mathcal{X}_q^K,\ \mathsf{c}\in\mathcal{X}_q.
 \]
Both \(U_s\) and \(U_c\) are continuous on their respective domains,  with \(U_s(\Delta_\infty;\mathsf c)=U_c(\underline{\Delta_\infty};\mathsf c)=0\). We will use their directional derivatives. Define the set of signed-measure differences
\[
\mathcal D_q
\coloneqq
\{\mathsf a-\mathsf b:\mathsf a,\mathsf b\in\mathcal X_q\}
\subseteq\mathcal M_q.
\]
Every \(\mathsf v\in\mathcal D_q\) is a finite signed measure of total mass zero. The linear extension of \(\mathrm H\) and the bilinear extensions of \(\circledast\) and \(\boxast\) extend \(U_s\) and \(U_c\) to polynomial functionals on \(\mathcal M_q\) and \(\mathcal M_q^K\), respectively. For \(\underline{\mathsf x}\in\mathcal X_q^K\) and \(\underline{\mathsf v},\underline{\mathsf z}\in\mathcal D_q^K\), with \(\mathsf c\in\mathcal X_q\) fixed, define the first- and second-order directional derivatives of \(U_c\) by
\[
\begin{aligned}
	\mathrm D U_c(\underline{\mathsf x};\mathsf c)
	[\underline{\mathsf v}]
	 \coloneqq
	\left.
	\frac{\partial}{\partial t}
	U_c(\underline{\mathsf x}+t\underline{\mathsf v};\mathsf c)
	\right|_{t=0},\qquad
	\mathrm D^2U_c(\underline{\mathsf x};\mathsf c)
	[\underline{\mathsf v},\underline{\mathsf z}]
	 \coloneqq
	\left.
	\frac{\partial^2}{\partial t\,\partial u}
	U_c(\underline{\mathsf x}
	+t\underline{\mathsf v}
	+u\underline{\mathsf z};\mathsf c)
	\right|_{t=u=0}.
\end{aligned} 
\]
The derivatives of \(U_s\) are defined analogously. These definitions use the signed-measure extensions, so the perturbed arguments, e.g., \(\underline{\mathsf x}+t\underline{\mathsf v}\) and \(\underline{\mathsf x}
+t\underline{\mathsf v}
+u\underline{\mathsf z}\), need not remain probability measures. We have the following results on \(U_s,U_c,\Delta E(\mathsf{c}_h)\).
\begin{lemma}
The following statements hold.
\begin{enumerate}[label=\roman*)]
\item For \(\mathsf v\in\mathcal D_q\) and
\(\underline{\mathsf v}\in\mathcal D_q^K\),
\begin{align*}
	\mathrm D U_s(\mathsf x;\mathsf c)[\mathsf v]
	={}&
	d_l(d_r-1)\,
	\mathrm H\!\left(
	\bigl(\mathsf T_s(\mathsf x;\mathsf c)-\mathsf x\bigr)
	\boxast
	\mathsf x^{\boxast(d_r-2)}
	\boxast\mathsf v
	\right),\\
	\mathrm D U_c(\underline{\mathsf x};\mathsf c)
	[\underline{\mathsf v}]
	={}&
	d_l(d_r-1)
	\sum_{i=1}^{K}
	\mathrm H\!\left(
	\bigl(
	[\mathsf T_c(\underline{\mathsf x};\mathsf c)]_i-\mathsf x_i
	\bigr)
	\boxast
	\mathsf x_i^{\boxast(d_r-2)}
	\boxast\mathsf v_i
	\right).
\end{align*}
\item  If \(\mathsf T_s(\mathsf x;\mathsf c)
\preceq(\succeq)\mathsf x\), then \(U_s(\mathsf T_s(\mathsf x;\mathsf c);\mathsf c)
\le U_s(\mathsf x;\mathsf c)\). Similarly, if \(\mathsf T_c(\underline{\mathsf x};\mathsf c)
\preceq(\succeq)\underline{\mathsf x}\), then \(U_c(\mathsf T_c(\underline{\mathsf x};\mathsf c);\mathsf c)
\le U_c(\underline{\mathsf x};\mathsf c)\). Hence, along any monotone DE sequence \(\{\mathsf{x}^{(\ell)}\}\) or \(\{\underline{\mathsf{x}}^{(\ell)}\}\),
\[
U_s(\mathsf x^{(\ell+1)};\mathsf c)
\le U_s(\mathsf x^{(\ell)};\mathsf c),
\qquad
U_c(\underline{\mathsf x}^{(\ell+1)};\mathsf c)
\le U_c(\underline{\mathsf x}^{(\ell)};\mathsf c).
\]
\item \(\Delta E(\mathsf c_{h_2})
\le
\Delta E(\mathsf c_{h_1})\) for any \(h_1<h_2\), with strict inequality if \(\mathcal F(\mathsf c_{h_1})\) is nonempty and compact. Moreover, if \(h<h^{\mathrm{FP}}\), then \(\Delta E(\mathsf c_h)>0\). If the uncoupled system is stable at \(h^{\mathrm{FP}}\) and \(h>h^{\mathrm{FP}}\), then \(\Delta E(\mathsf c_h)<0\). When \(d_l\ge3\), the   stability condition is automatic.  
\end{enumerate} 
\end{lemma}
Recall that \(\boldsymbol{J}=(J_1,\ldots,J_{d_l})\sim \pi_w\) denotes the random offset vector, where \(J_a\in \{0,\ldots,w-1\}\).
Regard \(\pi_w\) as a probability measure on \(\mathbb Z^{d_l}\). For each positive integer \(s\), define
\[\eta_w(s) \coloneqq
d_{\mathrm{TV}}\!\left(
\pi_w,\,
\mathrm{Law}(J_1+s,J_2,\ldots,J_{d_l})
\right),\]
and set \(\eta_w \coloneqq
\min_{s\ge1}\eta_w(s)\),
where
\[
d_{\mathrm{TV}}(\mu,\nu)
=
\frac12
\sum_{\boldsymbol j\in\mathbb Z^{d_l}}
|\mu(\boldsymbol j)-\nu(\boldsymbol j)|.
\]
In the definition of \(\eta_w(s)\), only one coordinate is shifted, without wraparound, where we used \(J_1\); by exchangeability, the\,\,choice of that coordinate is immaterial. The minimum is attained because \(\eta_w(s)=1\) for \(s\ge w\). As shown in Section V-C, the quantity \(\eta_w(s)\) controls the change in a DE output when its spatial index is shifted by \(s\), and allowing \(s>1\)   accommodates spreading laws supported on a sublattice. For example, if \(w\) is even and \(J_1,\ldots,J_{d_l}\) are i.i.d. uniform on \(\{0,2,\ldots,w-2\}\),\,\,then\,\,\(\eta_w=\eta_w(2)=2/w\to0\) as \(w\to\infty\), which, as we will see, is a sufficient edge-spreading condition    for threshold saturation.
\begin{theorem}[Threshold saturation for coupled ensemble]
Define the constant
\[
C_{q,d_l,d_r}
\coloneqq
d_l(d_r-1)
\bigl[d_l(d_r-1)-1\bigr]\log q. 
\]
The following statements hold for the coupled DE defined in \eqref{eq:coupled_DE}, with any exchangeable offset distribution \(\pi_w\).
\begin{enumerate}[label=\roman*)]
\item \textbf{Achievability}. For every \(h<h^{\mathrm{FP}}\), if \(\frac12C_{q,d_l,d_r}\eta_w
<
\Delta E(\mathsf c_h)\),
then, for every positive integer \(L\), \(\underline{\Delta_\infty}\) is the unique fixed point of coupled DE: for \(\underline{\mathsf{x}}\in\mathcal{X}_q^K\),
\[
\mathsf T_c(\underline{\mathsf x};\mathsf c_h)
=\underline{\mathsf x}
\quad\Longrightarrow\quad
\underline{\mathsf x}=\underline{\Delta_\infty}.
\]
If \(\mathcal F(\mathsf c_h)=\varnothing\), the above conclusion holds without a restriction on \(\eta_w\).
\item \textbf{Converse}. Suppose that the uncoupled system is stable at \(h^{\mathrm{FP}}\). For every \(h>h^{\mathrm{FP}}\) and every fixed \(w\), there is \(L_0\) such\,\,that
\[
L\ge L_0
\quad\Longrightarrow\quad
\mathsf T_c^{(\infty)}
(\underline{\Delta_0};\mathsf c_h)
\ne
\underline{\Delta_\infty}.
\]
\end{enumerate}
\end{theorem}
The proof of Theorem V.4 is provided in Sections V-C and V-D.
By Theorem V.4 i), the   edge-spreading condition \(\eta_w=o_w(1)\) is sufficient for the achievability   of threshold saturation. Together with Theorem V.4 ii), this yields threshold saturation under the   stability condition on the uncoupled system. 

\begin{lemma}[Asymptotic saturation]
For every fixed \(w\) and \(\pi_w\), \(h_c^{\mathrm{BP}}(w,L)\) is nonincreasing in \(L\), so the limit \(\lim_{L\to\infty}h_c^{\mathrm{BP}}(w,L)\) exists. If \(\eta_w\to0\) as \(w\to \infty\) and the uncoupled system is stable at \(h^{\mathrm{FP}}\), then
\begin{equation}
\lim_{w\to\infty}\lim_{L\to\infty}
h_c^{\mathrm{BP}}(w,L)
=
h^{\mathrm{FP}}. \label{eq:threshold_saturation}
\end{equation}
When \(h^{\mathrm{FP}}=\log q\), the stability condition may be omitted.
\end{lemma}  
\begin{lemma} 
 For the standard and improved coupled ensembles,
 \[\eta_w=\begin{cases}
 	1/w &\text{standard ensemble},\\
 	 {d_l/w} &\text{improved ensemble with }w\ge d_l.
 \end{cases}\]
Thus, if \(d_l\ge 3\), then both ensembles exhibit threshold saturation, i.e., \eqref{eq:threshold_saturation} holds. More precisely, for every \(h<h^{\mathrm{FP}}\), perfect decoding under coupled DE is guaranteed for all \(L\) whenever
\[
w>\begin{cases}
	\dfrac{C_{q,d_l,d_r}}{2\Delta E(\mathsf c_h)} &\text{standard ensemble},\\
	\dfrac{d_lC_{q,d_l,d_r}}{2\Delta E(\mathsf c_h)} &\text{improved ensemble with }w\ge d_l.
\end{cases} 
\]
The right-hand sides are interpreted as zero when \(\Delta E(\mathsf c_h)=+\infty\).
\end{lemma}

\subsection{Proof of Theorem V.4 i)}
Here, we prove the achievability part in Theorem V.4. 

Fix \(h<h^{\mathrm{FP}}\), write
\(\mathsf c=\mathsf c_h\), and retain \(N=2L\) and \(K=N+w-1\). By Lemma V.3 iii), \(\Delta E(\mathsf c)>0\).
The proof\,\,uses\,\,a\,\,one-sided auxiliary system, whose message distributions become more degraded from left to right and  whose message distributions are degraded with respect to those of the original terminated system. If its limiting profile were nontrivial, shifting it to the\,\,right by \(s\ge 1\) positions would decrease the potential by at least \(s\Delta E(\mathsf c)\). On the other hand, \(\eta_w(s)\) controls
the curvature along\,\,this shift and limits how much the potential can decrease. Comparing the  two bounds gives a contradiction under the\,stated\,condition.  

 \textit{A one-sided auxiliary system.}
We consider an auxiliary system indexed by positions \(i\in\mathbb Z\). It retains the perfect left boundary but has no right termination. When extending a finite profile, we copy its distribution at position \(K\) to all positions to its right. Specifically,   define, for \(\underline{\mathsf{z}}\in\mathcal{X}_q^{\mathbb{Z}}\), 
\begin{equation}
	\mathsf c_j^+=
	\begin{cases}
		\Delta_\infty,&j\le0,\\
		\mathsf c,&j\ge1,
	\end{cases}
	\qquad
	\mathsf F_i(\underline{\mathsf z})
	=\mathbb E_{\boldsymbol J}\!\left[
	\mathsf c^+_{i-J_1}\circledast
	\left(\mathop{\circledast}_{a=2}^{d_l}
	\mathsf z_{i-J_1+J_a}^{\boxast(d_r-1)}\right)
	\right],\quad i\in\mathbb Z.
	\label{eq:ach-one-sided-map}
\end{equation}
Starting from \(\mathsf x_i^{(0)}=\Delta_\infty\) for \(i\le0\) and
\(\mathsf x_i^{(0)}=\Delta_0\) for \(i\ge1\), iterate
\begin{equation}
	\mathsf x_i^{(\ell+1)}=
	\begin{cases}
		\Delta_\infty,&i\le0,\\
		\mathsf F_i(\underline{\mathsf x}^{(\ell)}),&1\le i\le K,\\
		\mathsf x_K^{(\ell+1)},&i>K.
	\end{cases}
	\label{eq:ach-auxiliary-DE}
\end{equation}
Thus, only the positions \(1,\ldots,K\) are updated by the DE map, while the right-hand extension remains constant in space and is updated at each iteration. This is similar  to the free right boundary condition of\cite[Def. 44]{SCLDPC}, formulated here in check-node-input coordinates. By removing the right termination and using the above extension, we can compare the auxiliary and\,\,original DE sequences through degradation monotonicity, without requiring reflection symmetry of \(\pi_w\).\footnote{The midpoint-based one-sided   modified systems
	in \cite{SCLDPC1,SCLDPC4} use the reflection symmetry induced by
	uniform edge spreading and symmetric termination to compare
	the auxiliary and original DE profiles.}

 By Proposition IV.16, \(\mathsf F_i\) is monotone in its profile: if \(\underline{\mathsf{z}} \preceq\underline{\mathsf{y}}\), then \(\mathsf F_i(\underline{\mathsf z})
 \preceq\mathsf F_i(\underline{\mathsf y})\). Moreover,   increasing \(i\) by one increases every profile index and
 replaces \(\mathsf c^+_{i-J_1}\) by the no-better channel
 \(\mathsf c^+_{i+1-J_1}\), so we have
 \begin{equation}
 	\mathsf z_j\preceq\mathsf z_{j+1}\   \forall j\in\mathbb Z 
 	\quad\Longrightarrow\quad
 	\mathsf F_i(\underline{\mathsf z})
 	\preceq\mathsf F_{i+1}(\underline{\mathsf z}).
 	\label{eq:ach-spatial-order}
 \end{equation}
 By the copying rule, the entire update in \eqref{eq:ach-auxiliary-DE} is also
 monotone, and its initialization is maximal. Thus, by induction,
 \begin{equation}
\mathsf x_{i-1}^{(\ell)}
\preceq\mathsf x_i^{(\ell)},
\qquad
\mathsf x_i^{(\ell+1)}
\preceq\mathsf x_i^{(\ell)},
\quad i\in\mathbb Z,\ \ell\ge0.
 	\label{eq:ach-two-orders}
 \end{equation}
By continuity and monotone convergence (see Propositions IV.11 and IV.18),
the sequence \(\{\underline{\mathsf{x}}^{(\ell)}\}\) converges componentwise\,\,to a limit
\(\underline{\mathsf x}\in\mathcal{X}_q^{\mathbb{Z}}\) satisfying
\begin{equation}
	\mathsf x_i=
	\begin{cases}
		\Delta_\infty,&i\le0,\\
		\mathsf F_i(\underline{\mathsf x}),&1\le i\le K,\\
		\mathsf a,&i>K,
	\end{cases}
	\qquad
	\mathsf a\coloneqq\mathsf x_K,
	\qquad
	\mathsf x_{i-1}\preceq\mathsf x_i\preceq\mathsf a.
	\label{eq:ach-limit-profile}
\end{equation}
We   compare the two systems. Denote the original forward-DE sequence
in \eqref{eq:coupled_DE_sequence} by
\(\underline{\mathsf x}_{\mathrm{orig}}^{(\ell)}\), extended by
\(\Delta_\infty\) outside \(\{1,\ldots,K\}\).
Initially, \(\underline{\mathsf x}_{\mathrm{orig}}^{(0)}
\preceq\underline{\mathsf x}^{(0)}\) on \(\mathbb{Z}\). If this order holds at iteration
\(\ell\), then by \(\mathsf c_j\preceq\mathsf c_j^+\) and
Proposition IV.16, for \(1\le i\le K\),
\begin{align}
	\mathsf x_{\mathrm{orig},i}^{(\ell+1)}
 =\mathbb E_{\boldsymbol J}\!\left[
	\mathsf c_{i-J_1}\circledast
	\left(\mathop{\circledast}_{a=2}^{d_l}
	(\mathsf x_{\mathrm{orig},i-J_1+J_a}^{(\ell)})
	^{\boxast(d_r-1)}\right)\right] \preceq\mathsf F_i(\underline{\mathsf x}^{(\ell)})
	=\mathsf x_i^{(\ell+1)}.
	\label{eq:ach-comparison-step}
\end{align}
Outside \(\{1,\ldots,K\}\), the original extension is \(\Delta_\infty\), so the same
order holds   as well. Induction and closedness of degradation\,give
\begin{equation}
	\underline{\mathsf x}_{\mathrm{orig}}^{(\ell)}
	\preceq\underline{\mathsf x}^{(\ell)}
	\quad(\ell\ge0),
	\qquad
	\underline{\mathsf x}_{\mathrm{orig}}^{(\infty)}
	\preceq\underline{\mathsf x}.\notag
\end{equation}
It thus suffices to show that the auxiliary limit \(\underline{\mathsf{x}}\) is perfect. Suppose otherwise. By \eqref{eq:ach-limit-profile},
\(\mathsf a\succ\Delta_\infty\). Since \(K=N+w-1\) and \(0\le J_1\le w-1\), we have \(K-J_1\ge N\ge1\), so \(\mathsf c_{K-J_1}^+=\mathsf c\). Moreover, \(\mathsf x_j\preceq\mathsf a\) for all \(j\in\mathbb Z\) by \eqref{eq:ach-limit-profile}. Hence,  
\[
\mathsf F_K(\underline{\mathsf x}) =\mathbb E_{\boldsymbol J}\left[
\mathsf c\circledast
\left(
\mathop{\circledast}_{a=2}^{d_l}
\mathsf x_{K-J_1+J_a}^{\boxast(d_r-1)}
\right)\right] \preceq
\mathsf c\circledast
\left(
\mathop{\circledast}_{a=2}^{d_l}
\mathsf a^{\boxast(d_r-1)}
\right) =
\mathsf c\circledast
\left(\mathsf a^{\boxast(d_r-1)}\right)^{\circledast(d_l-1)} =\mathsf T_s(\mathsf a;\mathsf c).
\]
Then, by \eqref{eq:ach-limit-profile},
\(\mathsf a\succ\Delta_\infty\), and Proposition IV.21, we have
\begin{equation}
	\Delta_\infty\prec\mathsf a
	=\mathsf F_K(\underline{\mathsf x})
	\preceq\mathsf T_s(\mathsf a;\mathsf c)
	\preceq\mathsf T_s^{(2)}(\mathsf a;\mathsf c)
	\preceq\cdots
	\xrightarrow{d_W}\mathsf T_s^{(\infty)}(\mathsf a;\mathsf c)
	\in\mathcal F(\mathsf c).\notag
\end{equation}
If \(\mathcal F(\mathsf c)=\varnothing\), this is already a
contradiction, proving Theorem V.4 i).
Otherwise, by Lemma V.3 ii) and continuity of \(U_s\),
\begin{equation}
	U_s(\mathsf a;\mathsf c)
	\ge\lim_{\ell\to\infty}
	U_s(\mathsf T_s^{(\ell)}(\mathsf a;\mathsf c);\mathsf c)=U_s(\mathsf T_s^{(\infty)}(\mathsf a;\mathsf c);\mathsf{c})
	\ge\Delta E(\mathsf c)>0.
	\label{eq:ach-endpoint-energy}
\end{equation}
We next compare the coupled potential at the auxiliary limiting profile \(\underline{\mathsf x}\) and at its right shift by \(s\) positions. Choose an integer \(s\ge1\) attaining \(\eta_w(s)=\eta_w\), and let \(M\coloneqq K+s-1\). Define the shift difference \(\underline{\mathsf d}\in\mathcal D_q^{\mathbb Z}\) by
\begin{equation} 
	\mathsf d_i\coloneqq\mathsf x_i-\mathsf x_{i-s},\qquad i\in\mathbb{Z}.\notag
\end{equation}
By the spatial monotonicity of \(\underline{\mathsf x}\) and its boundary conditions, \(\mathsf x_{i-s}\preceq\mathsf x_i\) for \(i\in\mathbb{Z}\) and \(\mathsf d_i=0\) for \(i\notin\{1,\ldots,M\}\).  Consider\,\,the same finite coupled potential as in Section V-B, but now with
\(M\) variable positions:
\begin{align}
	U_{c,M}(\underline{\mathsf z};\mathsf c) \coloneqq\sum_{i=1}^{M+w-1}
	\left[d_l\mathrm H(\mathsf z_i^{\boxast(d_r-1)})
	-\frac{d_l(d_r-1)}{d_r}
	\mathrm H(\mathsf z_i^{\boxast d_r})\right] -\sum_{j=1}^{M}\mathbb E_{\boldsymbol J}
	\mathrm H\!\left(\mathsf c\circledast
	\left(\mathop{\circledast}_{a=1}^{d_l}
	\mathsf z_{j+J_a}^{\boxast(d_r-1)}\right)\right).\notag
\end{align}
We will bound \(U_{c,M}(\underline{\mathsf x}-\underline{\mathsf d};\mathsf c)-U_{c,M}(\underline{\mathsf x};\mathsf c)\)   by two different methods: directly from its definition and via   interpolation.

First, for the check-node terms,   shifting the summation index and cancelling common terms give,   for \(m\in\{d_r-1,d_r\}\),
\begin{equation}
\sum_{i=1}^{M+w-1}
\left[
\mathrm H(\mathsf x_{i-s}^{\boxast m})
-\mathrm H(\mathsf x_i^{\boxast m})
\right] =
\sum_{i=1-s}^{0}\mathrm H(\mathsf x_i^{\boxast m})
-\sum_{i=M+w-s}^{M+w-1}
\mathrm H(\mathsf x_i^{\boxast m})=-s\mathrm H(\mathsf a^{\boxast m}),\label{eq:cn_term}
\end{equation}
where the last step follows since \(\mathsf{x}_i=\Delta_\infty\) for \(i\le 0\) and  \(\mathsf{x}_i=\mathsf{a}\) for \(i\ge K\). Similarly, for the variable-node terms,  
\begin{align}
	&\left(
	\sum_{j=1-s}^{M-s}-\sum_{j=1}^{M}
	\right)
	\mathbb E_{\boldsymbol J}
	\mathrm H\left(
	\mathsf c\circledast
	\left(
	\mathop{\circledast}_{a=1}^{d_l}
	\mathsf x_{j+J_a}^{\boxast(d_r-1)}
	\right)
	\right) =
	\left(
	\sum_{j=1-s}^{0}-\sum_{j=K}^{M}
	\right)
	\mathbb E_{\boldsymbol J}
	\mathrm H\left(
	\mathsf c\circledast
	\left(
	\mathop{\circledast}_{a=1}^{d_l}
	\mathsf x_{j+J_a}^{\boxast(d_r-1)}
	\right)
	\right)\notag\\
	&\ \ =
	\sum_{j=1-s}^{0}\mathbb E_{\boldsymbol J}
	\mathrm H\left(
	\mathsf c\circledast
	\left(
	\mathop{\circledast}_{a=1}^{d_l}
	\mathsf x_{j+J_a}^{\boxast(d_r-1)}
	\right)
	\right) -
	s\mathrm H\left(
	\mathsf c\circledast
	(\mathsf a^{\boxast(d_r-1)})^{\circledast d_l}
	\right),\label{eq:vn_term}
\end{align}  
where we used \(M-s+1=K\) and \(\mathsf x_{j+J_a}=\mathsf a\) for \(j\ge K\). Combining  \eqref{eq:cn_term} and \eqref{eq:vn_term} with the definition of \(U_s\), we obtain
\begin{equation}
U_{c,M}(\underline{\mathsf x}-\underline{\mathsf d};\mathsf c)
-U_{c,M}(\underline{\mathsf x};\mathsf c) =
-sU_s(\mathsf a;\mathsf c) -
\sum_{j=1-s}^{0}\mathbb E_{\boldsymbol J}
\mathrm H\left(
\mathsf c\circledast
\left(
\mathop{\circledast}_{a=1}^{d_l}
\mathsf x_{j+J_a}^{\boxast(d_r-1)}
\right)
\right) \le -sU_s(\mathsf a;\mathsf c)
\le -s\Delta E(\mathsf c).\label{eq:direct_bound}
\end{equation}
The last two steps follow from the nonnegativity of entropy and \eqref{eq:ach-endpoint-energy}. \eqref{eq:direct_bound} bounds the    potential change directly by its definition.

We now examine the potential change along the linear interpolation between \(\underline{\mathsf x}\) and its right shift \(\underline{\mathsf x}-\underline{\mathsf d}\). Define the function
\[
\phi(t)
\coloneqq
U_{c,M}(\underline{\mathsf x}-t\underline{\mathsf d};\mathsf c),
\qquad 0\le t\le1.
\] 
Since \(\mathsf x_i-t\mathsf d_i
=(1-t)\mathsf x_i+t\mathsf x_{i-s}\), the path consists of probability measures, and \(\phi\) is a polynomial in \(t\). We will show\,\,\(\phi'(0)\ge0\) and derive an expression for \(\phi''(t)\), whose lower bound will be obtained using \(\eta_w(s)\). For a profile \(\underline{\mathsf z}\), define
\[
\mathsf u_i=\mathsf u_i(\underline{\mathsf{z}})
\coloneqq
\mathsf z_i^{\boxast(d_r-2)}
\boxast\mathsf d_i,\qquad i\in\mathbb{Z},
\]
with the convention \(\mathsf z_i^{\boxast0}=\Delta_\infty\). Since \(\mathsf d_i=0\) for \(i\notin\{1,\ldots,M\}\), the same holds for \(\mathsf u_i\). By the same calculation as in Lemma V.3 i), with \(M\) variable positions, we have
\begin{equation}
\mathrm D U_{c,M}(\underline{\mathsf z};\mathsf c)
[\underline{\mathsf d}] =
d_l(d_r-1)
\sum_{i=1}^{M}
\mathrm H\left(
\bigl(\mathsf F_i(\underline{\mathsf z})-\mathsf z_i\bigr)
\boxast\mathsf u_i 
\right).\label{eq:U_cM_first_order_dd}
\end{equation}
Consequently,
\[
\begin{aligned}
	\phi'(0)
	&=
	-\mathrm D U_{c,M}(\underline{\mathsf x};\mathsf c)
	[\underline{\mathsf d}] =
	-d_l(d_r-1)
	\sum_{i=1}^{M}
	\mathrm H\left(
	\bigl(\mathsf F_i(\underline{\mathsf x})-\mathsf x_i\bigr)
	\boxast
	\mathsf x_i^{\boxast(d_r-2)}
	\boxast\mathsf d_i
	\right)\\
	&=
	-d_l(d_r-1)
	\sum_{i=K+1}^{M}
	\mathrm H\left(
	\bigl(\mathsf F_i(\underline{\mathsf x})-\mathsf a\bigr)
	\boxast
	\mathsf a^{\boxast(d_r-2)}
	\boxast\mathsf d_i
	\right) \ge0.
\end{aligned} 
\]
The third equality follows since \(\mathsf F_i(\underline{\mathsf x})=\mathsf x_i\) for \(1\le i\le K\) and \(\mathsf x_i=\mathsf a\) for \(K<i\le M\). The last step follows from\,\(\mathsf a\!=\!\mathsf{F}_K(\underline{\mathsf{x}})\!\preceq\mathsf F_i(\underline{\mathsf x})\) for \(i>K\), the degradation ordering of \(\mathsf d_i=\mathsf{x}_i-\mathsf{x}_{i-s}\), and \(\eqref{eq:entropy_dual_cons}\). We   differentiate  \eqref{eq:U_cM_first_order_dd} once more. By the product rule,
\[
\mathrm D\mathsf u_i(\underline{\mathsf z})
[\underline{\mathsf d}]
\coloneqq \left.\frac{d}{dt} \mathsf{u}_i(\underline{\mathsf{z}}+t\underline{\mathsf{d}})\right|_{t=0}=
(d_r-2)
\mathsf z_i^{\boxast(d_r-3)}
\boxast\mathsf d_i^{\boxast2}, 
\]
where the right-hand side is omitted when \(d_r=2\). Using the linearity of \(\mathrm H\) and the bilinearity of \(\boxast\), we have 
\begin{align}
	&\frac{
		\mathrm D^2U_{c,M}(\underline{\mathsf z};\mathsf c)
		[\underline{\mathsf d},\underline{\mathsf d}]
	}{
		d_l(d_r-1)
	} \notag\\
	&=
	\sum_{i=1}^{M}
	\mathrm H\left(
	\bigl(
	\mathrm D\mathsf F_i(\underline{\mathsf z})
	[\underline{\mathsf d}]
	-\mathsf d_i
	\bigr)
	\boxast\mathsf u_i 
	\right) +
	(d_r-2)
	\sum_{i=1}^{M}
	\mathrm H\left(
	\bigl(\mathsf F_i(\underline{\mathsf z})-\mathsf z_i\bigr)
	\boxast
	\mathsf z_i^{\boxast(d_r-3)}
	\boxast\mathsf d_i^{\boxast2}
	\right)\notag\\
	&=
	\sum_{i=1}^{M}
	\mathrm H\left(
	\mathrm D\mathsf F_i(\underline{\mathsf z})
	[\underline{\mathsf d}]
	\boxast\mathsf u_i 
	\right) +
	(d_r-2)
	\sum_{i=1}^{M}
	\mathrm H\left(
	\mathsf F_i(\underline{\mathsf z})
	\boxast\mathsf z_i^{\boxast(d_r-3)}
	\boxast\mathsf d_i^{\boxast2}
	\right) -
	(d_r-1)
	\sum_{i=1}^{M}
	\mathrm H\left(
	\mathsf z_i^{\boxast(d_r-2)}
	\boxast\mathsf d_i^{\boxast2}
	\right),\label{eq:P4}
\end{align}  
where differentiating the definition of \(\mathsf F_i\)  gives   
\begin{align}
 \mathrm D\mathsf F_i(\underline{\mathsf z})
	[\underline{\mathsf d}]\coloneqq \left.\frac{d}{dt} \mathsf{F}_i(\underline{\mathsf{z}}+t\underline{\mathsf{d}})\right|_{t=0}=
	(d_r-1)\mathbb E_{\boldsymbol J}
	\sum_{a=2}^{d_l}
	\mathsf c^+_{i-J_1}
	\circledast\mathsf u_{i-J_1+J_a}(\underline{\mathsf{z}}) \circledast
	\left(
	\mathop{\circledast}_{\substack{
			2\le b\le d_l\\b\ne a
	}}
	\mathsf z_{i-J_1+J_b}^{\boxast(d_r-1)}
	\right).\label{eq:P2}
\end{align}  
Both \(\mathsf u_i \) and \(\mathrm D\mathsf F_i(\underline{\mathsf z})[\underline{\mathsf d}]\) are signed measures  of total mass zero. Hence, by the entropy duality rule (Lemma IV.10),
\begin{equation}
\mathrm H\left(
\mathrm D\mathsf F_i(\underline{\mathsf z})
[\underline{\mathsf d}]
\boxast\mathsf u_i 
\right) =
-\mathrm H\left(
\mathrm D\mathsf F_i(\underline{\mathsf z})
[\underline{\mathsf d}]
\circledast\mathsf u_i 
\right).\label{eq:P1}
\end{equation}
Substituting \(\eqref{eq:P2}\) into the first sum of \(\eqref{eq:P4}\), applying \(\eqref{eq:P1}\), and reindexing with \(i=j+J_1\), we obtain
\begin{align}
	&\sum_{i=1}^{M}
	\mathrm H\left(
	\mathrm D\mathsf F_i(\underline{\mathsf z})
	[\underline{\mathsf d}]
	\boxast\mathsf u_i
	\right)\notag\\
	&=
	-(d_r-1)
	\sum_{j=1}^{M}\mathbb E_{\boldsymbol J}
	\sum_{a=2}^{d_l}
	\mathrm H\left(
	\mathsf c
	\circledast\mathsf u_{j+J_1}
	\circledast\mathsf u_{j+J_a}
	\circledast
	\left(
	\mathop{\circledast}_{\substack{
			2\le b\le d_l\\b\ne a
	}}
	\mathsf z_{j+J_b}^{\boxast(d_r-1)}
	\right)
	\right)\notag\\
	&=
	-(d_l-1)(d_r-1)
	\sum_{j=1}^{M}\mathbb E_{\boldsymbol J}
	\mathrm H\left(
	\mathsf c
	\circledast\mathsf u_{j+J_1}
	\circledast\mathsf u_{j+J_2}
	\circledast
	\left(
	\mathop{\circledast}_{a=3}^{d_l}
	\mathsf z_{j+J_a}^{\boxast(d_r-1)}
	\right)
	\right).\label{eq:P3}
\end{align} 
In the first equality, terms with \(j\le0\) vanish since \(\mathsf c_j^+=\Delta_\infty\) and \(\Delta_\infty\circledast\mathsf u_i=0\), while those with \(j>M\) vanish since\,\(\mathsf u_{j+J_1}=0\). The second equality follows from the exchangeability of \(J_2,\ldots,J_{d_l}\).  Combining  \eqref{eq:P4} and \eqref{eq:P3}, we obtain 
	\begin{align}
		&\frac{
			\mathrm D^2U_{c,M}(\underline{\mathsf z};\mathsf c)
			[\underline{\mathsf d},\underline{\mathsf d}]
		}{
			d_l(d_r-1)
		}\notag\\
		&=
		-(d_l-1)(d_r-1)
		\sum_{j=1}^{M}\mathbb E_{\boldsymbol J}
		\mathrm H\left(
		\mathsf c
		\circledast\mathsf u_{j+J_1}
		\circledast\mathsf u_{j+J_2}
		\circledast
		\left(
		\mathop{\circledast}_{a=3}^{d_l}
		\mathsf z_{j+J_a}^{\boxast(d_r-1)}
		\right)
		\right) +
		(d_r-2)
		\sum_{i=1}^{M}
		\mathrm H\left(
		\mathsf F_i(\underline{\mathsf z})
		\boxast\mathsf z_i^{\boxast(d_r-3)}
		\boxast\mathsf d_i^{\boxast2}
		\right)\label{eq:P5}\\
		&\quad-
		(d_r-1)
		\sum_{i=1}^{M}
		\mathrm H\left(
		\mathsf z_i^{\boxast(d_r-2)}
		\boxast\mathsf d_i^{\boxast2}
		\right).\notag
\end{align}  
In the above, an empty \(\circledast\)-product is  \(\Delta_0\), and the term multiplied by \(d_r-2\) vanishes when \(d_r=2\). Setting \(\underline{\mathsf z}=\underline{\mathsf x}-t\underline{\mathsf d}\) gives
\[ 
	\phi''(t)
	=
	\mathrm D^2U_{c,M}
	(\underline{\mathsf x}-t\underline{\mathsf d};\mathsf c)
	[\underline{\mathsf d},\underline{\mathsf d}],
	\qquad 0\le t\le1. 
\]
Since \(\mathsf{d}_i=\mathsf{x}_i-\mathsf{x}_{i-s}\) and \(\mathsf{x}_{i-s}\preceq \mathsf{x}_i\), by \eqref{eq:entropy_dual_cons}, \(\mathrm H (
\mathsf z_i^{\boxast(d_r-2)}
\boxast\mathsf d_i^{\boxast2}
)\le0\). Thus, the last term in \(\phi''(t)\) is nonnegative. To\,\,lower-bound \(\phi''(t)\), it remains to control the   two terms in \eqref{eq:P5}. The key observation is that \(\eta_w(s)\) controls the difference between the DE outputs at positions \(i\) and \(i-s\).  Let \(\pi_w^{(s)}
\coloneqq
\mathrm{Law}(J_1+s,J_2,\ldots,J_{d_l})\). By \(\eqref{eq:ach-one-sided-map}\), we have
\[
\begin{aligned}
	\mathsf F_i(\underline{\mathsf x}) =
	\mathbb E_{\boldsymbol{J}\sim \pi_w}
	\left[
	\mathsf c^+_{i-J_1}\circledast
	\left(
	\mathop{\circledast}_{a=2}^{d_l}
	\mathsf x_{i-J_1+J_a}^{\boxast(d_r-1)}
	\right)
	\right],\qquad
	\mathsf F_{i-s}(\underline{\mathsf x}) =
	\mathbb E_{\boldsymbol{J}\sim\pi_w^{(s)}}
	\left[
	\mathsf c^+_{i-J_1}\circledast
	\left(
	\mathop{\circledast}_{a=2}^{d_l}
	\mathsf x_{i-J_1+J_a}^{\boxast(d_r-1)}
	\right)
	\right].
\end{aligned} 
\]
Hence, the two outputs are expectations of the same expression under distinct offset distributions. For any linear functional \(f\) on \(\mathcal M_q\) that is bounded on \(\mathcal X_q\), by linearity,
\[
f\left(
\mathsf F_i(\underline{\mathsf x})
-\mathsf F_{i-s}(\underline{\mathsf x})
\right) =
\sum_{\boldsymbol j\in\mathbb Z^{d_l}}
\left[
\pi_w(\boldsymbol j)-\pi_w^{(s)}(\boldsymbol j)
\right] \cdot
f\left(
\mathsf c^+_{i-j_1}\circledast
\left(
\mathop{\circledast}_{a=2}^{d_l}
\mathsf x_{i-j_1+j_a}^{\boxast(d_r-1)}
\right)
\right).
\]
Since \(\sum_{\boldsymbol j}
 [
\pi_w(\boldsymbol j)-\pi_w^{(s)}(\boldsymbol j)
 ]=0\),  by the triangle inequality and  \(\eta_w(s)=d_{\mathrm{TV}}(\pi_w,\pi_w^{(s)})\),  
\begin{equation}
|f(
\mathsf F_i(\underline{\mathsf x})
-\mathsf F_{i-s}(\underline{\mathsf x})
)| \le
\frac12
\sum_{\boldsymbol j\in\mathbb{Z}^{d_l}}
\left|
\pi_w(\boldsymbol j)-\pi_w^{(s)}(\boldsymbol j)
\right| \cdot
\left[
\sup_{\mathsf p\in\mathcal X_q}f(\mathsf p)
-\inf_{\mathsf p\in\mathcal X_q}f(\mathsf p)
\right] =
\eta_w(s)
\left[
\sup_{\mathsf p\in\mathcal X_q}f(\mathsf p)
-\inf_{\mathsf p\in\mathcal X_q}f(\mathsf p)
\right].\label{eq:ach-offset-oscillation}
\end{equation}
We next relate this DE-output difference  to the shift difference \(\mathsf d_i\).  Put
\[\mathsf{b}_i\coloneqq \mathsf F_i(\underline{\mathsf x})
-\mathsf F_{i-s}(\underline{\mathsf x}),\qquad 1\le i\le M.\]
By \eqref{eq:ach-limit-profile},  \(\mathsf F_j(\underline{\mathsf x}) 
=\mathsf x_j\) for every \(j\le K\) (both are \(\Delta_\infty\) for \(j\le 0\)). Then, for every \(1\le i\le M\),    \(\mathsf F_{i-s}(\underline{\mathsf x})
=\mathsf x_{i-s}\) since \(i-s\le K-1\)  and thus
\[
\mathsf b_i -\mathsf d_i =
\mathsf F_i(\underline{\mathsf x})
-\mathsf x_{i-s}
-(\mathsf x_i-\mathsf x_{i-s}) =
\mathsf F_i(\underline{\mathsf x})-\mathsf x_i =
\begin{cases}
	0,&1\le i\le K,\\[2pt]
	\mathsf F_i(\underline{\mathsf x})-\mathsf a,
	&K<i\le M.
\end{cases}
\] 
Moreover, by \(\eqref{eq:ach-spatial-order}\), \(\mathsf a
=\mathsf F_K(\underline{\mathsf x}) 
\preceq
\mathsf F_i(\underline{\mathsf x})\) for \(i>K\), so \(\mathsf{b}_i-\mathsf{d}_i\) is either zero or a difference of degradation-ordered probability measures in \(\mathcal{X}_q\). Hence, \(\mathsf d_i\), \(\mathsf b_i\), and \(\mathsf b_i-\mathsf d_i\) are all differences of degradation-ordered probability measures. We now apply  \eqref{eq:ach-offset-oscillation} to the entropy terms in  \eqref{eq:P5}. For any \(1\le i,j\le M\) and \(\mathsf p,\mathsf p_1,\mathsf p_2,\mathsf r\in\mathcal X_q\), by Proposition IV.17,
\[
\begin{aligned}
	0 \le
	\mathrm H(\mathsf r\boxast\mathsf d_j\boxast\mathsf p)
	\le\mathrm H(\mathsf d_j),\qquad
	0 \le
	\mathrm H\left(
	\mathsf p\circledast
	(\mathsf p_1\boxast\mathsf r)
	\circledast
	(\mathsf p_2\boxast\mathsf d_j)
	\right) \le
	\mathrm H(\mathsf p_2\boxast\mathsf d_j)
	\le\mathrm H(\mathsf d_j).
\end{aligned} 
\]
Both expressions are linear in \(\mathsf r\). Applying \eqref{eq:ach-offset-oscillation} with \(f(\mathsf{r})=\mathrm H(\mathsf r\boxast\mathsf d_j\boxast\mathsf p)\) and \(f(\mathsf{r})=\mathrm H (
\mathsf p\circledast
(\mathsf p_1\boxast\mathsf r)
\circledast
(\mathsf p_2\boxast\mathsf d_j)
 )\),   we have 
\begin{equation}
\left|
\mathrm H(\mathsf b_i\boxast\mathsf d_j\boxast\mathsf p)
\right| \le
\eta_w(s)\mathrm H(\mathsf d_j),\qquad
\left|
\mathrm H\left(
\mathsf p\circledast
(\mathsf p_1\boxast\mathsf b_i)
\circledast
(\mathsf p_2\boxast\mathsf d_j)
\right)
\right| \le
\eta_w(s)\mathrm H(\mathsf d_j). \label{eq:P6}
\end{equation}
Since \(\mathsf b_i-\mathsf d_i\) is a degradation-ordered difference, using \eqref{eq:entropy_dual_cons} twice and  by \eqref{eq:P6}, we   have
\begin{equation}
0 \ge
\mathrm H(\mathsf d_i\boxast\mathsf d_j\boxast\mathsf p) =
\mathrm H(\mathsf b_i\boxast\mathsf d_j\boxast\mathsf p)
-
\mathrm H\left(
(\mathsf b_i-\mathsf d_i)
\boxast\mathsf d_j\boxast\mathsf p
\right) \ge
\mathrm H(\mathsf b_i\boxast\mathsf d_j\boxast\mathsf p) \ge
-\eta_w(s)\mathrm H(\mathsf d_j).\label{eq:P7}
\end{equation}
Similarly, we also have 
\begin{align}
	0
	&\le
	\mathrm H\left(
	\mathsf p\circledast
	(\mathsf p_1\boxast\mathsf d_i)
	\circledast
	(\mathsf p_2\boxast\mathsf d_j)
	\right) =
	\mathrm H\left(
	\mathsf p\circledast
	(\mathsf p_1\boxast\mathsf b_i)
	\circledast
	(\mathsf p_2\boxast\mathsf d_j)
	\right) -
	\mathrm H\left(
	\mathsf p\circledast
	\bigl(\mathsf p_1\boxast(\mathsf b_i-\mathsf d_i)\bigr)
	\circledast
	(\mathsf p_2\boxast\mathsf d_j)
	\right)\notag\\
	&\le
	\mathrm H\left(
	\mathsf p\circledast
	(\mathsf p_1\boxast\mathsf b_i)
	\circledast
	(\mathsf p_2\boxast\mathsf d_j)
	\right) \le
	\eta_w(s)\mathrm H(\mathsf d_j).\label{eq:P8}
\end{align} 
The bounds \eqref{eq:P7} and \eqref{eq:P8} remain valid for all   \(i,j\in\mathbb{Z}\), since \(\mathsf d_i=0\) outside \(\{1,\ldots,M\}\). We now apply the two bounds to\,\,the second derivative. Set \(\underline{\mathsf z}
=
\underline{\mathsf x}-t\underline{\mathsf d}\) for \(0\le t\le1\). When \(d_r\ge3\), taking \(j=i\) and \(\mathsf p
=
\mathsf F_i(\underline{\mathsf z})
\boxast\mathsf z_i^{\boxast(d_r-3)}\) in \eqref{eq:P7}, we obtain 
\begin{equation}
\mathrm H\left(
\mathsf F_i(\underline{\mathsf z})
\boxast
\mathsf z_i^{\boxast(d_r-3)}
\boxast\mathsf d_i^{\boxast2}
\right)
\ge
-\eta_w(s)\mathrm H(\mathsf d_i).\label{eq:P9}
\end{equation}
Similarly, taking \((i,j)\mapsto (j+J_1,j+J_2)\), \(\mathsf p =\mathsf c\circledast
 (
\mathop{\circledast}_{a=3}^{d_l}
\mathsf z_{j+J_a}^{\boxast(d_r-1)}
 )\), \(\mathsf p_1 =\mathsf z_{j+J_1}^{\boxast(d_r-2)}\) and \(\mathsf p_2=\mathsf z_{j+J_2}^{\boxast(d_r-2)}\) in \eqref{eq:P8}, we have
\begin{equation} 
\mathrm H\left(
\mathsf c
\circledast\mathsf u_{j+J_1}
\circledast\mathsf u_{j+J_2}
\circledast
\left(
\mathop{\circledast}_{a=3}^{d_l}
\mathsf z_{j+J_a}^{\boxast(d_r-1)}
\right)
\right) \le
\eta_w(s)\mathrm H(\mathsf d_{j+J_2}).\label{eq:P10}
\end{equation}
\eqref{eq:P10} holds for every realization of \(\boldsymbol{J}\), since \eqref{eq:P8} is uniform over all   \(\mathsf p,\mathsf p_1,\mathsf p_2\in\mathcal{X}_q\). Furthermore, for every realization of \(J_2\),
\begin{equation}
\sum_{j=1}^{M}
\mathrm H(\mathsf d_{j+J_2}) =
\sum_{i=1+J_2}^{M+J_2}
\mathrm H(\mathsf d_i) \le
\sum_{i=1}^{M}
\mathrm H(\mathsf d_i),\label{eq:P11}
\end{equation}
since \(J_2\ge0\), \(\mathrm H(\mathsf d_i)\ge0\), and \(\mathsf d_i=0\) outside \(\{1,\ldots,M\}\). Recall that \(\phi''(t)
=
\mathrm D^2U_{c,M}
(\underline{\mathsf z};\mathsf c)
[\underline{\mathsf d},\underline{\mathsf d}]\). Substituting \eqref{eq:P9} and \eqref{eq:P10} into \eqref{eq:P5}, and then using \eqref{eq:P11}, we obtain
\begin{align}
&\frac{\phi''(t)}{d_l(d_r-1)}\notag\\
&\ge
-(d_r-2)\eta_w(s)
\sum_{i=1}^{M}\mathrm H(\mathsf d_i) -
(d_l-1)(d_r-1)\eta_w(s)
\sum_{j=1}^{M}
\mathbb E_{\boldsymbol J}
\mathrm H(\mathsf d_{j+J_2}) \ge
- [d_l(d_r-1)-1 ]
\eta_w(s)
\sum_{i=1}^{M}\mathrm H(\mathsf d_i)\notag\\
&=- [d_l(d_r-1)-1 ]
\eta_w(s)\left(\sum_{i=M-s+1}^M\mathrm H(\mathsf x_i)
-\sum_{i=1-s}^{0}\mathrm H(\mathsf x_i) \right)=- [d_l(d_r-1)-1 ]
s\eta_w(s)\mathrm H(\mathsf a).\notag
\end{align}
Recalling that \(\eta_w(s)=\eta_w\), we conclude that
\[
\phi''(t) \ge
-d_l(d_r-1)
\bigl[d_l(d_r-1)-1\bigr]
s\eta_w(s)\mathrm H(\mathsf a) \ge
-C_{q,d_l,d_r}\,s\eta_w,
\qquad \forall 0\le t\le1,
\]
where \(C_{q,d_l,d_r}
=
d_l(d_r-1)
\bigl[d_l(d_r-1)-1\bigr]\log q\). Since \(\phi\) is a polynomial, by \eqref{eq:direct_bound}, Taylor’s formula with integral remainder, \(\phi'(0)\ge 0\), \(\phi''(t) \ge -C_{q,d_l,d_r} s\eta_w\) and the condition \(\frac{1}{2}C_{q,d_l,d_r}\eta_w< \Delta E(\mathsf{c})\), we obtain the contradiction
\[-s\Delta E(\mathsf{c})\ge \phi(1)-\phi(0)=\phi'(0)+\int_0^1(1-t)\phi''(t)\,dt  \ge-\frac12 C_{q,d_l,d_r}\,s\eta_w
>-s\Delta E(\mathsf c).\]
Thus, \(\underline{\mathsf{x}}=\underline{\Delta_\infty}\),  and the degradation comparison \(\underline{\mathsf x}_{\mathrm{orig}}^{(\infty)}
\preceq\underline{\mathsf x}\) established above implies that \(\underline{\Delta_\infty}\) is the unique fixed point of the original coupled DE.
\subsection{Proof of Theorem V.4 ii)}
We now prove the converse part in Theorem V.4. Fix \(h>h^{\mathrm{FP}}\) and \(w\), and put \(\mathsf c=\mathsf c_h\). By Lemma V.3 iii), \(\Delta E(\mathsf c)<0\).\,\,Thus, there exists \(\mathsf a\in\mathcal F(\mathsf c)\) such that \(U_s(\mathsf a;\mathsf c)<0\). Let \(\underline{\mathsf a}=(\mathsf a,\ldots,\mathsf a)\in\mathcal X_q^K\), where \(K=2L+w-1\). Since \(\mathsf c_j\preceq\mathsf c\) and \(\mathsf a_j\preceq\mathsf a\) for all \(j\in\mathbb Z\), where \(\underline{\mathsf a}\) is extended by \(\Delta_\infty\) outside \(\{1,\ldots,K\}\), by Proposition IV.16 and \(\eqref{eq:coupled_DE}\), for \(1\le i\le K\),
\[
[\mathsf T_c(\underline{\mathsf a};\mathsf c)]_i =\mathbb E_{\boldsymbol J}\left[
\mathsf c_{i-J_1}\circledast
\left(
\mathop{\circledast}_{a=2}^{d_l}
\mathsf a_{i-J_1+J_a}^{\boxast(d_r-1)}
\right)\right] \preceq
\mathsf c\circledast
\left(\mathsf a^{\boxast(d_r-1)}\right)^{\circledast(d_l-1)}
=\mathsf T_s(\mathsf a;\mathsf c)
=\mathsf a.
\]
Thus, \(\mathsf T_c(\underline{\mathsf a};\mathsf c)\preceq\underline{\mathsf a}\), and the coupled DE sequence initialized at \(\underline{\mathsf a}\) is decreasing and converges to a fixed point. By Lemma V.3 ii) and continuity of \(U_c\),
\[
U_c\!\left(
\mathsf T_c^{(\infty)}(\underline{\mathsf a};\mathsf c);
\mathsf c
\right)
\le U_c(\underline{\mathsf a};\mathsf c).
\]
Moreover, by the definitions of \(U_c\) and \(U_s\),
\[
U_c(\underline{\mathsf a};\mathsf c) =(2L+w-1)U_s(\mathsf a;\mathsf c) +(w-1)\mathrm H\left(
\mathsf c\circledast
\left(\mathsf a^{\boxast(d_r-1)}\right)^{\circledast d_l}
\right) \le (2L+w-1)U_s(\mathsf a;\mathsf c)
+(w-1)\log q<0
\]
for all sufficiently large \(L\), since \(w\) is fixed and \(U_s(\mathsf a;\mathsf c)<0\). Thus, since \(U_c(\underline{\Delta_\infty};\mathsf c)=0\), by the monotonicity of coupled DE,
\[
\mathsf T_c^{(\infty)}
(\underline{\Delta_0};\mathsf c)
\succeq
\mathsf T_c^{(\infty)}
(\underline{\mathsf a};\mathsf c)
\ne\underline{\Delta_\infty}.
\] 
This completes the proof.

\section{Conclusion and future work}

In this paper, we investigated random SC-LDPC
code ensembles over arbitrary finite fields
$\mathbb F_q$, constructed through general
variable-node edge spreading and independent
random monomial maps.
For both the standard and improved coupled ensembles, with fixed coupling parameters and degrees \(d_r\ge d_l\ge3\), we established asymptotically good
minimum distance and minimum stopping-set size,
together with precise polynomial estimates
for the corresponding failure probabilities.
We also derived\,\,explicit achievable bounds that
reveal the role of edge spreading in controlling
distance performance. The improved
ensemble has stronger explicit lower bounds
by directing the edges of each variable node
to distinct check positions. For iterative decoding over general QMSCs,
we developed a unified analytical framework
based on symmetric probability measures on
the probability simplex. This framework includes
the properties of BP updates and information
functionals, compact metric topologies,
degradation ordering, and the stability and
convergence of DE. Building on these results, we established threshold
saturation for the standard and improved coupled ensembles
over any finite field, under the stability condition of
the uncoupled system at $h^{\mathrm{FP}}$.
This condition holds automatically when $d_l\ge3$.
Specifically, for complete QMSC families ordered by
degradation, the coupled BP threshold converges to
$h^{\mathrm{FP}}$ as the coupling length $L\to\infty$,
followed by the coupling width $w\to\infty$.
The result also applies to more general exchangeable
edge-spreading laws satisfying $\eta_w\to0$ as
$w\to\infty$.

Beyond these theoretical results, our analysis
provides complementary criteria for evaluating
and designing edge-spreading rules. The explicit
distance bounds depend on the concentration of
variable-node edges at individual check positions,
whereas the sufficient condition for threshold
saturation depends on the shift-regularity of
their joint offset distribution. These criteria
may help guide the design of coupled codes
with favorable distance and decoding thresholds,
while accounting for the rate loss caused by
termination. 

Finally, although our theorem characterizes the limiting BP threshold through \(h^{\mathrm{FP}}\) for fixed degrees, its large-degree behavior is beyond the scope of this paper. In future work, we plan to establish that \(h^{\mathrm{FP}}\) approaches the Shannon threshold \(r\log q\)\,\,as \(d_l,d_r\to\infty\) with \(d_l/d_r\to r\in(0,1)\) for general   QMSC families. We also aim to investigate finite-length decoding performance, particularly the number of BP iterations required for reliable decoding and the trade-offs among decoding complexity, latency, and coupling parameters.



{\appendices
\counterwithin*{theorem}{section}
\renewcommand{\thetheorem}{A\arabic{section}.\arabic{theorem}}
\section{Minimum Distance and Stopping Set Size Analysis}
\subsection{Proof of Proposition III.14}
Fix a feasible support type \(\underline\ell\). The number of physical-variable supports of this type is
\[
\prod_{k=1}^{N}\prod_{\underline t\in\mathcal T}
\binom{np(\underline t)}{\ell_{k,\underline t}}.
\]
For any fixed vector \(\boldsymbol c\in\mathbb{F}_q^{Nn}\) with this support type, exactly \(e_i\) of its sockets carry nonzero values into check position \(i\).\,\,Since the monomial maps are uniform and independent across check positions, these values are mapped uniformly onto the weight-\(e_i\) vectors in \(\mathbb F_q^{nd_l}\). Consequently,
\[
\Pr\{\boldsymbol c\in\mathcal C_{d_l,d_r,w,L,n}\}
=
\prod_{i=1}^{K}
\frac{[z^{e_i}]W_{q,d_r}(z)^m}
{\binom{nd_l}{e_i}(q-1)^{e_i}},
\]
where the numerator counts the weight-\(e_i\) vectors satisfying the \(m\) parity checks at position \(i\). Each support of size \(\ell\) admits \((q-1)^\ell\) vectors with that support. Hence, using \(\sum_{i=1}^{K}e_i=d_l\ell\), we have
\[
	\mathbb E A_{\underline\ell} =
\left[
\prod_{k=1}^{N}\prod_{\underline t\in\mathcal T}
\binom{np(\underline t)}{\ell_{k,\underline t}}
\right]
(q-1)^\ell
\prod_{i=1}^{K}
\frac{[z^{e_i}]W_{q,d_r}(z)^m}
{\binom{nd_l}{e_i}(q-1)^{e_i}} =
(q-1)^{-(d_l-1)\ell}
\left[
\prod_{k=1}^{N}\prod_{\underline t\in\mathcal T}
\binom{np(\underline t)}{\ell_{k,\underline t}}
\right]
\prod_{i=1}^{K}
\frac{[z^{e_i}]W_{q,d_r}(z)^m}
{\binom{nd_l}{e_i}}.
\]
This proves \eqref{AVE_WTD1}. For stopping sets, fix a physical-variable support \(S\) of type \(\underline\ell\). At check position \(i\), its \(e_i\) sockets are matched uniformly to an \(e_i\)-subset of the \(nd_l=md_r\) check sockets. Since
\[
\widetilde W_{d_r}(z)
=(1+z)^{d_r}-d_rz
=\sum_{ {0\le j\le d_r:j\ne1}}
\binom{d_r}{j}z^j,
\]
the coefficient \([z^{e_i}]\widetilde W_{d_r}(z)^m\) counts precisely those subsets for which no check node is incident to exactly one selected edge,\,so
\[
\Pr\{S\text{ is a stopping set in }\mathcal G_{d_l,d_r,w,L,n}\}
=
\prod_{i=1}^{K}
\frac{[z^{e_i}]\widetilde W_{d_r}(z)^m}
{\binom{nd_l}{e_i}}.
\]
Summing over all supports of type \(\underline\ell\) proves \eqref{AVE_STD1}:
\[
\mathbb E\widetilde A_{\underline\ell}
=
\left[
\prod_{k=1}^{N}\prod_{\underline t\in\mathcal T}
\binom{np(\underline t)}{\ell_{k,\underline t}}
\right]
\prod_{i=1}^{K}
\frac{[z^{e_i}]\widetilde W_{d_r}(z)^m}
{\binom{nd_l}{e_i}},
\]  
\subsection{Proof of Proposition III.15} 
First, we establish continuity. For either single-check enumerator \(P\in\{W_{q,d_r},\widetilde W_{d_r}\}\), write \(P(z)=\sum_{r=0}^{D}c_rz^r\), where\,\,\(D=\deg P\), \(c_r\ge0\), and \(c_0,c_D>0\). Then
\[
\Phi_P(x)
\coloneqq
\log\inf_{z>0}\frac{P(z)}{z^x}
=
\inf_{s\in\mathbb R}
\{\log P(e^s)-sx\}.
\]
The function \(\Phi_P\) is finite precisely on \([0,D]\). As an infimum of continuous affine functions, it is concave and upper semicontinuous, hence continuous on \([0,D]\). Now restrict the normalized support types to
\[
\mathcal D_P
=
\left\{
\underline a:
0\le a_{k,\underline t}\le p(\underline t),\
d_r\beta_i\le D,\ 1\le i\le K
\right\}.
\] 
The sets \(\mathcal D_P\cap\mathcal A(\alpha)\) are compact and vary continuously with \(\alpha\) wherever nonempty, by convex interpolation. Since the objectives defining \(g_{q,d_l,d_r,w,L}\) and \(\widetilde g_{d_l,d_r,w,L}\) are continuous on these sets, by Berge's maximum theorem, both functions are continuous on their respective finite domains. At \(\alpha=0\), only $\underline a=\underline0$ is feasible, so both functions equal \(0\). Next, using
\[
\frac{e^{MH_2(r/M)}}{M+1}
\le
\binom Mr
\le e^{MH_2(r/M)},
\qquad
[z^r]P(z)^m
\le
\inf_{z>0}\frac{P(z)^m}{z^r},
\] 
together with the exact enumerators in \(\eqref{AVE_WTD1}\) and \(\eqref{AVE_STD1}\), we obtain
\[
\begin{aligned}
	\mathbb E A_\ell
	&\le
	(n+1)^{N|\mathcal T|}
	(nd_l+1)^K
	\exp\left\{
	Nn\,g_{q,d_l,d_r,w,L}
	\left(\frac{\ell}{Nn}\right)
	\right\},\\
	\mathbb E\widetilde A_\ell
	&\le
	(n+1)^{N|\mathcal T|}
	(nd_l+1)^K
	\exp\left\{
	Nn\,\widetilde g_{d_l,d_r,w,L}
	\left(\frac{\ell}{Nn}\right)
	\right\}.
\end{aligned}
\] 
Since \(N,K,|\mathcal T|\) and \(d_l\) are fixed, these give the desired uniform spectral bounds. By the concavity of \(H_q\) and \(\sum_{b=0}^{w-1}t_b=d_l\),
\[
\begin{aligned}
 \sum_{k,\underline t}
	p(\underline t)
	H_q\left(
	\frac{a_{k,\underline t}}{p(\underline t)}
	\right) =
	\sum_{i=1}^{K}
	\sum_{b=0}^{w-1}
	\sum_{\underline t\in\mathcal T}
	\frac{t_bp(\underline t)}{d_l}
	H_q\left(
	\frac{a_{i-b,\underline t}}{p(\underline t)}
	\right) \le
	\sum_{i=1}^{K}H_q(\beta_i).
\end{aligned}
\] 
The same inequality holds for \(H_2\). Consequently, 
	\begin{align}
		Ng_{q,d_l,d_r,w,L}(\alpha) \le\max_{\underline a\in\mathcal A(\alpha)}
		\sum_{i=1}^K g_{q,d_l,d_r}(\beta_i),\qquad
		N\widetilde g_{d_l,d_r,w,L}(\alpha) \le\max_{\underline a\in\mathcal A(\alpha)}
		\sum_{i=1}^K \widetilde g_{d_l,d_r}(\beta_i).
		\label{eq:sec3_spectral_comparison}
	\end{align}
	Moreover, note that
	\[
	\sum_{i=1}^K\beta_i=N\alpha,
	\qquad
	0\le\beta_i\le\frac{Nt_{\max}}{d_l}\alpha.
	\]
	Since both single-check enumerators satisfy \(P(z)=1+O(z^2)\), choosing \(z=\sqrt{x}\) gives
	\[
	g(x)
	\le
	\left(\frac{d_l}{2}-1\right)
	x\log x+O(x)<0,
	\qquad x\downarrow0,
	\]
 for \(d_l\ge3\), where \(g\) represents either \(g_{q,d_l,d_r}\) or \(\widetilde g_{d_l,d_r}\). Thus, for all sufficiently small \(\alpha>0\), every \(\beta_i\) lies in the interval\,\,where \(g_{q,d_l,d_r}\) and \(\widetilde g_{d_l,d_r}\) are nonpositive. Since \(\max_{1\le i\le K}\beta_i
 \ge\frac{N\alpha}{K}>0\),	at least one of these terms is strictly negative, uniformly\,\,over the compact feasible set, so by \eqref{eq:sec3_spectral_comparison}, both \(\alpha_{q,d_l,d_r,w,L}\) and \(\widetilde{\alpha}_{d_l,d_r,w,L}\) are strictly positive. Furthermore, by the zero properties in Theorem III.6, if
	\[
	0<\alpha<
	\min\left\{
	1,\frac{d_l}{Nt_{\max}}\alpha_{q,d_l,d_r}
	\right\},
	\]
then \(0\le\beta_i<\alpha_{q,d_l,d_r}\) for every \(i\), and at least one \(\beta_i>0\). Therefore, \(\sum_{i=1}^{K}g_{q,d_l,d_r}(\beta_i)<0\), which proves the desired  lower bound on \(\alpha_{q,d_l,d_r,w,L}\). The same argument with \(\widetilde g_{d_l,d_r}\) proves the corresponding stopping-set bound on \(\widetilde\alpha_{d_l,d_r,w,L}\).  
\subsection{Proof of Theorem III.16}
We first establish the upper bounds simultaneously. Let \(A_\ell=A_\ell(\mathcal C_{d_l,d_r,w,L,n})\) and \(\widetilde A_\ell=\widetilde A_\ell(\mathcal G_{d_l,d_r,w,L,n})\). Since the support of every nonzero codeword is a stopping set,
\begin{equation}
A_\ell\le(q-1)^\ell\widetilde A_\ell.\label{eq:weight_ss_compare}
\end{equation}
Moreover, for either \(B_\ell=A_\ell\) or \(B_\ell=\widetilde A_\ell\), by the union bound and Markov's inequality,
\begin{equation}
\Pr\{d\le N\alpha n\}
\le
\sum_{\ell=1}^{\lfloor N\alpha n\rfloor}
\mathbb E B_\ell,\label{eq:P_lb}
\end{equation}
where \(d\) and \(\alpha\) denote the corresponding minimum-size quantity and spectral parameter. We first bound the contributions from small \(\ell\). Put \(M=nd_l=md_r\) and \(C_0=2^{d_r}-d_r-1\). Since
\[
\widetilde W_{d_r}(z)
=1+\sum_{j=2}^{d_r}\binom{d_r}{j}z^j
\le e^{C_0z^2},
\qquad 0<z\le1,
\]
choosing \(z=\sqrt{r/M}\) and using \(\binom Mr\ge(M/r)^r\), we obtain
\[ 
	\frac{[z^r]\widetilde W_{d_r}(z)^m}{\binom Mr} \le
	\frac{e^{C_0mz^2}}{z^r\binom Mr} \le
	\left(
	e^{C_0/d_r}\sqrt{\frac rM}
	\right)^r,
	\qquad 1\le r\le M. 
\]
For a fixed subset \(S\) of \(\ell\) physical variable nodes, let \(e_i\) be its number of sockets entering check position \(i\), as   in \eqref{eq:ell_e_i}. Since
\[
\sum_{i=1}^{K}e_i=d_l\ell,
\qquad
0\le e_i\le d_l\ell,
\]
by
the independence of the check-position matchings, we have
\[
\begin{aligned}
	\Pr\{S\text{ is a stopping set in }\mathcal G_{d_l,d_r,w,L,n}\} =
	\prod_{i=1}^{K}
	\frac{[z^{e_i}]\widetilde W_{d_r}(z)^m}
	{\binom{M}{e_i}} \le
	\prod_{i=1}^{K}
	\left(C_1\sqrt{\frac{e_i}{M}}\right)^{e_i} \le
	\left(C_1\sqrt{\frac{\ell}{n}}\right)^{d_l\ell},
\end{aligned}
\]
where \(C_1=e^{C_0/d_r}\) and factors with \(e_i=0\) are interpreted as one. Consequently,
\[ 
	\mathbb E\widetilde A_\ell \le
	\binom{Nn}{\ell}
	\left(C_1\sqrt{\frac{\ell}{n}}\right)^{d_l\ell} \le
	\left[
	C_2\left(\frac{\ell}{n}\right)^{d_l/2-1}
	\right]^\ell, 
\] 
and by \eqref{eq:weight_ss_compare}, upon increasing \(C_2\) if necessary, for either \(B_\ell=A_\ell\) or \(B_\ell=\widetilde A_\ell\),
\[
\mathbb E B_\ell
\le
\left[
C_2\left(\frac{\ell}{n}\right)^{d_l/2-1}
\right]^\ell=:a_\ell.
\] 
Here and below, all constants are independent of \(n\) and \(\ell\). Set \(\gamma\coloneqq d_l/2-1>0\). Choose a sufficiently small constant \(\eta>0\) such that
\[
C_2(e \eta)^\gamma<\frac12.
\]
For \(a_\ell=[C_2(\ell/n)^\gamma]^\ell\), we have \(a_2=O(n^{2-d_l})\) and \( {a_{\ell+1}/a_\ell}
=
C_2 (\frac{\ell+1}{n} )^\gamma
 (1+\frac1\ell )^{\gamma\ell}
\le\frac12\) for \(2\le\ell<\lfloor \eta n\rfloor\).
Hence,
\[
\sum_{\ell=2}^{\lfloor\eta n\rfloor}
\mathbb E B_\ell
\le
2a_2
=O(n^{2-d_l}).
\] 
For \(\ell=1\), consider a singleton at variable position \(k\) with edge type \(\underline t=(t_0,\ldots,t_{w-1})\). By \(\eqref{AVE_STD1}\),
\begin{equation}
\mathbb E\widetilde A_1
=
\sum_{k=1}^{N}\sum_{\underline t\in\mathcal T}
np(\underline t)
\prod_{b=0}^{w-1}
\frac{[z^{t_b}]\widetilde W_{d_r}(z)^m}
{\binom{nd_l}{t_b}}.\label{eq:EA_1}
\end{equation}
Since \(\widetilde W_{d_r}(z)
=1+\sum_{j=2}^{d_r}\binom{d_r}{j}z^j\) has no linear term, each contribution to the coefficient of \(z^r\) in \(\widetilde W_{d_r}(z)^m\) contains at most \(\lfloor r/2\rfloor\) nonconstant factors. Thus, for every fixed \(r\ge0\),
\[
[z^r]\widetilde W_{d_r}(z)^m
=O\!\left(m^{\lfloor r/2\rfloor}\right).
\]
Since \(m=d_ln/d_r\) and \(t_b\le d_l\) are fixed, using \(\binom{nd_l}{t_b}\ge (nd_l/t_b)^{t_b}\), we have
\[ 
	\frac{[z^{t_b}]\widetilde W_{d_r}(z)^m}
	{\binom{nd_l}{t_b}} =
	O\!\left(n^{\lfloor t_b/2\rfloor-t_b}\right) =
	O\!\left(n^{-\lceil t_b/2\rceil}\right). 
\] 
Substituting the above estimate into \eqref{eq:EA_1}, we obtain
\[
\mathbb E\widetilde A_1 \le
\sum_{k=1}^{N}\sum_{\underline t\in\mathcal T}
O\!\left(
n^{1-\sum_{b=0}^{w-1}\lceil t_b/2\rceil}
\right) =
O\!\left(n^{1-\lceil d_l/2\rceil}\right),
\] 
where we use \(\sum_{b=0}^{w-1}
 \lceil\frac{t_b}{2} \rceil
\ge
\big\lceil\frac12\sum_{b=0}^{w-1}t_b\big\rceil
=
 \lceil\frac{d_l}{2} \rceil\) and the fact that \(N\) and \(|\mathcal T|\) are fixed. Since \(A_1\le(q-1)\widetilde A_1\), we obtain
\[ 
	\mathbb E A_1,  \mathbb E\widetilde A_1
	=O\!\left(n^{1-\lceil d_l/2\rceil}\right).
\]
For the improved ensemble, every edge type belongs to \(\{0,1\}^w\), so parallel edges are excluded, and thus no weight-\(1\) codeword or size-\(1\) stopping set can exist, i.e., \(A_1=\widetilde A_1=0\). For the standard ensemble, when \(q=2\) and \(d_l\) is odd, every codeword has even Hamming weight, so \(A_1=0\). Next, by the definitions of \(\alpha_{q,d_l,d_r,w,L}\) and \(\widetilde{\alpha}_{d_l,d_r,w,L}\) and the continuity in Proposition\,III.15, for any fixed \(0<\alpha<\alpha_{q,d_l,d_r,w,L}\) and \(0<\eta<N\alpha\), it follows that
\[
\max_{x\in[\eta/N,\alpha]}
g_{q,d_l,d_r,w,L}(x)<0.
\]
The same inequality holds for \(\widetilde g_{d_l,d_r,w,L}\) whenever \(0<\widetilde\alpha<\widetilde\alpha_{d_l,d_r,w,L}\). Thus, by \(\eqref{eq:sec3_uniform_spectral_bound}\),
\[ 
	\sum_{\ell=\lceil\eta n\rceil}^{\lfloor N\alpha n\rfloor}
	\mathbb E A_\ell \le
	\exp\left\{
	Nn\max_{x\in[\eta/N,\alpha]}
	g_{q,d_l,d_r,w,L}(x)+O(\log n)
	\right\} =e^{-\Omega(n)}, 
\] 
and likewise, \(\sum_{\ell=\lceil\eta n\rceil}^{\lfloor N\widetilde\alpha n\rfloor}
\mathbb E\widetilde A_\ell
=e^{-\Omega(n)}\).
Substituting the above bounds  for \(\ell=1\), \(2\le\ell\le\eta n\), and \(\ell\ge\eta n\) into \eqref{eq:P_lb}  proves all the stated probability upper bounds.

We now prove the matching lower bounds.  For the standard ensemble, choose the edge type
\[
\hat{\underline t}=(d_l,0,\ldots,0).
\]
At a fixed physical position, there are \(np(\hat{\underline t})=\Theta(n)\) variables of this type. For each such variable node \(v\), let \(E_v\) denote either the event that \(\{v\}\) is a stopping set or the event that it supports a weight-one codeword. By \(\eqref{AVE_WTD1}\) and \(\eqref{AVE_STD1}\),
\[
\Pr(E_v)=
\begin{cases}
	\displaystyle
	\frac{[z^{d_l}]\widetilde W_{d_r}(z)^m}
	{\binom M{d_l}},
	&\text{stopping set},\\[4mm]
	\displaystyle
	\frac{[z^{d_l}]W_{q,d_r}(z)^m}
	{\binom M{d_l}(q-1)^{d_l}},
	&\text{codeword}.
\end{cases}
\] 
Both \(\widetilde W_{d_r}(z)\) and \(W_{q,d_r}(z)\) have positive quadratic coefficients; their cubic coefficients are also positive except for \(W_{2,d_r}\). For even \(d_l\), the leading contribution to the coefficient of \(z^{d_l}\) comes from \(d_l/2\) quadratic factors; for odd \(d_l\), it comes from one cubic factor and \((d_l-3)/2\) quadratic factors. Thus,
\[
[z^{d_l}]P(z)^m
=
\Theta(m^{\lfloor d_l/2\rfloor})
\] 
for \(P=\widetilde W_{d_r}\), and also for \(P=W_{q,d_r}\) when \(q>2\) or \(d_l\) is even. Since \(\binom M{d_l}=\Theta(n^{d_l})\), 
\[
\Pr(E_v)
=
\Theta(n^{-\lceil d_l/2\rceil})
\] 
in these cases. For distinct variable nodes \(u,v\) of type \(\hat{\underline t}\), the event \(E_u\cap E_v\) requires the \(d_l\) sockets of each variable to occupy at most \(\lfloor d_l/2\rfloor\) check nodes. Under the uniform matching,
\[ 
	\Pr(E_u\cap E_v)
 \le
	\frac{O(m^{2\lfloor d_l/2\rfloor})}
	{(M)_{2d_l}} =
	O(n^{-2\lceil d_l/2\rceil}), 
\] 
where \((M)_r=M(M-1)\cdots(M-r+1)\). Since there are \(\Theta(n)\) candidate variables, by Bonferroni's inequality, 
\[
\begin{aligned}
	\Pr\{d=1\}\ge \Pr\left(\bigcup_v E_v\right)
	 \ge
	\sum_v\Pr(E_v)
	-\sum_{u<v}\Pr(E_u\cap E_v) \ge
	\Theta\!\left(n^{1-\lceil d_l/2\rceil}\right)
	-O\!\left(n^{2-2\lceil d_l/2\rceil}\right) =\Omega\!\left(n^{1-\lceil d_l/2\rceil}\right),
\end{aligned} 
\]
where \(d\) denotes the corresponding minimum distance or minimum stopping-set size. This establishes the matching lower\,\,bounds for the standard ensemble's stopping-set probability and its minimum-distance probability when \(q>2\) or \(d_l\) is even. Consider the remaining cases. Choose the edge type
\[
\underline t^*
=(\underbrace{1,\ldots,1}_{d_l},0,\ldots,0),
\] 
which has positive probability under both edge-spreading rules. At a fixed physical position, there are \(np(\underline t^*)=\Theta(n)\) variables of this edge type. For two distinct such variables \(u,v\), let \(F_{uv}\) denote either the event that \(\{u,v\}\) is a stopping set or the\,\,event that it supports a weight-two codeword. Since each variable has exactly one socket at each of \(d_l\) distinct check positions,\,\(\{u,v\}\) forms a stopping set precisely when their sockets meet the same check node at every such position. Conditioned on the matching of the first socket, the second must occupy one of the remaining \(d_r-1\) sockets of the same check node, among \(M-1\) available sockets. These matchings are independent across check positions. For a weight-\(2\) codeword, the two nonzero variable symbols must additionally satisfy all \(d_l\) parity-check equations. Conditioned on the socket matchings, each equation specifies a ratio between the two symbols. The ratios are independent and uniform on \(\mathbb F_q^\times\), so the probability that all \(d_l\) ratios coincide is \((q-1)^{1-d_l}\). Therefore,
\[
\Pr(F_{uv})=
\begin{cases}
	\displaystyle
	\left(\frac{d_r-1}{M-1}\right)^{d_l},
	&\text{stopping set} \\ 
	\displaystyle
	(q-1)^{1-d_l}
	\left(\frac{d_r-1}{M-1}\right)^{d_l},
	&\text{codeword} 
\end{cases}=\Theta(n^{-d_l}).
\]
Consider two distinct unordered pairs \(\{u,v\}\ne\{x,y\}\). Since every weight-\(2\) codeword induces a size-\(2\) stopping set, it suffices to bound the probability that both pairs satisfy the required socket coincidences. At each of the \(d_l\) check positions, either the two pairs share one variable, in which case three sockets must meet the same check node, or they are disjoint, in which case two prescribed pairs of sockets must separately meet check nodes. The respective probabilities are bounded by
\[
\begin{cases}
	\displaystyle
	\frac{(d_r-1)(d_r-2)}
	{(M-1)(M-2)},
	&\text{one shared variable} \\ 
	\displaystyle
	\frac{(d_r-1)^2}
	{(M-1)(M-3)},
	&\text{disjoint pairs} 
\end{cases}=O(n^{-2}).
\]
By independence across the \(d_l\) check positions, we obtain
\[
\Pr(F_{uv}\cap F_{xy})
=
O(n^{-2d_l}),
\qquad
\{u,v\}\ne\{x,y\}.
\] 
There are \(\binom{np(\underline t^*)}{2}=\Theta(n^2)\) candidate pairs \(\{u,v\}\) with \(u<v\). In the remaining cases, weight-\(1\) codewords or\,size-\(1\)\,stopping\,sets are impossible, as established above. Hence, the occurrence of \(F_{uv}\) implies \(d=2\), where \(d\) denotes the corresponding minimum distance or minimum stopping-set size. By Bonferroni's inequality,
\[ 
\Pr\{d=2\}\ge \Pr\left(\bigcup_{u<v}F_{uv}\right)
 \ge
	\sum_{u<v}\Pr(F_{uv})
	-\sum_{\{u,v\}<\{x,y\}}
	\Pr(F_{uv}\cap F_{xy}) \ge
	\Theta(n^{2-d_l})-O(n^{4-2d_l}) =\Omega(n^{2-d_l}), 
\] where the last step follows from \(d_l\ge3\). Together with the upper bounds, this completes the proof.
 \subsection{Proof of Corollary III.17}
 Fix \(0<\alpha<\alpha_{q,d_l,d_r,w,L}\) and
 \(0<\widetilde\alpha<\widetilde\alpha_{d_l,d_r,w,L}\).
 Under the stated degree conditions, both probability bounds in Theorem III.16 are \(O(n^{-2})\). Thus,
 \begin{align*}
 	\sum_{\substack{n\ge1:n\text{ admissible}}}
 	\left[ \Pr\!\left\{
 	\frac{d_{\min}(\mathcal C_{d_l,d_r,w,L,n})}{Nn}\le\alpha
 	\right\} +\Pr\!\left\{
 	\frac{d_{\mathrm{ss}}(\mathcal G_{d_l,d_r,w,L,n})}{Nn}
 	\le\widetilde\alpha
 	\right\}\right]<\infty.
 \end{align*}
Then, by the first Borel--Cantelli lemma, with probability one,
 neither event occurs for infinitely many admissible \(n\).
 Applying this conclusion to countable sequences
 \(\alpha\uparrow\alpha_{q,d_l,d_r,w,L}\) and
 \(\widetilde\alpha\uparrow\widetilde\alpha_{d_l,d_r,w,L}\) 
 proves  the desired achievability result.
\section{Iterative Decoding Analysis}
\subsection{Proof of Proposition IV.4} 
Since \(Y\sim m\) and the APP operator \(\psi\) is \(G\)-equivariant, \(\bar{\mathsf x}\) is \(G\)-invariant, so \(\bar{\mathsf x}\in\overline{\mathcal X}_q\). By the definition of \(\psi\),  
\[ 
	\mathsf x_i(B) =W(\psi^{-1}B\,|\, i) =q\int_{\psi^{-1}B}\psi_i(y)\,m(dy) =q\int_B y_i\,\bar{\mathsf x}(d\underline y)\qquad i\in\mathbb F_q, \ B\in\mathcal B(\mathcal S_q). 
\]
By the additive invariance of \(\bar{\mathsf x}\),  \(\mathsf x_0(B^{+i})
=
q\int_B y_i\,\bar{\mathsf x}(d\underline y)\) for  any \(i\in\mathbb{F}_q\) and Borel set \(B\subseteq\mathcal{S}_q\). Thus, with \(\mathsf{x}\coloneqq \mathsf{x}_0\), \(\mathsf x=\Phi\bar{\mathsf x}\in\mathcal X_q\).
Next, for any \(\mu\in\overline{\mathcal X}_q\), by the affine invariance of \(\mu\),  \((\Phi\mu)(B^{+a})
=
q\int_B y_a\,\mu(d\underline y)\). Thus, 
\[ 
	(\Psi\Phi\mu)(B) =\frac1q\sum_{a\in\mathbb F_q}
	(\Phi\mu)(B^{+a}) =\int_B\sum_{a\in\mathbb F_q}y_a\,\mu(d\underline y) =\mu(B). 
\]
So \(\Psi\Phi\) is the identity on \(\overline{\mathcal X}_q\). Since \(\mathcal X_q=\Phi(\overline{\mathcal X}_q)\), it follows that \(\Phi\Psi\) is also the identity on \(\mathcal X_q\). Finally, given any \(\mu\in\overline{\mathcal X}_q\),\,\,put
\[
K_\mu(B\mid i)
\coloneqq
q\int_B y_i\,\mu(d\underline y).
\]
Since \(\int y_i\,d\mu=1/q\), this defines a probability kernel. The \(G\)-symmetry of \(K_\mu\) follows from the affine invariance of \(\mu\). Under uniform input, the output distribution of \(K_\mu\) is \(\mu\), and its APP operator satisfies
\[
[\psi(\underline y)]_i
=
\frac1q
\frac{dK_\mu(\cdot\,|\, i)}{d\mu}(\underline y)
=
y_i
\qquad \mu\text{-a.e.}
\]
Thus every \(\mu\in\overline{\mathcal X}_q\) is realized as the marginal APP distribution of the QMSC \(K_\mu\), completing the proof.
\subsection{Proof of Lemma \ref{Closure,  Edge-Label Invariance, and Reference Distributions of BP Updates}}
 By Proposition IV.4, \(\bar{\mathsf x}_j\) is \(G\)-invariant, and for every \(i\in\mathbb F_q\),
\begin{equation}
\mathsf x_j(d\underline y)
=
qy_0\,\bar{\mathsf x}_j(d\underline y),
\qquad
\int y_i\,\bar{\mathsf x}_j(d\underline y)
=
\frac1q \label{eq:A1}.
\end{equation} 
We first prove \(\mathsf{x}_1\circledast \mathsf{x}_2\in\mathcal{X}_q\) and \eqref{eq:f_dPsi_circledast}. Define a Borel measure \(\nu\) on \(\mathcal{S}_q\) by
\[ 
	\nu(B)
	\coloneqq q\int_{\mathcal S_q\times\mathcal{S}_q} D(\underline y,\underline z)
	\mathbf1\{\mathcal V(\underline y,\underline z)\in B\} \cdot
	\bar{\mathsf x}_1(d\underline y)
	\bar{\mathsf x}_2(d\underline z),\qquad B\in\mathcal{B}(\mathcal{S}_q). 
\] 
By \eqref{eq:A1}, \(\nu(\mathcal S_q) =
q\sum_{i\in\mathbb F_q}
\left(\int y_i\,d\bar{\mathsf x}_1\right)
\left(\int z_i\,d\bar{\mathsf x}_2\right)
=1\), so \(\nu\) is a probability measure.
Moreover, for every \(g\in G\),
\[ 
	D(g\cdot\underline y,g\cdot\underline z)
	 =D(\underline y,\underline z),\qquad
	\mathcal V(g\cdot\underline y,g\cdot\underline z) =g\cdot\mathcal V(\underline y,\underline z). 
\]
Since both \(\bar{\mathsf x}_1\) and \(\bar{\mathsf x}_2\) are \(G\)-invariant, \(\nu\) is also \(G\)-invariant and thus \(\nu\in\overline{\mathcal X}_q\). For every bounded Borel \(f:\mathcal{S}_q\to\mathbb{R}\),
\begin{align*}
\int f\,d(\Phi\nu)& =
q\int w_0f(\underline w)\,
\nu(d\underline w) \overset{(\mathrm{a})}{=}
q^2\int_{\mathcal S_q\times\mathcal{S}_q}
D(\underline y,\underline z)
[\mathcal V(\underline y,\underline z)]_0  
f(\mathcal V(\underline y,\underline z))
\,\bar{\mathsf x}_1(d\underline{y})\bar{\mathsf x}_2(d\underline{z}) \overset{(\mathrm{b})}{=}
q^2\int_{\mathcal S_q\times\mathcal{S}_q}
y_0z_0f(\mathcal V(\underline y,\underline z))
\,d\bar{\mathsf x}_1d\bar{\mathsf x}_2\\
& \overset{(\mathrm{c})}{=}\int f(\mathcal{V}(\underline{y},\underline{z}))\mathsf x_1(d\underline y)\mathsf x_2(d\underline z)=
\int f\,d(\mathsf x_1\circledast\mathsf x_2).
\end{align*}
(a) follows from the definition of \(\nu\); (b) follows from \(D(\underline y,\underline z)
[\mathcal V(\underline y,\underline z)]_0
=y_0z_0\) (this   also holds when \(D=0\), which implies \(y_0z_0=0\)); (c) follows from \eqref{eq:A1}. Thus, it follows that
\[
\mathsf x_1\circledast\mathsf x_2
=
\Phi\nu \in\mathcal X_q.
\] 
Since \(\Psi\Phi=\mathrm{id}\), we obtain \(\Psi(\mathsf x_1\circledast\mathsf x_2)
=
\nu\), and \eqref{eq:f_dPsi_circledast} follows from the definition of \(\nu\). We now prove \(\mathsf{x}_1\boxast \mathsf{x}_2\in\mathcal{X}_q\) and \eqref{eq:f_dPsi_boxast}. Define   \(\mu 
\coloneqq
\mathcal C_\#
(\bar{\mathsf x}_1\otimes\bar{\mathsf x}_2)\). Given any \(a\in\mathbb{F}_q\) and \(b\in\mathbb{F}_q^{\times}\), let \(g,h\in G\) be the affine maps such that \(g(i)=a+bi\) and \(h(i)=bi\) for \(i\in\mathbb{F}_q\). Then, for all \(i\in\mathbb{F}_q\) and \(\underline{y},\underline{z}\in\mathcal{S}_q\),
\[ 
	[\mathcal C(g\cdot\underline y,h\cdot\underline z)]_i =
	\sum_{\substack{u,v\in\mathbb F_q\\u+v=i}}
	(g\cdot\underline y)_u
	(h\cdot\underline z)_v =
	\sum_{\substack{u,v\in\mathbb F_q\\u+v=i}}
	y_{(u-a)/b}z_{v/b} =
	\sum_{\substack{r,s\in\mathbb F_q\\r+s=(i-a)/b}}
	y_rz_s =
	[\mathcal C(\underline y,\underline z)]_{g^{-1}i}  =
	[g\cdot\mathcal C(\underline y,\underline z)]_i. 
\]
Thus, for every \(\underline{y},\underline{z}\in\mathcal{S}_q\),
\begin{equation}
\mathcal C(g\cdot\underline y,h\cdot\underline z)
=
g\cdot\mathcal C(\underline y,\underline z).\label{eq:A2}
\end{equation}
Let \(\underline Y\sim\bar{\mathsf x}_1\) and \(\underline Z\sim\bar{\mathsf x}_2\) be independent random probability vectors. Then \(\mu=\operatorname{Law}(\mathcal C(\underline Y,\underline Z))\), and since \(\bar{\mathsf x}_1,\bar{\mathsf x}_2\) are \(G\)-invariant, we have  \((g\cdot\underline Y,h\cdot\underline Z)
\overset d=
(\underline Y,\underline Z)\). Thus, by \eqref{eq:A2}, \(g\cdot\mathcal C(\underline Y,\underline Z)
 =
\mathcal C(g\cdot\underline Y,h\cdot\underline Z) \overset d=
\mathcal C(\underline Y,\underline Z)\), which implies \(g_\#\mu
=
\mu\). Since \(g\in G\) can be arbitrary, \(\mu\) is a \(G\)-invariant probability measure and thus \(\mu\in \overline{\mathcal{X}}_q\). For any bounded Borel \(f:\mathcal{S}_q\to\mathbb{R}\), we have
\begin{align*}
	\int f\,d(\Phi\mu )
	&=
	q\int_{\mathcal S_q\times\mathcal S_q}
	[\mathcal C(\underline y,\underline z)]_0
	f(\mathcal C(\underline y,\underline z))
	\,\bar{\mathsf x}_1(d\underline{y})\bar{\mathsf x}_2(d\underline{z})=
	q\sum_{r\in\mathbb F_q}
	\int_{\mathcal S_q\times \mathcal{S}_q}
	y_rz_{-r}f(\mathcal C(\underline y,\underline z))
	\,\bar{\mathsf x}_1(d\underline{y})\bar{\mathsf x}_2(d\underline{z})\\
	&\overset{(\mathrm{a})}{=}
	q^2\int_{\mathcal S_q\times\mathcal{S}_q}
	y_0z_0f(\mathcal C(\underline y,\underline z))
	\,\bar{\mathsf x}_1(d\underline{y})\bar{\mathsf x}_2(d\underline{z}) =
	\int f\,d(\mathsf x_1\boxast\mathsf x_2).
\end{align*} 
(a) follows since, for each \(r\in\mathbb{F}_q\), with \(g,h\in G\) defined by \(g(i)=i-r\) and \(h(i)=i+r\), we have \(\mathcal C(\underline y,\underline z)=\mathcal C(g\cdot\underline y,
h\cdot\underline z)\)\,and
\[ 
 \int y_rz_{-r}f(\mathcal C(\underline y,\underline z))
	\,d\bar{\mathsf x}_1d\bar{\mathsf x}_2 =\int
	[g\cdot\underline y]_0
	[h\cdot\underline z]_0
	f\bigl(\mathcal C(g\cdot\underline y,
	h\cdot\underline z)\bigr)
	\,d\bar{\mathsf x}_1d\bar{\mathsf x}_2 =\int y_0z_0f(\mathcal C(\underline y,\underline z))
	\,d\bar{\mathsf x}_1d\bar{\mathsf x}_2,
\]
where the last step follows since  \(\bar{\mathsf x}_1,\bar{\mathsf x}_2\) are \(G\)-invariant. Hence, \(\mathsf x_1\boxast\mathsf x_2
=
\Phi\mu \in\mathcal X_q\) and thus \(\Psi(\mathsf x_1\boxast\mathsf x_2)
=
\mu\). By the definition of \(\mu\), this proves \eqref{eq:f_dPsi_boxast}. Finally, for any \(a,b,c\in\mathbb F_q^\times\), a change of variables gives
\[
\mathcal C_{a,b,c}(\underline y,\underline z)
=
\mathcal C\left(
\underline y^{\times(-c/a)},
\underline z^{\times(-c/b)}
\right). 
\] 
Since  \(\mathsf{x}_1,\mathsf{x}_2\in \mathcal X_q\) are invariant under multiplicative coordinate reindexings, the above   implies \((\mathcal C_{a,b,c})_\#
(\mathsf x_1\otimes\mathsf x_2) =
\mathcal C_\#
(\mathsf x_1\otimes\mathsf x_2) =
\mathsf x_1\boxast\mathsf x_2\).
This completes the proof.
\subsection{Algebraic Properties of BP Convolutions}
The commutativity of \(\circledast\) and \(\boxast\) follows from \(\mathcal V(\underline y,\underline z)
=
\mathcal V(\underline z,\underline y),\mathcal C(\underline y,\underline z)
=
\mathcal C(\underline z,\underline y)\) for all \(\underline{y},\underline{z}\in\mathcal{S}_q\). To establish associativity, consider independent random probability vectors \(\underline Y,\underline Z,\underline W\) with distributions \(\mathsf x_1,\mathsf x_2,\mathsf x_3\in\mathcal X_q\). Then, \(Y_0Z_0W_0>0\) a.s. Hence,\,\,the variable-node update is well defined at every intermediate step, and
\[ 
	[\mathcal V(\mathcal V(\underline Y,\underline Z),\underline W)]_i
	 =
	\frac{Y_iZ_iW_i}
	{\sum_{j\in\mathbb F_q}Y_jZ_jW_j} =
	[\mathcal V(\underline Y,\mathcal V(\underline Z,\underline W))]_i. 
\]
Thus, \((\mathsf x_1\circledast\mathsf x_2)
\circledast\mathsf x_3
=
\mathsf x_1\circledast
(\mathsf x_2\circledast\mathsf x_3)\), i.e., \(\circledast\) is associative. For check-node updates, by the definition of \(\mathcal{C}\),
\[ 
	[\mathcal C(\mathcal C(\underline y,\underline z),\underline w)]_i =
	\sum_{ {r,s,t\in\mathbb F_q:r+s+t=i}}
	y_rz_sw_t =
	[\mathcal C(\underline y,\mathcal C(\underline z,\underline w))]_i\qquad \forall\underline{y},\underline{z},\underline{w}\in\mathcal{S}_q, 
\] 
so \(\boxast\) is associative as well. Finally, put \(  e_0=(\mathbf1\{i=0\})_{i\in\mathbb F_q}\) and \(u=(1/q)_{i\in\mathbb F_q}\). For every \(\underline y\in\mathcal S_q\) with \(y_0>0\), we have
\[ 
	\mathcal V(\underline y,  u) =\underline y, \qquad
	\mathcal V(\underline y,  e_0) =  e_0, \qquad
	\mathcal C(\underline y,  e_0) =\underline y, \qquad
	\mathcal C(\underline y,  u) =  u. 
\]
Since \(\Delta_0=\delta_{  u}\), \(\Delta_\infty=\delta_{  e_0}\), and \(y_0>0\) holds \(\mathsf x\)-a.e. for every \(\mathsf x\in\mathcal X_q\), these identities establish \eqref{eq:algebra}. 

\subsection{Signed Extensions of BP Convolutions}
For every \(\alpha\in\mathcal M_q\), \(|\alpha|(\{\underline y:y_0=0\})=0\) since \(\alpha\) is a finite linear combination of measures in \(\mathcal X_q\). Thus, for \(\alpha,\beta\in\mathcal M_q\),
\[
D(\underline y,\underline z)
\ge y_0z_0>0
\qquad
(|\alpha|\otimes|\beta|)\text{-a.e.}
\]
Hence, \(\mathcal V_\#(\alpha\otimes\beta)\) is well-defined independently of the convention at \(D=0\). Together with \(\mathcal C_\#(\alpha\otimes\beta)\), this defines the bilinear extensions of the two convolutions. Indeed, if \(\alpha=\sum_r a_r\mathsf x_r\) and \(\beta=\sum_s b_s\mathsf y_s\), then, for \(*\in\{\circledast,\boxast\}\), by Lemma IV.7,
\[
\alpha*\beta
=
\sum_{r,s}a_rb_s(\mathsf x_r*\mathsf y_s)
\in\mathcal M_q.
\] 
\subsection{Proof of Lemmas IV.9 and IV.10}
By double transitivity of \(G\) and \(G\)-invariance of \(\bar{\mathsf x}\), for every bounded Borel function \(f:[0,1]^2\to\mathbb R\) and distinct \(i,j\in\mathbb F_q\),
\[
\int f(y_i,y_j)\,d\bar{\mathsf x}
=
\int f(y_0,y_1)\,d\bar{\mathsf x}.
\]
This gives the symmetrized representations of \(\mathfrak B\) and \(\mathfrak P\). For every \(\underline y\in\mathcal S_q\), we have the bounds
\[ 
	0
	 \le\sum_{i\ne j}\sqrt{y_iy_j}
	\le\frac12\sum_{i\ne j}(y_i+y_j)
	=q-1,\qquad
	0 \le\sum_{i\ne j}\min\{y_i,y_j\}
	\le\frac12\sum_{i\ne j}(y_i+y_j)
	=q-1. 
\] 
For both sums, the lower bound holds with equality exactly at the vertices \(e_i\), and the upper bound exactly at \(u\). Likewise,
\[ 
	0\le-\sum_i y_i\log y_i\le\log q, \qquad
	\frac1q\le\sum_i y_i^2\le1, 
\]
 with the corresponding equality conditions. Since \(\bar{\mathsf x}\) is \(G\)-invariant,
\[ 
	\bar{\mathsf x}\bigl(\{e_i:i\in\mathbb F_q\}\bigr)=1
	 \iff
	\bar{\mathsf x}=\frac1q\sum_{i\in\mathbb{F}_q}\delta_{e_i}
	\iff\mathsf x=\Delta_\infty,\qquad
	\bar{\mathsf x}(\{u\})=1
	 \iff\mathsf x=\Delta_0. 
\]
This establishes all extremal values in Lemma IV.9. We now prove Lemma IV.10. For every \(\mathsf x\in\mathcal X_q\), we have
\[ 
	\mathrm H(\mathsf x)
	 =-\int\sum_i y_i\log y_i\,d\bar{\mathsf x} 
	 =-q\int y_0\log y_0\,d\bar{\mathsf x} =-\int\log y_0\,d\mathsf x. 
\]
Let \(\mathsf x_1,\mathsf x_2\in\mathcal X_q\). By the definition of \(\circledast\),
\[
\mathrm H(\mathsf x_1\circledast\mathsf x_2)
=
-\int_{\mathcal S_q\times\mathcal{S}_q}
\log\frac{y_0z_0}{D(\underline y,\underline z)}
\,d\mathsf x_1d\mathsf x_2.
\]
By Lemma IV.7, \((\mathcal C_{1,-1,1})_\#
(\mathsf x_1\otimes\mathsf x_2)
=
\mathsf x_1\boxast\mathsf x_2\), and since  \([\mathcal C_{1,-1,1}(\underline y,\underline z)]_0
=
D(\underline y,\underline z)\) for all \(\underline y,\underline z\in\mathcal{S}_q\), we have
\[
\mathrm H(\mathsf x_1\boxast\mathsf x_2)
=
-\int_{\mathcal S_q\times\mathcal{S}_q}
\log D(\underline y,\underline z)
\,d\mathsf x_1d\mathsf x_2.
\]
Thus, \(\mathrm H(\mathsf x_1\circledast\mathsf x_2)
+\mathrm H(\mathsf x_1\boxast\mathsf x_2) =-\int 
\log(y_0z_0)\,d\mathsf x_1d\mathsf x_2 =\mathrm H(\mathsf x_1)+\mathrm H(\mathsf x_2)\). All integrals are finite since \(0<y_0z_0\le D(\underline y,\underline z)\le1\)  \((\mathsf x_1\otimes\mathsf x_2)\)-a.e. and \(\mathrm H(\mathsf x_j)\le\log q\). For \(\alpha=\sum_r a_r\mathsf x_r\) and \(\beta=\sum_s b_s\mathsf y_s\), by bilinearity of \(\circledast,\boxast\) and linearity of \(\mathrm{H}\),
\[ 
 \mathrm H(\alpha\circledast\beta)
	+\mathrm H(\alpha\boxast\beta) =\sum_{r,s}a_rb_s
	\bigl(\mathrm H(\mathsf x_r)+\mathrm H(\mathsf y_s)\bigr) =\beta(\mathcal S_q)\mathrm H(\alpha)
	+\alpha(\mathcal S_q)\mathrm H(\beta). 
\]
Thus, when \(\alpha(\mathcal S_q)=\beta(\mathcal S_q)=0\), \(\mathrm H(\alpha\circledast\beta)
+\mathrm H(\alpha\boxast\beta)=0\). This proves Lemma IV.10.
\subsection{Properties of \(\mathfrak{B}\) and \(\mathfrak{P}\)}
\begin{lemma}[Bhattacharyya multiplicativity and bounds]
For every \(\mathsf x,\mathsf x_1,\mathsf x_2\in\mathcal X_q\),  \(\mathfrak B(\mathsf x_1\circledast\mathsf x_2)
=
\mathfrak B(\mathsf x_1)\mathfrak B(\mathsf x_2)\) and
\[ 2\mathfrak P(\mathsf x)
\le\mathfrak B(\mathsf x)
\le
2\sqrt{\mathfrak P(\mathsf x)
	\bigl(1-\mathfrak P(\mathsf x)\bigr)}.
\]
\end{lemma}
\begin{proof}
Write \(\bar{\mathsf x}_j=\Psi\mathsf x_j\), \(j=1,2\). By \eqref{eq:f_dPsi_circledast},
\[ 
	 \mathfrak B(\mathsf x_1\circledast\mathsf x_2) =q^2\int_{\mathcal S_q\times\mathcal S_q}
	D(\underline y,\underline z)
	\sqrt{
		\frac{y_0z_0}{D(\underline y,\underline z)}
		\frac{y_1z_1}{D(\underline y,\underline z)}
	}
	\,d\bar{\mathsf x}_1d\bar{\mathsf x}_2 =q^2\int_{\mathcal S_q\times \mathcal S_q}
	\sqrt{y_0y_1}\sqrt{z_0z_1}\,
	d\bar{\mathsf x}_1d\bar{\mathsf x}_2 =\mathfrak B(\mathsf x_1)\mathfrak B(\mathsf x_2). 
\] 
Here, the integrand is defined as zero when \(D(\underline y,\underline z)=0\). Let \(\bar{\mathsf x}=\Psi\mathsf{x}\). Since \(\min\{y_0,y_1\}\le\sqrt{y_0y_1}\), we have
\[ 
	2\mathfrak P(\mathsf x)
	 =q\int\min\{y_0,y_1\}\,d\bar{\mathsf x} \le q\int\sqrt{y_0y_1}\,d\bar{\mathsf x}
	=\mathfrak B(\mathsf x). 
\]
On the other hand, by the Cauchy–Schwarz inequality,
\[ 
	\mathfrak B(\mathsf x) =q\int\sqrt{y_0y_1}\,d\bar{\mathsf x} =q\int
	\sqrt{\min\{y_0,y_1\}\max\{y_0,y_1\}}
	\,d\bar{\mathsf x} \leq q
	\sqrt{
		\int\min\{y_0,y_1\}\,d\bar{\mathsf x}
		\int\max\{y_0,y_1\}\,d\bar{\mathsf x}
	} =2\sqrt{\mathfrak P(\mathsf x)
		\bigl(1-\mathfrak P(\mathsf x)\bigr)}, 
\]
where the last equality follows since \(	\int\min\{y_0,y_1\}\,d\bar{\mathsf x}=\frac2q\mathfrak P(\mathsf x)\) and \(\int\max\{y_0,y_1\}\,d\bar{\mathsf x}
 =\frac2q\bigl(1-\mathfrak P(\mathsf x)\bigr)\).
\end{proof}
\begin{lemma}[Check-node Bhattacharyya bounds]
For every \(\mathsf x_1,\mathsf x_2\in\mathcal X_q\),
\[
\mathfrak B(\mathsf x_1\boxast\mathsf x_2)
\le 
\mathfrak B(\mathsf x_1)+\mathfrak B(\mathsf x_2) +(q-2)\mathfrak B(\mathsf x_1)\mathfrak B(\mathsf x_2).
\] 
Consequently, for every \(\mathsf x\in\mathcal X_q\), \(m\ge1\), and \(\beta=\mathfrak B(\mathsf x)\),
\[
\mathfrak B(\mathsf x^{\boxast m})
\le
\begin{cases}
	\min\{1,m\beta\},&q=2,\\[2mm]
	\displaystyle
	\min\left\{
	1,\frac{(1+(q-2)\beta)^m-1}{q-2}
	\right\},&q>2.
\end{cases}
\]
\end{lemma}
\begin{proof}
For every \(\mathsf x\in\mathcal X_q\), \(\mathfrak B(\mathsf x)
 =q\int\sqrt{y_0y_1}\,d\Psi\mathsf x =\int\sqrt{ {y_1/y_0}}\,d\mathsf x\), where \(y_0>0\) holds \(\mathsf x\)-a.e. Let \(\underline w=\mathcal C_{1,1,1}(\underline y,\underline z)\). Then
\[
w_0
 =\sum_{u+v=0}y_uz_v
\ge y_0z_0,\qquad
w_1 =y_0z_{-1}+y_{-1}z_0
+\sum_{ {u,v\ne0:u+v=-1}}y_uz_v.
\]
Thus, whenever \(y_0,z_0>0\),
\[
\sqrt{\frac{w_1}{w_0}}
 \le
\sqrt{
	\frac{z_{-1}}{z_0}
	+\frac{y_{-1}}{y_0}
	+\sum_{ {u,v\ne0:u+v=-1}}
	\frac{y_u}{y_0}\frac{z_v}{z_0}
} \le
\sqrt{\frac{z_{-1}}{z_0}}
+\sqrt{\frac{y_{-1}}{y_0}}
+\sum_{ {u,v\ne0:u+v=-1}}
\sqrt{\frac{y_u}{y_0}}\sqrt{\frac{z_v}{z_0}}.
\]
By multiplicative invariance,  \(\mathfrak B(\mathsf x)=\int\sqrt{ {y_a/y_0}}\,\mathsf x(d\underline y)\) for every \(a\in\mathbb F_q^\times\), and since \(|\{(u,v)\in(\mathbb F_q^\times)^2:u+v=-1\}|
=q-2\), integrating the preceding inequality against \(\mathsf x_1\otimes\mathsf x_2\), and using the edge-label invariance of \(\boxast\), we obtain
\[
	\mathfrak B(\mathsf x_1\boxast\mathsf x_2)
\le 
\mathfrak B(\mathsf x_1)+\mathfrak B(\mathsf x_2) +(q-2)\mathfrak B(\mathsf x_1)\mathfrak B(\mathsf x_2).
\]
 For the second inequality, let \(\beta=\mathfrak B(\mathsf x)\) and \(b_m=\mathfrak B(\mathsf x^{\boxast m})\). Then by the first inequality, we have
\[
b_{m+1}\le
(1+(q-2)\beta)b_m+\beta,
\qquad b_1=\beta.
\]
Therefore, by induction,
\[
b_m \le
\beta\sum_{j=0}^{m-1}(1+(q-2)\beta)^j =
\begin{cases}
	m\beta,&q=2,\\[2mm]
	\displaystyle
	\frac{(1+(q-2)\beta)^m-1}{q-2},&q>2.
\end{cases}
\] 
Combining this with \(b_m\le1\) proves the claimed bounds. 
\end{proof}
By Lemma A2.2, for \(\mathsf{x}\in\mathcal{X}_2\), \(\mathfrak B(\mathsf x^{\boxast m})\le m\mathfrak{B}(\mathsf{x})\); for \(q>2\) and \(\mathsf{x}\in\mathcal{X}_q\), 
\[
m\mathfrak{B}(\mathsf{x})\le\frac1{2(q-2)}
\quad\Longrightarrow\quad
\mathfrak B(\mathsf x^{\boxast m})\le2m\mathfrak{B}(\mathsf{x}),
\]
since  \(\mathfrak B(\mathsf x^{\boxast m})
 \le m\mathfrak{B}(\mathsf{x})(1+(q-2)\mathfrak{B}(\mathsf{x}))^{m-1} 
 \le m\mathfrak{B}(\mathsf{x}) e^{(q-2)m\mathfrak{B}(\mathsf{x})}
\le e^{1/2}m\mathfrak{B}(\mathsf{x})
\le 2m\mathfrak{B}(\mathsf{x})\).
\begin{lemma}[Strict concavity of the Bhattacharyya kernel] 
The kernel function
\[
f(\underline y)
\coloneqq
\frac1{q-1}
\sum_{ {i,j\in\mathbb F_q:i\ne j}}
\sqrt{y_i y_j},
\qquad \underline y\in\mathcal S_q,
\]
of \(\mathfrak{B}\) is strictly concave on \(\mathcal S_q\). 
\end{lemma}
\begin{proof}
For any \(\underline y,\underline z\in\mathcal S_q\), \(\theta\in(0,1)\), and distinct \(i, j\in\mathbb{F}_q\), by the Cauchy–Schwarz inequality,
\[ 
 \sqrt{
		(\theta y_i+(1-\theta)z_i)
		(\theta y_j+(1-\theta)z_j)
	} \ge
	\theta\sqrt{y_iy_j}
	+(1-\theta)\sqrt{z_iz_j}, 
\]
with equality if and only if \(y_i z_j=y_j z_i\). If \(\underline y\ne\underline z\), then, since \(\sum_i y_i=\sum_i z_i=1\), there exist \(i\ne j\) such that \(y_i z_j\ne y_j z_i\). Summing the above inequalities   yields \(f(\theta\underline y+(1-\theta)\underline z)
>
\theta f(\underline y)
+(1-\theta)f(\underline z)\), so \(f\) is strictly concave on \(\mathcal{S}_q\).
\end{proof}
\begin{lemma}[Repetition error exponent]
For every \(\mathsf x\in\mathcal X_q\setminus\{\Delta_\infty\}\),
\[
\lim_{n\to\infty}
\frac1n\log\mathfrak P(\mathsf x^{\circledast n})
=\log\mathfrak B(\mathsf x).
\]
\end{lemma}
\begin{proof}
Let \(\underline Y\sim\mathsf x\). Since \(Y_0>0\) a.s., the log-likelihood ratio random variable \(L=\log (Y_0/Y_1) \) is well-defined in \(\mathbb R\cup\{+\infty\}\),\,with \(L=+\infty\) when \(Y_1=0\). Let \(\mathsf a\coloneqq \mathrm{Law}(L)\) denote its distribution. By the definitions of \(\mathfrak B\) and \(\mathfrak P\),
\[
\mathfrak B(\mathsf x)
 =\int e^{-\ell/2}\,\mathsf a(d\ell),\qquad
\mathfrak P(\mathsf x)
 =\frac12\int\min\{1,e^{-\ell}\}\,\mathsf a(d\ell).
\]
Here, \(e^{-\infty}=0\). We first establish the LLR symmetry
\begin{equation}
\mathsf a(-E)
=
\int_E e^{-\ell}\,\mathsf a(d\ell),
\qquad \forall E\in\mathcal B(\mathbb R).\label{eq:LLR_symmetry}
\end{equation}
Let \(\tau(i)=1-i\), which interchanges \(0\) and \(1\). Since \(\tau\in G\) and \(\bar{\mathsf x}=\Psi\mathsf x\) is \(G\)-invariant, for every Borel \(E\subseteq\mathbb R\),
\[
	\mathsf a(-E) =q\int_{\{L\in-E\}}y_0\,\bar{\mathsf x}(d\underline y) =q\int_{\{L\in E\}}y_1\,\bar{\mathsf x}(d\underline y) =q\int_{\{L\in E\}}\frac{y_1}{y_0}y_0\,
\bar{\mathsf x}(d\underline y) =\int_E e^{-\ell}\,\mathsf a(d\ell).
\]
The sets \(\{L\in E\}\) in the above calculation contain only vectors \(\underline{y}\in\mathcal{S}_q\) with \(y_0,y_1>0\). Put
\[
\beta\coloneqq\mathfrak B(\mathsf x)>0
\]
(recall that \(\mathsf{x}\in\mathcal{X}_q\setminus\{\Delta_\infty\}\)) and define a Borel probability measure \(\nu\) on \(\mathbb R\) by
\[
\nu(d\ell)
\coloneqq
\frac{e^{-\ell/2}}{\beta}\,\mathsf a(d\ell).
\]
By \eqref{eq:LLR_symmetry}, \(\nu(-E) = \beta^{-1}
\int_E e^{\ell/2}\,\mathsf a(-d\ell) = \beta^{-1}
\int_E e^{-\ell/2}\,\mathsf a(d\ell)
=\nu(E)\) \(\forall E\in\mathcal{B}(\mathbb{R})\). Thus, \(\nu\) is symmetric about zero, and
\[
\int e^{\ell/2}\,\nu(d\ell)=\int e^{-\ell/2}\,\nu(d\ell)
=
\frac{\mathsf a(\mathbb R)}{\beta}
\le\frac1\beta<+\infty.
\]
Thus, \(\int|\ell|\,\nu(d\ell)<2\int (e^{\ell/2}+e^{-\ell/2})\,\nu(d\ell)<\infty\) and \(\int\ell\,\nu(d\ell)=0\).
Let \(L_1,L_2,\ldots\) be i.i.d. with distribution \(\nu\), and put
\[
S_n\coloneqq\sum_{j=1}^nL_j.
\]
By the law of large numbers, \(S_n/n\longrightarrow0\) a.s. Under variable-node convolution, the LLRs add. Since contributions with any \(L_j=+\infty\) vanish in the expression for \(\mathfrak P\),  changing measure from \(\mathsf{a}\) to \(\nu\), we have
\[
\mathfrak P(\mathsf x^{\circledast n}) =\frac12
\int_{\mathbb R^n}
\min\left\{1,e^{-\sum_{j=1}^n\ell_j}\right\}
\,\mathsf a^{\otimes n}(d\boldsymbol\ell) =\frac{\beta^n}{2}
\int_{\mathbb R^n}
e^{\frac12\sum_{j=1}^n\ell_j}
\min\left\{1,e^{-\sum_{j=1}^n\ell_j}\right\}
\,\nu^{\otimes n}(d\boldsymbol\ell) =\frac{\beta^n}{2}
\mathbb E\left[e^{-|S_n|/2}\right].
\]
Thus, for every \(\varepsilon>0\), restricting the expectation to the event \(\{|S_n|\le\varepsilon n\}\) gives
\[
\frac{\beta^n}{2}e^{-\varepsilon n/2}
\Pr\{|S_n|\le\varepsilon n\}
 \le
\mathfrak P(\mathsf x^{\circledast n}) \le\frac{\beta^n}{2}.
\]
Since  \(S_n/n\longrightarrow0\) a.s., we have \(\Pr\{|S_n|\le\varepsilon n\}\longrightarrow1\), and thus
\[
\log\beta-\frac{\varepsilon}{2}
\le
\liminf_{n\to\infty}
\frac1n\log\mathfrak P(\mathsf x^{\circledast n})
\le
\limsup_{n\to\infty}
\frac1n\log\mathfrak P(\mathsf x^{\circledast n})
\le\log\beta.
\]
Letting \(\varepsilon\downarrow0\), we obtain \(\lim_{n\to\infty}
\frac1n\log\mathfrak P(\mathsf x^{\circledast n})
=
\log\mathfrak B(\mathsf x)\).
\end{proof}
\subsection{Proof of Proposition IV.11}
Since \(\mathcal S_q\) is compact, \((\mathcal P(\mathcal S_q),W_2)\) is compact, and \(W_2\)-convergence is equivalent to weak convergence. If \(\mu_n\in\overline{\mathcal X}_q\) and \(\mu_n\xrightarrow{W_2}\mu\), then, for every \(g\in G\) and continuous \(f\),
\[ 
	\int f\,d(g_\#\mu)
 =\lim_n\int f\,d(g_\#\mu_n) 
 =\lim_n\int f\,d\mu_n
	=\int f\,d\mu. 
\]
Hence, the limit  \(\mu\in\overline{\mathcal X}_q\), proving compactness of \(\overline{\mathcal X}_q\). Moreover, for \(\mu_n\xrightarrow{W_2}\mu\),
\[ 
	\int f\,d(\Phi\mu_n) =q\int y_0f(\underline y)\,\mu_n(d\underline y) \longrightarrow
	q\int y_0f(\underline y)\,\mu(d\underline y)
	=\int f\,d(\Phi\mu). 
\]
Thus, \(\Phi\) is a continuous bijection from compact \(\overline{\mathcal X}_q\) onto \(\mathcal X_q\), viewed as a subspace of \(\mathcal P(\mathcal S_q)\) with its weak topology. Since\,\,the latter is Hausdorff, \(\Phi\) is a homeomorphism. Hence,   \(\mathsf x_n\xrightarrow{d_W}\mathsf x\) if and only if \(\mathsf x_n\) converges weakly to \(\mathsf{x}\), and compactness of \((\mathcal X_q,d_W)\) follows. For continuous \(f\),
\[
\mathsf x_n\xrightarrow{d_W}\mathsf x
\Longrightarrow
F_f(\mathsf x_n)
=\int f\,d\Psi\mathsf x_n
\longrightarrow
\int f\,d\Psi\mathsf x
=F_f(\mathsf x).
\]
Finally, suppose \(\mathsf x_{j,n}\xrightarrow{d_W}\mathsf x_j\) for \(j=1,2\), and write \(\bar{\mathsf x}_{j,n}=\Psi\mathsf x_{j,n}\) and \(\bar{\mathsf x}_j=\Psi\mathsf x_j\). Then \(\bar{\mathsf x}_{1,n}\otimes\bar{\mathsf x}_{2,n}\) converges weakly to \(\bar{\mathsf x}_1\otimes\bar{\mathsf x}_2\). For every continuous \(f:\mathcal{S}_q\to\mathbb{R}\), the function \((\underline y,\underline z)
\mapsto
D(\underline y,\underline z)
f(\mathcal V(\underline y,\underline z))\) extended by zero on \(\{D=0\}\) is continuous, since \(|Df(\mathcal V)|\le\|f\|_\infty D\). The function \(f\circ\mathcal C\) is also continuous. Therefore, by \(\eqref{eq:f_dPsi_circledast}\) and \(\eqref{eq:f_dPsi_boxast}\),
\begin{align*}
&\int f\,d\Psi(\mathsf x_{1,n}\circledast\mathsf x_{2,n}) =q\int Df(\mathcal V)\,
d\bar{\mathsf x}_{1,n}d\bar{\mathsf x}_{2,n} \longrightarrow
q\int Df(\mathcal V)\,
d\bar{\mathsf x}_1d\bar{\mathsf x}_2 =\int f\,d\Psi(\mathsf x_1\circledast\mathsf x_2),\\
&\int f\,d\Psi(\mathsf x_{1,n}\boxast\mathsf x_{2,n}) =\int f(\mathcal C)\,
d\bar{\mathsf x}_{1,n}d\bar{\mathsf x}_{2,n} \longrightarrow
\int f(\mathcal C)\,
d\bar{\mathsf x}_1d\bar{\mathsf x}_2 =\int f\,d\Psi(\mathsf x_1\boxast\mathsf x_2).
\end{align*} 
Since weak convergence and \(W_2\)-convergence are equivalent on \(\mathcal P(\mathcal S_q)\), \(\mathsf x_{1,n}*\mathsf x_{2,n}
\xrightarrow{d_W}
\mathsf x_1*\mathsf x_2\) for \(*\in\{\circledast,\boxast\}\).
\subsection{Proof of Theorem IV.13}
i) \(\Longrightarrow\) ii). Under uniform input \(X\), consider the Markov chain
 \(X\to\underline Y\to\underline Z\), where \(\underline{Y}\sim W_{\mathsf x_1}(\cdot\,|X)\) and \(\underline{Z}\sim W_{\mathsf x_2}(\cdot\,|\,X)\)\,are APP vectors. Then, for every \(i\in\mathbb{F}_q\), \(\mathbb E[Y_i\,|\,\underline Z]
 =
 \mathbb E\!\left[
 \Pr(X=i\,|\,\underline Y)
 \,|\,\underline Z
 \right] =
 \Pr(X=i\,|\,\underline Z)
 =Z_i\) a.s. Hence \(\mathbb E[\underline Y\mid\underline Z]
 =\underline Z\) a.s.
 
ii) \(\Longrightarrow\) i). Assume that \((\underline Y,\underline Z)\) satisfies ii), and let \(\mu=\operatorname{Law}(\underline Y,\underline Z)\). Since \(\mathcal S_q\) is a compact metric space, the disintegration theorem yields a Markov kernel \(K\) on \(\mathcal S_q\) such that \(\mu(d\underline y,d\underline z)
=
\bar{\mathsf x}_1(d\underline y)\,
K(d\underline z\,|\,\underline y)\).
Then, for every \(i\in\mathbb F_q\) and \(B\in\mathcal B(\mathcal S_q)\),
\[
	\int_{\mathcal S_q}
K(B\,|\,\underline y)\,
W_{\mathsf x_1}(d\underline y\,|\, i) =q\,\mathbb E[Y_i\mathbf1 {\{\underline Z\in B\}}] =q\,\mathbb E[Z_i\mathbf1 {\{\underline Z\in B\}}] =W_{\mathsf x_2}(B\,|\, i).
\]
Thus, \(\mathsf x_1\preceq\mathsf x_2\). Moreover, \(K\) may be replaced by \(K_G(B\,|\,\underline y)
=
\frac1{|G|}
\sum_{g\in G}
K(gB\,|\, g\cdot\underline y)\). \(K_G\) is \(G\)-equivariant. By the symmetry of \(W_{\mathsf x_1}\) and \(W_{\mathsf x_2}\),  \(\int K_G(B\mid\underline y)
W_{\mathsf x_1}(d\underline y\,|\, i) =
\frac1{|G|}\sum_{g\in G}
W_{\mathsf x_2}(gB\,|\, gi) =
W_{\mathsf x_2}(B\,|\, i) \) for every \(i\in\mathbb F_q\) and \(B\in\mathcal B(\mathcal S_q)\).

ii) \(\Longleftrightarrow\) iii). By Strassen's   theorem\cite[Theorem 8]{Strassen}, for any probability measures \(\mu,\nu\) on a compact convex set \(C\subset\mathbb R^d\),\,\,the following are equivalent:
\begin{enumerate}
\item  There exists a coupling \((Y,Z)\) with marginals \(\mu,\nu\) such that \(\mathbb E[Y\,|\, Z]=Z\) a.s.
\item \(\int_C f\,d\mu\le\int_C f\,d\nu\) for every continuous concave \(f:C\to\mathbb R\).
\end{enumerate} 
Applying this theorem with \(C=\mathcal S_q\), \(\mu=\bar{\mathsf x}_1\), and \(\nu=\bar{\mathsf x}_2\), we obtain (iii) \(\Longrightarrow\) (ii), since every continuous concave function on \(\mathcal S_q\) is bounded. Conversely,  ii) implies iii), since by   Jensen's inequality,   for every bounded concave \(f:\mathcal S_q\to\mathbb R\),
\[
\int f\,d\bar{\mathsf x}_1=\mathbb E[f(\underline Y)]=\mathbb E [\mathbb{E}[f ( \underline Y ) \mid\underline Z]]
\le
\mathbb E\!\left[f\!\left(\mathbb E[\underline Y\mid\underline Z]\right)\right]
=
\mathbb E[f(\underline Z)]=\int f\,d\bar{\mathsf x}_2.
\] 

For probability measures \(\mu,\nu\in\mathcal P_2(\mathbb R^q)\),  the quadratic 2-Wasserstein distance is given by
\[
W_2(\mu,\nu)^2
=
\inf_{ {\underline Z\sim\mu,
		\underline U\sim\nu}}
\mathbb E\|\underline Z-\underline U\|_2^2,
\] 
where the infimum is over all couplings of \(\mu\) and \(\nu\). A coupling attaining this infimum is called \textit{optimal}. Such a coupling\,\,always exists by weak compactness of the set of couplings and lower semicontinuity of the quadratic cost.

ii) \(\Longrightarrow\) iv). Let \(\eta\in\mathcal P_2(\mathbb R^q)\), and choose an optimal coupling \((\underline Z,\underline U)\) of \(\bar{\mathsf x}_2\) and \(\eta\). Conditioned on \(\underline Z\), generate \(\underline Y\) independently of \(\underline U\) according to the coupling in  ii). Then \(\underline{U}\to\underline{Z}\to\underline{Y}\) forms a Markov chain, where  \(\mathbb E[\underline Y\mid\underline Z,\underline U]
=\mathbb E[\underline Y\mid\underline Z]
=\underline Z\). Thus,
\[
\begin{aligned}
	W_2(\bar{\mathsf x}_1,\eta)^2
	&\le \mathbb E\|\underline Y-\underline U\|_2^2 =\mathbb E\|\underline Z-\underline U\|_2^2
	+\mathbb E\|\underline Y-\underline Z\|_2^2
	+2\mathbb E\langle\underline Y-\underline Z,\underline Z-\underline U\rangle \overset{(\mathrm{a})}{=}\mathbb E\|\underline Z-\underline U\|_2^2
	+\mathbb E\|\underline Y-\underline Z\|_2^2\\
	&\!\overset{(\mathrm{b})}{=}W_2(\bar{\mathsf x}_2,\eta)^2
	+\mathbb E\|\underline Y\|_2^2
	-2\mathbb E\langle\mathbb E[\underline Y\mid\underline Z],\underline Z\rangle
	+\mathbb E\|\underline Z\|_2^2 =W_2(\bar{\mathsf x}_2,\eta)^2
	+\mathbb E\|\underline Y\|_2^2
	-\mathbb E\|\underline Z\|_2^2\\
	&=W_2(\bar{\mathsf x}_2,\eta)^2
	+\mathrm Q(\mathsf x_1)-\mathrm Q(\mathsf x_2).
\end{aligned}
\]
(a) follows since  \(\mathbb E\langle\underline Y-\underline Z,\underline Z-\underline U\rangle =\mathbb E\left[
\left\langle
\mathbb E[\underline Y-\underline Z\,|\,\underline Z,\underline U],
\underline Z-\underline U
\right\rangle\right]=0\). (b) follows from the  optimality of \((\underline Z,\underline U)\) and   linearity of inner product. Since \(\eta\in\mathcal P_2(\mathbb R^q)\) can be arbitrary, this proves iv).

iv) \(\Longrightarrow\) ii). Let \(\Pi(\mu,\eta)\) denote the set of couplings of \(\mu\) and \(\eta\). Expanding the   quadratic 2-Wasserstein distance gives
\[
W_2(\mu,\eta)^2
= \int\|\underline y\|_2^2\,\mu(d\underline y)
+\int\|\underline u\|_2^2\,\eta(d\underline u) -2\sup_{\pi\in\Pi(\mu,\eta)}
\int\langle\underline y,\underline u\rangle\,d\pi.
\] 
Hence,  iv) implies
\begin{equation}
\sup_{\pi\in\Pi(\bar{\mathsf x}_1,\eta)}
\int\langle\underline y,\underline u\rangle\,d\pi
\ge
\sup_{\pi\in\Pi(\bar{\mathsf x}_2,\eta)}
\int\langle\underline y,\underline u\rangle\,d\pi\qquad \forall \eta\in\mathcal P_2(\mathbb R^q).\label{eq:A3}
\end{equation}
Recall that, for a differentiable convex function \(f:\mathbb R^q\to\mathbb R\), its convex conjugate \(f^*:\mathbb{R}^q\to\mathbb{R}\cup\{+\infty\}\), defined by
\[
f^*(\underline u)\coloneqq 
\sup_{\underline y\in\mathbb R^q}
\{\langle\underline y,\underline u\rangle-f(\underline y)\},
\]
satisfies the Fenchel–Young inequality
\[
f(\underline y)+f^*(\underline u)
\ge\langle\underline y,\underline u\rangle,
\]
with equality if and only if \(\underline u=\nabla f(\underline y)\). Now, let \(f\in C^1(\mathbb R^q)\) be convex and set \(\eta=(\nabla f)_\#\bar{\mathsf x}_2
\in\mathcal P_2(\mathbb R^q)\). By Fenchel–Young inequality and \eqref{eq:A3}, 
\[
\int f\,d\bar{\mathsf x}_1+\int f^*\,d\eta \ge
\sup_{\pi\in\Pi(\bar{\mathsf x}_1,\eta)}
\int\langle\underline y,\underline u\rangle\,d\pi \ge
\sup_{\pi\in\Pi(\bar{\mathsf x}_2,\eta)}
\int\langle\underline y,\underline u\rangle\,d\pi =
\int f\,d\bar{\mathsf x}_2+\int f^*\,d\eta.
\]
The last equality is attained by the coupling \((\underline Z,\underline U)\), where \(\underline Z\sim\bar{\mathsf x}_2\) and \(\underline U=\nabla f(\underline Z)\sim\eta\). Moreover,
\[
f^*(\nabla f(\underline z))
=
\langle\underline z,\nabla f(\underline z)\rangle-f(\underline z)
\]
is bounded on \(\mathcal S_q\), since \(f,\nabla f\) are continuous and \(\mathcal S_q\) is compact. Thus, \(\int f^*\,d\eta\) is finite, and subtracting this common term\,\,gives
\begin{equation}
\int f\,d\bar{\mathsf x}_1
\ge
\int f\,d\bar{\mathsf x}_2.\label{eq:A4}
\end{equation}
We now extend \eqref{eq:A4} to continuous convex functions on \(\mathcal S_q\). Let \(\varphi:\mathcal S_q\to\mathbb R\) be continuous and convex. We show that \(\varphi\) can\,\,be uniformly approximated on \(\mathcal S_q\) by convex functions in \(C^1(\mathbb R^q)\). For \(\lambda>0\), define its Moreau envelope
\[
\varphi_\lambda(\underline y)\coloneqq
\min_{\underline z\in\mathcal S_q}
\left\{
\varphi(\underline z)
+\frac{\|\underline y-\underline z\|_2^2}{2\lambda}
\right\},
\qquad \underline y\in\mathbb R^q,
\]
and denote the unique minimizer by \(P_\lambda(\underline y)\). By the Moreau–Yosida regularization theorem\cite[Sec.~12.4]{BauschkeCombettes2017}, \(\varphi_\lambda\) is convex and belongs to \(C^1(\mathbb R^q)\), with
\[
\nabla\varphi_\lambda(\underline y)
=
\frac{\underline y-P_\lambda(\underline y)}{\lambda}.
\]
Moreover, for every \(\underline y\in\mathcal S_q\), the minimizing property yields
\[
0
\le\varphi(\underline y)-\varphi_\lambda(\underline y)
\le\varphi(\underline y)-\varphi(P_\lambda(\underline y)),\qquad
\|\underline y-P_\lambda(\underline y)\|_2^2
 \le
2\lambda\left(
\max_{\mathcal S_q}\varphi-\min_{\mathcal S_q}\varphi
\right).
\] 
Since \(\varphi\) is uniformly continuous on the compact set \(\mathcal S_q\), these inequalities imply \(\lim_{\lambda\downarrow0}
\|\varphi_\lambda-\varphi\|_{\infty,\mathcal S_q}=0\). Applying \eqref{eq:A4} to \(\varphi_\lambda\) and letting \(\lambda\downarrow0\), we obtain
\[
\int\varphi\,d\bar{\mathsf x}_1
\ge
\int\varphi\,d\bar{\mathsf x}_2
\]
for every continuous convex \(\varphi:\mathcal S_q\to\mathbb R\). Equivalently, the reverse inequality holds for every continuous concave function. By Strassen's theorem mentioned above, this yields the coupling in ii).
\subsection{Proof of  Proposition IV.14}
Let \(\bar{\mathsf x}_i=\Psi\mathsf{x}_i\), \(i=1,2\). Taking \(\eta=\bar{\mathsf x}_2\) in Theorem IV.13 iv), we immediately obtain
\[
d_W(\mathsf x_1,\mathsf x_2)^2 =W_2(\bar{\mathsf x}_1,\bar{\mathsf x}_2)^2 \le \mathrm Q(\mathsf x_1)-\mathrm Q(\mathsf x_2).
\]
Let \((\underline Y,\underline Z)\) be the coupling in Theorem IV.13 ii), satisfying \(\underline Y\sim \bar{\mathsf x}_1\), \(\underline{Z}\sim \bar{\mathsf x}_2\) and \(\mathbb E[\underline Y\,|\,\underline Z]=\underline Z\). By Pinsker's inequality,
\[
\mathrm H(\mathsf x_2)-\mathrm H(\mathsf x_1)
 =\mathbb E\left[
\sum_{i\in\mathbb F_q}Y_i\log\frac{Y_i}{Z_i}
\right] \ge\frac12\mathbb E\|\underline Y-\underline Z\|_1^2 \ge\mathbb E\|\underline Y-\underline Z\|_2^2=\mathbb E\|\underline Y\|_2^2
-2\mathbb E\langle\mathbb E[\underline Y\mid\underline Z],\underline Z\rangle
+\mathbb E\|\underline Z\|_2^2=\mathrm Q(\mathsf x_1)-\mathrm Q(\mathsf x_2).
\]
The second inequality follows from
\(\|\underline v\|_1^2\ge2\|\underline v\|_2^2\) whenever \(\sum_i v_i=0\). So \eqref{eq:dQH} holds. For bounded strictly concave \(f:\mathcal{S}_q\to\mathbb{R}\), by Jensen's inequality, \(f(\underline Z)=f(\mathbb E[\underline Y\,|\,\underline Z])
\ge
\mathbb E[f(\underline Y)\,|\,\underline Z]\) a.s. Equality of expectations forces \(\underline Y=\underline Z\) a.s. Hence, \(\mathsf x_1\prec\mathsf x_2\) implies   \(F_f(\mathsf x_1)=\mathbb{E}[f(\underline{Y})]<\mathbb{E}[f(\underline{Z})]=F_f(\mathsf x_2)\).
\subsection{Proof of Lemma IV.15}
Let \(p=2\mathfrak P(\mathsf x)\) and \(\bar{\mathsf x}=\Psi\mathsf x\). If \(p=0\) or \(p=1\), then Lemma IV.9 gives \(\mathsf{x}=\Delta_\infty\) or \(\Delta_0\), respectively, and the result follows trivially. Assume \(0<p<1\). By the double transitivity of \(G\),
\[
\int y_i\,d\bar{\mathsf x}=\frac1q,
\qquad
\int\min\{y_i,y_j\}\,d\bar{\mathsf x}=\frac pq,
\quad i\ne j.
\]
For every Borel   \(B\subseteq\mathcal{S}_q\), define
\[
K(B\mid i)
=
\frac{q}{(q-1)(1-p)}
\sum_{j\ne i}
\int_B
\bigl(y_i-\min\{y_i,y_j\}\bigr)
\,d\bar{\mathsf x},\qquad
K(B\mid\{i,j\})
=
\frac qp
\int_B\min\{y_i,y_j\}\,d\bar{\mathsf x},
\qquad i\ne j.
\]
Both are probability measures by the preceding identities. For the size-\(2\) QPEC with erasure probability \(p\), input \(i\) produces \(i\) with probability \(1-p\), and each pair \(\{i,j\}\), \(j\ne i\), with probability \(p/(q-1)\). Consequently,
\[
(1-p)K(B\mid i)
+\frac p{q-1}\sum_{j\ne i}K(B\mid\{i,j\}) =
\frac q{q-1}
\sum_{j\ne i}\int_B y_i\,d\bar{\mathsf x} =
q\int_B y_i\,d\bar{\mathsf x}
=
W_{\mathsf x}(B\mid i).
\]
Thus \(W_{\mathsf x}\) is degraded with respect to the size-\(2\) QPEC, proving \(\mathsf e_{2\mathfrak P(\mathsf x)}
\preceq\mathsf x\).
\subsection{Proof of Proposition IV.16}
Since \(\mathsf x_j\preceq\mathsf y_j\), \(j=1,2\), there exist Markov kernels \(K_j\) on \(\mathcal{S}_q\) satisfying
\[
W_{\mathsf y_j}(B\mid i)
=
\int_{\mathcal S_q}
K_j(B\mid\underline y)\,
W_{\mathsf x_j}(d\underline y\mid i),\qquad \forall i\in\mathbb{F}_q,\ B\in\mathcal{B}(\mathcal{S}_q).
\]
We first consider two experiments. For the variable-node update, let \(X\) be uniform on \(\mathbb F_q\), and let \(\underline Y_1\) and  \(\underline Y_2\) be conditionally independent given \(X\), with \(\underline Y_j \mid X \sim
W_{\mathsf x_j}(\cdot\mid X)\) for   \(j=1,2\). Then, by Bayes' rule,
\[
\Pr(X=\cdot\mid\underline Y_1,\underline Y_2)
=
\mathcal V(\underline Y_1,\underline Y_2)\qquad \text{a.s.}
\]
For the check-node update, let \(U,V\) be independent and uniform on \(\mathbb F_q\), and set \(X=U+V\). Independently generate \(\underline{Y}_1,\underline{Y}_2\)\,by \(\underline Y_1\mid U \sim
W_{\mathsf x_1}(\cdot\mid U)\) and \(\underline Y_2\mid V \sim
W_{\mathsf x_2}(\cdot\mid V)\).
Since the posterior distributions of \(U\) and \(V\) are conditionally independent, \(\Pr(X=i\mid\underline Y_1,\underline Y_2)
 =\sum_{u+v=i}
\Pr(U=u\mid\underline Y_1)
\Pr(V=v\mid\underline Y_2) =[\mathcal C(\underline Y_1,\underline Y_2)]_i\) a.s. for every \(i\in\mathbb{F}_q\). Thus,
\[
\Pr(X=\cdot \mid\underline Y_1,\underline Y_2) =  \mathcal C(\underline Y_1,\underline Y_2)\qquad\text{a.s.}
\]
In either experiment, independently generate \(\underline Z_j\mid\underline Y_j\sim K_j(\cdot\mid\underline Y_j)\), \(j=1,2\), and define the random APP vector
\[
\underline M
=\Pr(X=\cdot\mid\underline Y_1,\underline Y_2),
\qquad
\underline N
=\Pr(X=\cdot\mid\underline Z_1,\underline Z_2).
\]
By construction, \(\underline M\sim \Psi(\mathsf x_1*\mathsf x_2)\) and \(\underline N\sim\Psi(\mathsf y_1*\mathsf y_2)\), where \(*\in\{\circledast,\boxast\}\), and \(X\to
(\underline Y_1,\underline Y_2)
\to
(\underline Z_1,\underline Z_2)\) forms  a Markov chain. Set \(\underline e_X=(\mathbf{1}\{X=i\})_{i\in\mathbb{F}_q}\). Then, by conditional independence and the tower property,
\[
\mathbb E[\underline M\mid\underline Z_1,\underline Z_2]
 =\mathbb E\!\left[
\mathbb E[\underline e_X\mid
\underline Y_1,\underline Y_2,
\underline Z_1,\underline Z_2]
\,\middle|\,
\underline Z_1,\underline Z_2
\right] =\mathbb E[\underline e_X\mid
\underline Z_1,\underline Z_2] =\underline N\qquad \text{a.s.}
\]
Since \(\underline N=\Pr(X=\cdot\,|\,\underline Z_1,\underline Z_2)\) is a measurable function of \((\underline Z_1,\underline Z_2)\), we obtain \(\mathbb E[\underline M\,|\,\underline N]=\mathbb E [
\mathbb E[\underline M\,|\,\underline Z_1,\underline Z_2]
\,|\,\underline N ] =\mathbb E[\underline N\,|\,\underline N]=\underline N\). By Theorem IV.13 ii),  \(\mathsf x_1*\mathsf x_2
\preceq
\mathsf y_1*\mathsf y_2\)  for \(*\in\{\circledast,\boxast\}\). 

Finally, suppose that \(\mathsf x_1\prec\mathsf x_2\) and \(\mathsf z\ne\Delta_\infty\). By Lemma A2.3, \(\mathfrak{B}\) admits a strictly concave kernel on \(\mathcal{S}_q\), so by Proposition IV.14, \(\mathfrak B(\mathsf x_1)
<
\mathfrak B(\mathsf x_2)\) and \(\mathfrak B(\mathsf z)>0\). By Lemma\,A2.1,\,\(\mathfrak B(\mathsf x_1\circledast\mathsf z) =\mathfrak B(\mathsf x_1)\mathfrak B(\mathsf z) <\mathfrak B(\mathsf x_2)\mathfrak B(\mathsf z) =\mathfrak B(\mathsf x_2\circledast\mathsf z)\), so \(\mathsf x_1\circledast\mathsf z\neq \mathsf x_2\circledast\mathsf z\). Together with \(\mathsf x_1\circledast\mathsf z
\preceq
\mathsf x_2\circledast\mathsf z\), this proves \(\mathsf x_1\circledast\mathsf z
\prec
\mathsf x_2\circledast\mathsf z\).
\subsection{Proof of Proposition IV.17}
Let \(\mathsf x\preceq\mathsf x'\) and \(\delta=\mathsf x'-\mathsf x\). By Proposition IV.16 and the concavity of the entropy kernel,
\[
\mathrm H(\delta\circledast\mathsf z)\ge0,
\qquad
\mathrm H(\delta\boxast\mathsf z)\ge0.
\] 
Since \(\delta(\mathcal S_q)=0\), by Lemma IV.10,  \(\mathrm H(\delta\circledast\mathsf z)
+
\mathrm H(\delta\boxast\mathsf z)
=
\mathrm H(\delta)\). Thus
\[
0\le\mathrm H(\delta*\mathsf z)\le\mathrm H(\delta).
\]
Since \(\mathrm H (\delta*(\mathsf z_1-\mathsf z_2) ) =
\mathrm H(\delta*\mathsf z_1)
-
\mathrm H(\delta*\mathsf z_2)\),  and  both terms are in \([0,\mathrm H(\delta)]\),
\[
\left|
\mathrm H\bigl(\delta*(\mathsf z_1-\mathsf z_2)\bigr)
\right|
\le\mathrm H(\delta).
\]
Finally, replacing \(\mathsf z_j\) with \(\mathsf z_j*\mathsf z_3\), \(j=1,2\), proves the third inequality.
\subsection{Proof of Proposition IV.18}
\textit{Proof of i):} Let \(\bar{\mathsf x}_n=\Psi\mathsf{x}_n\), \(\bar{\mathsf y}_n=\Psi\mathsf y_n\), \(\bar{\mathsf x}=\Psi\mathsf{x}\) and \(\bar{\mathsf y}=\Psi\mathsf y\).
Since \(\mathsf x_n\preceq\mathsf y_n\), by Theorem IV.13 iv), for every \(\eta\in\mathcal P_2(\mathbb R^q)\),
\[
W_2(\bar{\mathsf x}_n,\eta)^2
-
W_2(\bar{\mathsf y}_n,\eta)^2
\le
\mathrm Q(\mathsf x_n)-\mathrm Q(\mathsf y_n).
\]
By the triangle inequality for \(W_2\),  \(|
W_2(\bar{\mathsf x}_n,\eta)
-
W_2(\bar{\mathsf x},\eta)|
\le d_W(\mathsf x_n,\mathsf x)\to0\),
and similarly for \(\mathsf y_n\). Since \(\mathrm Q\) is continuous,\,passing to the limit gives  \[
W_2(\bar{\mathsf x},\eta)^2
-
W_2(\bar{\mathsf y},\eta)^2
\le
\mathrm Q(\mathsf x)-\mathrm Q(\mathsf y).
\] 
Since \(\eta\in\mathcal P_2(\mathbb R^q)\) is arbitrary, by Theorem IV.13 iv), \(\mathsf x\preceq\mathsf y\).

\emph{Proof of ii):} Suppose \(\mathsf x_n\preceq\mathsf x_{n+1}\). Then \(\{\mathrm Q(\mathsf x_n)\}\) is nonincreasing and bounded below by \(1/q\). For \(m\ge n\), by \eqref{eq:dQH},
\[
d_W(\mathsf x_n,\mathsf x_m)^2 \le
\mathrm Q(\mathsf x_n)-\mathrm Q(\mathsf x_m) \longrightarrow0,
\qquad m,n\to\infty.
\]
Thus \(\{\mathsf x_n\}\) is Cauchy. By completeness of \((\mathcal X_q,d_W)\), \(\mathsf x_n\xrightarrow{d_W}\mathsf x\) for some \(\mathsf x\in\mathcal X_q\). By Part i), \(\mathsf x_n\preceq\mathsf x\ \forall n\). 
The case \(\mathsf x_n\succeq\mathsf x_{n+1}\) follows identically, with the order reversed.
\subsection{Proof of Proposition IV.21}
Since degradation is preserved under convex combinations, by Proposition IV.16, \(\mathsf T_s\) is nondecreasing in both arguments.\,By Proposition IV.11, it is jointly continuous. Thus, i) follows by induction. For ii), suppose that \(\mathsf T_s(\mathsf x;\mathsf c)\preceq\mathsf x\). By monotonicity, \(\mathsf T_s^{(\ell+1)}(\mathsf x;\mathsf c)
\preceq
\mathsf T_s^{(\ell)}(\mathsf x;\mathsf c)\) for  \(\ell\ge0\).  Proposition IV.18 ii) gives convergence in \(d_W\), and continuity of \(\mathsf T_s\) shows that the limit is a fixed point. The case with the reversed order is identical. For iii), since \(\mathsf T_s(\Delta_0;\mathsf c)
=
\mathsf c
\preceq\Delta_0\), by part ii),  \(\mathsf x_\ell
\xrightarrow{d_W}
\mathsf x_\infty(\mathsf c)\)\,\,and \(\mathsf T_s(\mathsf x_\infty(\mathsf c);\mathsf c)
=
\mathsf x_\infty(\mathsf c)\). If \(\mathsf x\) is a DE fixed point, then by monotonicity,  \(\mathsf x
 =\mathsf T_s^{(\ell)}(\mathsf x;\mathsf c) \preceq
\mathsf T_s^{(\ell)}(\Delta_0;\mathsf c)
=\mathsf x_\ell\).
Passing to the limit and applying Proposition IV.18 i), we obtain \(\mathsf x\preceq\mathsf x_\infty(\mathsf c)\). Finally, since \(\Delta_\infty\) is always a fixed point, \(\mathsf x_\infty(\mathsf c)=\Delta_\infty\) if and only if \(\Delta_\infty\)  is the unique fixed point. This completes the proof.
\subsection{Proof of Theorem IV.22}
\emph{Proof of i).}  Assume  \(\alpha:=\mathfrak B(\mathsf c)\lambda'(0)\rho'(1)<1\).
For \(\mathsf x\in\mathcal X_q\), put \(b=\mathfrak B(\mathsf x)\in[0,1]\). By Lemma A2.2, for every integer\,\(m\ge1\),
\[
\mathfrak B(\mathsf x^{\boxast m}) \le mb+
\sum_{j=2}^{m}
\binom mj(q-2)^{j-1}b^j \le mb+C_{q,m}b^2,\qquad C_{q,m}
\coloneqq
\sum_{j=2}^{m}\binom mj(q-2)^{j-1}.
\] 
This bound also holds for \(q=2\), in which case \(C_{2,m}=0\). Consequently,
\[
\mathfrak B\bigl(\rho^{\boxast}(\mathsf x)\bigr)
 =\sum_{r\ge2}\rho_r
\mathfrak B(\mathsf x^{\boxast(r-1)}) \le \sum_{r\ge2}\rho_r
\bigl[(r-1)b+C_{q,r-1}b^2\bigr] =\rho'(1)b+O(b^2).
\]
The remainder is uniform over \(\mathsf x\in\mathcal X_q\) because \(0\le b\le1\) and the polynomial \(\rho\) has finite degree. More precisely, the quadratic coefficient can be bounded by the finite constant \(\sum_{r\ge2}\rho_r C_{q,r-1}\), which depends only on \(q\) and \(\rho\). By the multiplicativity\,\,of\,\,\(\mathfrak B\) under \(\circledast\), we further have
\begin{equation}
\mathfrak B\bigl(\mathsf T_s(\mathsf x;\mathsf c)\bigr) =\mathfrak B(\mathsf c)\,
\lambda\!\left(
\mathfrak B\bigl(\rho^{\boxast}(\mathsf x)\bigr)
\right) \le
\mathfrak B(\mathsf c)
\left[
\lambda'(0)\rho'(1)b+O(b^2)
\right] =\alpha b+O(b^2),
\qquad b\downarrow0,\label{eq:A13}
\end{equation}
where the remainder is again uniform over \(\mathsf x\in\mathcal X_q\), since \(\lambda\) is also a polynomial with finite degree. As \(\alpha<1\), there exist\,\,\(\kappa>0\) and \(\theta\in(0,1)\) such that \(\mathfrak B\bigl(\mathsf T_s(\mathsf x;\mathsf c)\bigr)
\le\theta\mathfrak B(\mathsf x)\) whenever \(\mathfrak B(\mathsf x)\le\kappa\).
Thus, if \(\mathfrak B(\mathsf x_{\ell_0})\le\kappa\) for some \(\ell_0\ge0\), then
\[
\mathfrak B(\mathsf x_\ell)
\le
\theta^{\ell-\ell_0}
\mathfrak B(\mathsf x_{\ell_0})
\longrightarrow0.
\]
Since \(\mathfrak B\) is continuous on the compact space \((\mathcal X_q,d_W)\) and vanishes only at \(\Delta_\infty\), it follows that \(\mathsf x_\ell\xrightarrow{d_W}\Delta_\infty\).

\emph{Proof of ii).}  Assume \(\mathfrak B(\mathsf c)\lambda'(0)\rho'(1)>1\), and put \(a\coloneqq \lambda'(0)\rho'(1)\). Set
\[
\Delta_{\{0,*\}}\coloneqq
\frac1{q-1}
\sum_{j\in\mathbb F_q^\times}
\Delta_{\{0,j\}}\in\mathcal{X}_q,
\]
where \(\Delta_{\{0,j\}}\) is concentrated on the APP vector assigning probability \(1/2\) to each of \(0\) and \(j\). For \(0\le\varepsilon\le1/2\), define
\[
\mathsf z_\varepsilon\coloneqq 
(1-2\varepsilon)\Delta_\infty
+2\varepsilon\Delta_{\{0,*\}}.
\]
Then, \(\mathsf z_\varepsilon\) is the conditional APP distribution of a size-2 QPEC with erasure probability \(2\varepsilon\), given that \(0\in\mathbb{F}_q\) is transmitted.\,\,By Lemma IV.15, it follows that
\begin{equation}
\mathsf x\succeq
\mathsf z_{\mathfrak P(\mathsf x)}
\qquad \forall\mathsf x\in\mathcal X_q.\label{eq:A5}
\end{equation}
Moreover, \(\mathsf z_{\varepsilon_2}\succeq\mathsf z_{\varepsilon_1}\) whenever \(\varepsilon_2\ge\varepsilon_1\). We first derive the linearization of the DE operator around \(\Delta_\infty\). For any \(\mathsf u\in\mathcal X_q\), let
\[
\mathsf u_t\coloneqq 
(1-t)\Delta_\infty+t\mathsf u,\qquad t\in [0,1].
\]
Since \(\Delta_\infty\) is the identity under \(\boxast\) and is absorbing under \(\circledast\), for every integer \(m\ge1\),
\[
\mathsf u_t^{\boxast m} =
\Delta_\infty
+mt(\mathsf u-\Delta_\infty)
+O(t^2),\qquad \mathsf u_t^{\circledast m} =
\Delta_\infty
+t^m(\mathsf u^{\circledast m}-\Delta_\infty).
\]
Since \(\lambda\) and \(\rho\) are polynomials with  finite degrees, the above estimates imply
\[
\rho^{\boxast}(\mathsf u_t)
 =
\Delta_\infty
+t\rho'(1)(\mathsf u-\Delta_\infty)
+O(t^2),\qquad \lambda^{\circledast}
\bigl(\rho^{\boxast}(\mathsf u_t)\bigr)
 =
\Delta_\infty
+ta(\mathsf u-\Delta_\infty)
+O(t^2).
\] 
Hence, for every \(\mathsf u\in\mathcal{X}_q\), as \(t\downarrow 0\), we have
\begin{equation}
\mathsf T_s(\mathsf u_t;\mathsf c)
=
\Delta_\infty
+ta(\mathsf c\circledast\mathsf u-\Delta_\infty)
+O(t^2)\label{eq:A6}.
\end{equation}
Here, the remainders are understood in total variation and are uniform over \(\mathsf u\in\mathcal X_q\). Write \(\|\mu\|_{\mathrm{TV}}=|\mu|(\mathcal S_q)\). The pushforward definitions give
\[
\|\mu*\nu\|_{\mathrm{TV}}
\le
\|\mu\|_{\mathrm{TV}}\|\nu\|_{\mathrm{TV}},
\qquad
*\in\{\circledast,\boxast\}.
\]
Thus, by the signed extensions in Appendix II-D, \(\mathsf T_s\) is a polynomial map in this norm. For zero-mass signed measures\,\,\(\nu\in\mathcal M_q\),
\[
\mathsf T_s(\Delta_\infty+\nu;\mathsf c)
=
\Delta_\infty
+a\,\mathsf c\circledast\nu
+O(\|\nu\|_{\mathrm{TV}}^2).
\]
Iterating this expansion with \(t=2\varepsilon\) gives, for every fixed integer \(n\ge1\), 
\begin{equation}
\mathsf T_s^{(n)}
(\mathsf z_\varepsilon;\mathsf c)
= 
\Delta_\infty +
2\varepsilon a^n
\bigl(
\mathsf c^{\circledast n}
\circledast\Delta_{\{0,*\}}
-\Delta_\infty
\bigr)
+O_n(\varepsilon^2).\label{eq:A7}
\end{equation}
For every \(\mathsf{u}\in\mathcal{X}_q\), we have
\begin{equation}
\mathfrak P
(\mathsf u\circledast\Delta_{\{0,*\}})
 =
\frac1{2(q-1)}
\sum_{j\in\mathbb F_q^\times}
\int 
\mathbf1 {\{j=1\}}
\min\left\{1,\frac{y_1}{y_0}\right\}
\,\mathsf u(d\underline y) =
\frac1{2(q-1)}
\int 
\min\left\{1,\frac{y_1}{y_0}\right\}
\,\mathsf u(d\underline y) =
\frac{\mathfrak P(\mathsf u)}{q-1}.\label{eq:A8}
\end{equation}
Combining \eqref{eq:A7} and \eqref{eq:A8}, we obtain
\begin{equation}
	\mathfrak P\!\left(
\mathsf T_s^{(n)}
(\mathsf z_\varepsilon;\mathsf c)
\right)
=
\frac{2a^n}{q-1}
\mathfrak P(\mathsf c^{\circledast n})
\,\varepsilon
+O_n(\varepsilon^2).\label{eq:A9}
\end{equation} 
By the repetition-error exponent established in Lemma A2.4 and the assumption \(a\mathfrak B(\mathsf c)>1\), we have
\[
\lim_{n\to\infty}
\frac1n
\log\left[
a^n\mathfrak P(\mathsf c^{\circledast n})
\right] =
\log\bigl(a\mathfrak B(\mathsf c)\bigr) >0,
\]
so we may fix an integer \(N\ge1\) such that \(\frac{2}{q-1}a^N
\mathfrak P(\mathsf c^{\circledast N})
>1\). By \eqref{eq:A9}, there exist constants \(\gamma>1\) and \(\varepsilon_0\in(0,1/2)\) such\,that
\begin{equation}
\mathfrak P\!\left(
\mathsf T_s^{(N)}
(\mathsf z_\varepsilon;\mathsf c)
\right)
\ge\gamma\varepsilon \qquad \forall  0<\varepsilon\le\varepsilon_0.\label{eq:A10}
\end{equation} 
We may further decrease \(\varepsilon_0\), if necessary, so that
\begin{equation}
\mathfrak P\!\left(
\mathsf T_s^{(r)}
(\mathsf z_{\varepsilon_0};\mathsf c)
\right)>0,
\qquad 0\le r<N.\label{eq:A12}
\end{equation}
This follows from \eqref{eq:A9} for each fixed \(1\le r<N\), since \(a>0\) and \(\mathfrak P(\mathsf c^{\circledast r})>0\); for \(r=0\), we have \(\mathfrak P(\mathsf z_{\varepsilon_0})=\varepsilon_0/(q-1)>0\). Given any \(\mathsf x\in\mathcal{X}_q\setminus\{\Delta_\infty\}\),   set \(p=\mathfrak P(\mathsf x)>0\). By \eqref{eq:A5}, degradation monotonicity of DE, and  \eqref{eq:A10},
\begin{equation}
\mathfrak P\bigl(
\mathsf T_s^{(N)}(\mathsf x;\mathsf c)
\bigr) \ge
\mathfrak P\!\left(
\mathsf T_s^{(N)}
(\mathsf z_{\min\{p,\varepsilon_0\}};\mathsf c)
\right) \ge
\gamma\min\{p,\varepsilon_0\}.\label{eq:A11}
\end{equation}
For an arbitrary initialization \(\mathsf x_0\in\mathcal{X}_q\setminus\{\Delta_\infty\}\), set \(p_k\coloneqq\mathfrak P(\mathsf x_{kN})\) for \(k\ge0\). It follows from \eqref{eq:A11} that
\[
p_{k+1}
\ge\gamma\min\{p_k,\varepsilon_0\}.
\]
Since \(p_0>0\) and \(\gamma>1\), there exists \(k_0\ge0\) such that \(p_k\ge\varepsilon_0\) \(\forall k\ge k_0\). Then by \eqref{eq:A5}, \(\mathsf x_{kN}\succeq
\mathsf z_{\varepsilon_0}\) for all  \(k\ge k_0\).
Consequently, for all \(k\ge k_0\) and \(0\le r<N\), \(\mathfrak P(\mathsf x_{kN+r}) \ge
\mathfrak P  (
\mathsf T_s^{(r)}
(\mathsf z_{\varepsilon_0};\mathsf c)
 )\). By \eqref{eq:A12}, we have
\[
\xi\coloneqq
\min_{0\le r<N}
\mathfrak P\!\left(
\mathsf T_s^{(r)}
(\mathsf z_{\varepsilon_0};\mathsf c)
\right)>0.
\] 
Thus, \(\liminf_{\ell\to\infty}
\mathfrak P(\mathsf x_\ell)
\ge\xi>0\). This completes the proof.
\subsection{Proof of Corollary IV.23}
It suffices to prove the general case, since the special case follows by taking \(\mathcal C=\{\mathsf c\}\). Put
\[
\alpha_{\mathcal C}
\coloneqq
\sup_{\mathsf c\in\mathcal C}
\mathfrak B(\mathsf c)\lambda'(0)\rho'(1)<1.
\] 
By \eqref{eq:A13}, there exist constants \(K\ge0\) and \(b_0>0\) such that, for every \(\mathsf c\in\mathcal C\) and \(\mathsf x\in\mathcal X_q\) with \(b=\mathfrak B(\mathsf x)\le b_0\),
\begin{equation}
\mathfrak B\bigl(\mathsf T_s(\mathsf x;\mathsf c)\bigr)
 \le
\mathfrak B(\mathsf c)
\lambda'(0)\rho'(1)b+Kb^2 \le
(\alpha_{\mathcal C}+Kb)b.\label{eq:A14}
\end{equation} 
Here, \(K\) can be chosen independently of \(\mathsf c\), since \(\mathfrak B(\mathsf c)\le1\) and \((\lambda,\rho)\) have finite degrees. Choose \(\kappa_{\mathcal C}\in(0,b_0]\) such that
\[
\alpha_{\mathcal C}+K\kappa_{\mathcal C}<1.
\]
If \(\mathsf x\in\mathcal F(\mathsf c)\) for some \(\mathsf c\in\mathcal C\), then \(b=\mathfrak B(\mathsf x)>0\); if \(b\le\kappa_{\mathcal C}\), then by \eqref{eq:A14}, \(b
 =\mathfrak B\bigl(\mathsf T_s(\mathsf x;\mathsf c)\bigr) \le
(\alpha_{\mathcal C}+K\kappa_{\mathcal C})b
<b\), a contradiction. Therefore, \(\mathfrak B(\mathsf x)\ge\kappa_{\mathcal C}\) for all \(\mathsf c\in\mathcal C\) and \(\mathsf x\in\mathcal F(\mathsf c)\). Since \(\mathfrak B(\Delta_\infty)=0<\kappa_{\mathcal C}\), we have 
\[
 \{
(\mathsf c,\mathsf x):
\mathsf c\in\mathcal C,\
\mathsf x\in\mathcal F(\mathsf c)
 \} =
 \{
(\mathsf c,\mathsf x)\in
\mathcal C\times\mathcal X_q:
\mathsf T_s(\mathsf x;\mathsf c)=\mathsf x,\
\mathfrak B(\mathsf x)\ge\kappa_{\mathcal C}
 \}.
\]
The right-hand side is compact by the compactness of \(\mathcal C\times\mathcal X_q\) and the continuity of \(\mathsf T_s\) and \(\mathfrak B\). For the special case  \(\mathcal C=\{\mathsf c\}\), the set \(\mathcal F(\mathsf c)\) is compact and, since \(\Delta_\infty\notin\mathcal F(\mathsf c)\), is separated from \(\Delta_\infty\).  
\section{Proofs of Results in Section V}
\subsection{Proof of Theorem V.2}
The proof is similar to that of\cite[Thm. 2]{LDPC1}. We first establish the   computation-tree limit
\[\mathbb{E} P_{\ell,n}=p_\ell+O(n^{-1}).\]
Let \(p(\underline t)\) be the edge-type distribution specified in Section III-B. At each variable-node position, including the virtual positions, there are \(np(\underline t)\) nodes of type \(\underline t\). For a uniformly selected check-node-input edge at position \(i\), the probability that its variable-node endpoint is at position \(i-k\) and has edge type \(\underline t\) is
\[
\frac{np(\underline t)t_k}{nd_l}
=\frac{t_k}{d_l}p(\underline t) =
\Pr\{J_1=k,  \boldsymbol J
\text{ has edge type }\underline t\},
\]
where the last equality follows from the exchangeability of \(\boldsymbol J\). Thus, \(\boldsymbol J\sim\pi_w\) indeed describes the offset distribution associated with a uniformly selected check-node-input edge in the finite Tanner graph, consistent with the coupled DE recursion. For fixed \(\ell\), the computation neighborhood of a uniformly selected edge contains only \(O_{d_l,d_r,\ell}(1)\) sockets. During the exploration of this neighborhood, each newly revealed socket is matched uniformly among the remaining available sockets at its check-node position. Since all node degrees are bounded and each position has \(nd_l\) sockets, the probability of encountering a previously visited node is \(O(n^{-1})\). Therefore, for an edge \(e\) selected uniformly among those originating from transmitted variable nodes,
\[
\Pr(\mathcal T_e^c)=O(n^{-1}), 
\]
where \(\mathcal T_e\) denotes the event that its depth-\(\ell\) computation neighborhood is cycle-free. Conditioned on \(\mathcal T_e\), the neighborhood is generated by sampling finitely many sockets without replacement. Coupling these samples with independent samples from the corresponding limiting distributions shows that their joint laws differ by \(O(n^{-1})\) in total variation. Consequently, the conditional distribution of the BP message associated with \(e\), given \(\mathcal T_e\), differs by \(O(n^{-1})\) from its ideal computation-tree distribution.\,The\\ check-node and variable-node updates on the ideal tree yield the recursion defining \(\underline{\mathsf x}^{(\ell)}\). For the edge \(e\), its variable-node\,position is uniform on \(\{1,\ldots,N\}\), and its ordered offset vector has distribution \(\pi_w\). Hence, the corresponding error probability on the limiting computation tree is \(p_\ell\). Let \(Z_\ell(e)\) denote the random error indicator for the  edge \(e\), then
\begin{equation}
\mathbb E[Z_\ell(e)\,|\,\mathcal T_e]
=p_\ell+O(n^{-1}).\notag
\end{equation}
By   the definition of \(P_{\ell,n}\), we obtain
\[
\mathbb E P_{\ell,n} =\mathbb E Z_\ell(e) =\Pr(\mathcal T_e)
\mathbb E[Z_\ell(e)\,|\,\mathcal T_e] +
\Pr(\mathcal T_e^c)
\mathbb E[Z_\ell(e)\,|\,\mathcal T_e^c] =\bigl(1-O(n^{-1})\bigr)
\bigl(p_\ell+O(n^{-1})\bigr)
+O(n^{-1}) =p_\ell+O(n^{-1}).
\]

We next prove the concentration inequality \eqref{eq:concentration}.   Let \(Z=Nd_ln\,P_{\ell,n}\) be the number of erroneous variable-to-check messages after \(\ell\) BP updates. By the ensemble construction in Section III-B, the random monomial map at each check-node position consists of a uniformly random socket matching and independent, uniformly distributed nonzero edge labels. These random choices are independent across check-node positions. We reveal the randomness in the following order: the \(K\) socket matchings, one matched socket at a time; the \(Kd_ln\) edge labels; the \(Nn\) channel outputs; and, if necessary, the \(Nd_ln\) independent tie-breaking variables. The total number of exposures is at most
\[
T=\bigl(2Kd_l+N+Nd_l\bigr)n. 
\]
Since BP is a local algorithm, changing at most two graph edges, one edge label, one channel output, or one tie-breaking\,variable can affect only a bounded number of messages after \(\ell\) iterations. More precisely, there exists a constant \(\Gamma_\ell>0\), depending only on \(d_l,d_r,\ell\), such that any such modification changes \(Z\) by at most \(\Gamma_\ell\), e.g., \(\Gamma_\ell=16\max\{2,d_l,d_r\}^{2\ell+4}\) is a valid conservative bound, obtained by counting directed messages whose computation neighborhoods intersect the modified edges or vertices. To handle the dependence within a uniform matching, observe that, conditioned on previously exposed matches, any two possible partners of the next socket can be coupled by a transposition of the remaining matching. This bijection preserves the conditional uniform distribution and changes at most two edges. Let \((M_t)_{t=0}^{T}\) denote the Doob martingale associated with these successive exposures. By the preceding bounded-difference argument,
\[
M_0=\mathbb E Z,\qquad M_T=Z,
\qquad
|M_t-M_{t-1}|\le\Gamma_\ell
\quad(1\le t\le T). 
\]
Then, by the Azuma–Hoeffding inequality, for every \(\delta>0\),
\[
\Pr\left\{
|P_{\ell,n}-\mathbb E P_{\ell,n}|>\delta
\right\} =
\Pr\left\{
|Z-\mathbb E Z|>\delta Nd_ln
\right\} \le
2\exp\left(
-\frac{\delta^2N^2d_l^2n^2}
{2T\Gamma_\ell^2}
\right) =
2\exp\left(
-\frac{\delta^2N^2d_l^2}
{2(2Kd_l+N+Nd_l)\Gamma_\ell^2}\,n
\right).
\] 
Hence, the concentration inequality \eqref{eq:concentration} holds with 
\[
\beta=
\frac{\delta^2N^2d_l^2}
{2(2Kd_l+N+Nd_l)\Gamma_\ell^2}>0,
\] 
which is independent of \(n\).  Since  \(\mathbb E P_{\ell,n}=p_\ell+O(n^{-1})\), for any fixed \(\varepsilon>0\) and all sufficiently large \(n\),  \(|\mathbb E P_{\ell,n}-p_\ell|<\frac{\varepsilon}{2}\).\,Thus
\[
\Pr\{|P_{\ell,n}-p_\ell|>\varepsilon\} \le
\Pr\left\{
|P_{\ell,n}-\mathbb E P_{\ell,n}|
>\frac{\varepsilon}{2}
\right\} \le 2e^{-\beta_\varepsilon n}
\]
for some constant \(\beta_\varepsilon >0\) and all sufficiently large \(n\). Then, \(\sum_n \Pr\{|P_{\ell,n}-p_\ell|>\varepsilon\}<\infty\) for every \(\varepsilon>0\). By the Borel–Cantelli lemma, \(P_{\ell,n}\longrightarrow p_\ell\) a.s.
 \subsection{Proof of Lemma V.3}
 \textit{Proof of i).} We provide the detailed calculations for the directional derivatives of \(U_c\).\footnote{Directional derivatives of polynomial functionals are computed by the usual rules of polynomial differentiation, using the linearity of \(\mathrm H\) and the bilinearity of \(\circledast\) and \(\boxast\). See\cite[Props.~14 and~16]{SCLDPC4} for the basic differentiation rules and\cite[Lemmas~23 and~38]{SCLDPC4} for explicit calculations of potential derivatives.} Fix \(\underline{\mathsf x}\in\mathcal X_q^K\), \(\underline{\mathsf v}\in\mathcal D_q^K\), and define
 \[
 \mathsf u_i
 \coloneqq
 \mathsf x_i^{\boxast(d_r-2)}
 \boxast\mathsf v_i,
 \qquad 1\le i\le K,
 \]
 with the convention \(\mathsf x_i^{\boxast0}=\Delta_\infty\). Each \(\mathsf u_i\in\mathcal{D}_q\) is a signed measure of total mass zero. Differentiating the first sum in \(U_c\) gives
 \[
 d_l(d_r-1)
 \sum_{i=1}^{K}
 \left[
 \mathrm H(\mathsf u_i)
 -\mathrm H(\mathsf x_i\boxast\mathsf u_i)
 \right].
 \]
 For the second sum, write  \(\mathsf m_i\coloneqq
 \mathsf x_i^{\boxast(d_r-1)}\). Then by the exchangeability of \(\boldsymbol J\), its derivative can be given by
 \[
-(d_r-1)\sum_{j=1}^{N}
\mathbb E_{\boldsymbol J}
\Bigg[
\sum_{a=1}^{d_l}
\mathrm H\left(
\mathsf c\circledast\mathsf u_{j+J_a}
\circledast\left(\mathop{\circledast}_{b\ne a}
\mathsf m_{j+J_b}\right)
\right)
\Bigg] =-d_l(d_r-1)
\sum_{j=1}^{N}
\mathbb E_{\boldsymbol J}
\left[
\mathrm H\left(
\mathsf c\circledast\mathsf u_{j+J_1}
\circledast
\left(\mathop{\circledast}_{a=2}^{d_l}
\mathsf m_{j+J_a}\right)
\right)
\right],
 \]
 Reindexing by \(i=j+J_1\) and recalling that \(\mathsf c_j=\Delta_\infty\) outside \(\{1,\ldots,N\}\), we can rewrite the above quantity as
\[
-d_l(d_r-1)
\sum_{i=1}^{K}
\mathbb E_{\boldsymbol J}
\left[
\mathrm H\left(
\mathsf c_{i-J_1}
\circledast
 \left(\mathop{\circledast}_{a=2}^{d_l}
\mathsf m_{i-J_1+J_a}\right) 
\circledast\mathsf u_i
\right)
\right] =-d_l(d_r-1)
\sum_{i=1}^{K}
\mathrm H\left(
[\mathsf T_c(\underline{\mathsf x};\mathsf c)]_i
\circledast\mathsf u_i
\right),
\]
by the definition of \(\mathsf T_c\) and the linearity of the signed-measure extensions. Combining the two derivatives, we obtain
\[
\frac{\mathrm D U_c(\underline{\mathsf x};\mathsf c)
	[\underline{\mathsf v}]}
{d_l(d_r-1)} =
\sum_{i=1}^{K}
\left[
\mathrm H(\mathsf u_i)
-\mathrm H(\mathsf x_i\boxast\mathsf u_i) 
-\mathrm H\bigl(
[\mathsf T_c(\underline{\mathsf x};\mathsf c)]_i
\circledast\mathsf u_i
\bigr)
\right].
\] 
By the entropy duality rule and the zero total mass of \(\mathsf u_i\), \(\mathrm H(\mathsf u_i)
-\mathrm H(\mathsf x_i\boxast\mathsf u_i)
=
\mathrm H(\mathsf x_i\circledast\mathsf u_i)\). Thus,
\[
\frac{\mathrm D U_c(\underline{\mathsf x};\mathsf c)
	[\underline{\mathsf v}]}
{d_l(d_r-1)} =
\sum_{i=1}^{K}
\mathrm H\left(
\bigl(\mathsf x_i-
[\mathsf T_c(\underline{\mathsf x};\mathsf c)]_i\bigr)
\circledast\mathsf u_i
\right) =
\sum_{i=1}^{K}
\mathrm H\left(
\left([\mathsf T_c(\underline{\mathsf x};\mathsf c)]_i
-\mathsf x_i\right)
\boxast\mathsf u_i
\right).
\]
The last equality again follows from entropy duality.  By similar calculations, we can verify the directional derivatives of \(U_s\).

\textit{Proof of ii).} We first prove a consequence of entropy duality and degradation: for \(\mathsf a\preceq\mathsf b\), \(\mathsf c\preceq\mathsf d\), and \(\mathsf z\in\mathcal X_q\), 
\begin{equation}
\mathrm H\bigl(
(\mathsf b-\mathsf a)
\circledast(\mathsf d-\mathsf c)
\bigr) \ge0,\qquad
\mathrm H\bigl(
(\mathsf b-\mathsf a)
\boxast(\mathsf d-\mathsf c)
\boxast\mathsf z
\bigr) \le0.\label{eq:entropy_dual_cons}
\end{equation}
The proof follows those of\cite[Lem. E. 10]{ModernCode},\cite[Lem. 5]{GEXIT}. Represent \(\mathsf a,\mathsf c\) by conditionally independent observations \(Y,Z\) of a uniform input \(X\), and let \(Y',Z'\) be their respective degraded observations representing \(\mathsf b,\mathsf d\). Then $X\rightarrow Y \rightarrow Y^{\prime}$, $X\rightarrow Z \rightarrow Z^{\prime}$ and $(Y,Y^{\prime})\rightarrow X\rightarrow (Z,Z^{\prime})$ form Markov chains, and by the entropy representation of \(\circledast\),
\[
\mathrm H (
(\mathsf b-\mathsf a)
\circledast(\mathsf d-\mathsf c)
 ) =
H(X\mid Y',Z')
-H(X\mid Y,Z') -H(X\mid Y',Z)
+H(X\mid Y,Z) =I(Y;Z\mid Y',Z')\ge0.
\]
 Since both signed differences have total mass zero, by the entropy duality rule,   \(\mathrm H (
 (\mathsf b-\mathsf a)
 \boxast(\mathsf d-\mathsf c)
  )
 =
 -\mathrm H (
 (\mathsf b-\mathsf a)
 \circledast(\mathsf d-\mathsf c)
  )
 \le0\). The second inequality in \eqref{eq:entropy_dual_cons} follows by replacing \(\mathsf a\preceq\mathsf b\) with \(\mathsf a\boxast\mathsf z\preceq\mathsf b\boxast\mathsf z\). Now, let
 \[
 \underline{\mathsf y}
 =
 \mathsf T_c(\underline{\mathsf x};\mathsf c),
 \]
and suppose that \(\underline{\mathsf y}\preceq\underline{\mathsf x}\) or \(\underline{\mathsf y}\succeq\underline{\mathsf x}\). Consider the interpolation \(\underline{\mathsf z}(t)
=
(1-t)\underline{\mathsf x}
+t\underline{\mathsf y}\) for \(0\le t\le1\). If \(\underline{\mathsf y}\preceq\underline{\mathsf x}\), then \(\underline{\mathsf y}
\preceq\underline{\mathsf z}(t)
\preceq\underline{\mathsf x}\),\,\,so by the monotonicity of \(\mathsf T_c\), we have \(\mathsf T_c(\underline{\mathsf z}(t);\mathsf c)
\preceq
\mathsf T_c(\underline{\mathsf x};\mathsf c)
=
\underline{\mathsf y}
\preceq\underline{\mathsf z}(t)\). The \(\underline{\mathsf y}\succeq\underline{\mathsf x}\) case follows similarly, with all degradation inequalities reversed. Now define \(\phi(t)
\coloneqq
U_c(\underline{\mathsf z}(t);\mathsf c)\). By the definition of the directional derivative of \(U_c\),
 \[
\phi'(t)=\mathrm{D} U_c(\underline{\mathsf{z}}(t);\mathsf{c})[\underline{\mathsf y}-\underline{\mathsf{x}}]
= d_l(d_r-1)
\sum_{i=1}^{K}
\mathrm H\left(
\bigl(
[\mathsf T_c(\underline{\mathsf z}(t);\mathsf c)]_i
-\mathsf z_i(t)
\bigr)  \boxast
\mathsf z_i(t)^{\boxast(d_r-2)}
\boxast(\mathsf y_i-\mathsf x_i)
\right).
 \]
For each \(1\le i\le K\), the signed differences \( [\mathsf T_c(\underline{\mathsf z}(t);\mathsf c)]_i
-\mathsf z_i(t)\) and \(\mathsf y_i-\mathsf x_i\)  have the same orientation under degradation.\,Thus, by (\ref{eq:entropy_dual_cons}), \( \phi'(t)\le0\ \forall 0\le t\le1\), and hence
 \[
	U_c(\mathsf T_c(\underline{\mathsf x};\mathsf c);\mathsf c)=\phi (1)=\phi(0)+\int_0^1 \phi'(t)\, dt\le \phi(0)=
U_c(\underline{\mathsf x};\mathsf c).
 \]
The same argument applies to \(U_s\). Consequently, along any monotone DE sequence,
 \[
U_s(\mathsf x^{(\ell+1)};\mathsf c)
 \le U_s(\mathsf x^{(\ell)};\mathsf c),\qquad
U_c(\underline{\mathsf x}^{(\ell+1)};\mathsf c) \le U_c(\underline{\mathsf x}^{(\ell)};\mathsf c).
 \]
 
 \textit{Proof of iii).} Let \(0\le h_1<h_2\le\log q\), and put \(\mathsf c_j=\mathsf c_{h_j}\) for \(j=1,2\). Then \(\mathsf c_1\prec\mathsf c_2\), and by Propositions IV.14 and IV.16,
\begin{equation} 
U_s(\mathsf a;\mathsf c_2)
-U_s(\mathsf a;\mathsf c_1) =
\mathrm H\left(
\mathsf c_1\circledast
(\mathsf a^{\boxast(d_r-1)})^{\circledast d_l}
\right) -
\mathrm H\left(
\mathsf c_2\circledast
(\mathsf a^{\boxast(d_r-1)})^{\circledast d_l}
\right)
<0\qquad \forall \mathsf{a}\in\mathcal{X}_q\setminus\{\Delta_\infty\}.\label{eq:aux}
\end{equation}
Indeed, for \(\mathsf{a}\in\mathcal{X}_q\setminus\{\Delta_\infty\}\),  \(\mathsf a\preceq\mathsf a^{\boxast(d_r-1)}\), so \(\mathsf a^{\boxast(d_r-1)}\ne\Delta_\infty\), and then by Lemmas IV.9 and A2.1, \(\mathfrak B (
 (\mathsf a^{\boxast(d_r-1)} )^{\circledast d_l}
 )
=
\mathfrak B  (\mathsf a^{\boxast(d_r-1)} )^{d_l}
>0\), so \(
(\mathsf a^{\boxast(d_r-1)} )^{\circledast d_l}
\neq\Delta_\infty\), and thus \eqref{eq:aux} follows from strict degradation and the fact that  \(\mathrm{H}\) admits a\,\,strictly concave kernel. If \(\mathcal F(\mathsf c_1)=\varnothing\), then \(\Delta E(\mathsf c_1)=+\infty\), and the desired   monotonicity \(\Delta E(\mathsf c_2)\le \Delta E(\mathsf c_1)\) is immediate. Otherwise, fix any \(\mathsf a\in\mathcal F(\mathsf c_1)\). By Proposition IV.21,
 \[
\mathsf a =\mathsf T_s(\mathsf a;\mathsf c_1)
\preceq\mathsf T_s(\mathsf a;\mathsf c_2) \preceq\mathsf T_s^{(2)}(\mathsf a;\mathsf c_2)
\preceq\cdots
\longrightarrow\mathsf y
\in\mathcal F(\mathsf c_2).
 \]
Here \(\mathsf y\ne\Delta_\infty\), since \(\mathsf a\ne\Delta_\infty\) and \(\mathsf a\preceq\mathsf y\). By Lemma V.3 ii) and \eqref{eq:aux}, 
\begin{equation}
\Delta E(\mathsf c_2) \le U_s(\mathsf y;\mathsf c_2)  \le U_s(\mathsf a;\mathsf c_2) <U_s(\mathsf a;\mathsf c_1).\label{eq:dfsa}
\end{equation}
Taking the infimum over \(\mathsf a\in\mathcal F(\mathsf c_1)\) gives \(\Delta E(\mathsf c_2)
\le\Delta E(\mathsf c_1)\). Moreover, if \(\mathcal F(\mathsf c_1)\) is nonempty and compact, continuity of \(U_s\) ensures the existence of \(\mathsf a^*\in\mathcal F(\mathsf c_1)\) such that \(U_s(\mathsf a^*;\mathsf c_1)
=\Delta E(\mathsf c_1)\). Applying \eqref{eq:dfsa} with \(\mathsf a=\mathsf a^*\), we obtain
\[
\Delta E(\mathsf c_2)
<
U_s(\mathsf a^*;\mathsf c_1)
=\Delta E(\mathsf c_1).
\]
This establishes the claimed monotonicity and strict monotonicity of \(h\mapsto\Delta E(\mathsf c_h)\). It remains to establish the sign of \(\Delta E(\mathsf{c}_h)\) on either side of \(h^{\mathrm{FP}}\). If \(h<h^{\mathrm{FP}}\), then, by the definition of \(h^{\mathrm{FP}}\), there exists \(h'>h\) such that \(\Delta E(\mathsf c_{h'})>0\). Hence
\[
\Delta E(\mathsf c_h)
\ge\Delta E(\mathsf c_{h'})
>0.
\]
Suppose that the uncoupled system is stable at \(h^{\mathrm{FP}}\), and let \(h>h^{\mathrm{FP}}\). By the continuity of \(h\mapsto\mathfrak B(\mathsf c_h)\), the stability condition \(\mathfrak B(\mathsf c_{h^{\mathrm{FP}}})
\lambda'(0)\rho'(1)<1\)
continues to hold for all \(h'\) sufficiently close to \(h^{\mathrm{FP}}\). Choose such an \(h'\in(h^{\mathrm{FP}},h)\). By the definition of \(h^{\mathrm{FP}}\), \(\Delta E(\mathsf c_{h'})\le0\), which implies \(\mathcal F(\mathsf c_{h'})\ne\varnothing\). Since the uncoupled system is stable at \(h'\), by Corollary IV.23, \(\mathcal F(\mathsf c_{h'})\) is compact. The strict monotonicity established above thus yields \(\Delta E(\mathsf c_h)
<
\Delta E(\mathsf c_{h'})
\le0\). 
\subsection{Proof of Lemma V.5}
Fix \(w\) and \(\pi_w\). For each \(L\), let \(\mathsf x_i^{(\ell,L)}\) denote the DE message distribution at position \(i\) after \(\ell\) iterations of \(\eqref{eq:coupled_DE_sequence}\), extended by \(\Delta_\infty\) outside \(\{1,\ldots,2L+w-1\}\). Let \(\mathsf c_j^{(L)}=\mathsf c_h\) for \(1\le j\le2L\) and \(\mathsf c_j^{(L)}=\Delta_\infty\) otherwise.  Then, for \(L_1\le L_2\),  we have
\[
\mathsf c_j^{(L_1)}
\preceq\mathsf c_j^{(L_2)},\qquad \mathsf x_i^{(0,L_1)}
\preceq
\mathsf x_i^{(0,L_2)},
\qquad \forall i,j\in\mathbb Z.
\]
Thus, by \(\eqref{eq:coupled_DE}\) and BP monotonicity, induction gives \(\mathsf x_i^{(\ell,L_1)}
\preceq
\mathsf x_i^{(\ell,L_2)}\) for all \(i\in\mathbb Z\) and \(\ell\ge0\), so  perfect DE decoding for \(L_2\) implies perfect DE decoding for \(L_1\). Hence, \(h_c^{\mathrm{BP}}(w,L_1)
\ge h_c^{\mathrm{BP}}(w,L_2)\) and the limit
\[
b_w\coloneqq
\lim_{L\to\infty}h_c^{\mathrm{BP}}(w,L)
\]
exists since \(0\le h_c^{\mathrm{BP}}(w,L)\le\log q\). Now suppose that \(\eta_w\to0\) as \(w\to \infty\). For any \(h<h^{\mathrm{FP}}\), by Lemma V.3 iii), \(\Delta E(\mathsf c_h)>0\). Thus, for all sufficiently large \(w\),
\[
\frac12 C_{q,d_l,d_r}\eta_w
<
\Delta E(\mathsf c_h).
\]
By Theorem V.4 i), \(h_c^{\mathrm{BP}}(w,L)\ge h\) for every \(L\). Therefore, \(b_w\ge h\) for all sufficiently large \(w\), and letting \(h\uparrow h^{\mathrm{FP}}\) gives
\[
\liminf_{w\to\infty}b_w\ge h^{\mathrm{FP}}.
\]
If \(h^{\mathrm{FP}}<\log q\) and the uncoupled system is stable at \(h^{\mathrm{FP}}\), then, for every \(h>h^{\mathrm{FP}}\) and fixed \(w\), by Theorem V.4 ii),\,\(h_c^{\mathrm{BP}}(w,L)\le h\) for all sufficiently large \(L\). Consequently, \(b_w\le h\)  for every \(w\), and  letting \(h\downarrow h^{\mathrm{FP}}\) yields
\[
\limsup_{w\to\infty}b_w\le h^{\mathrm{FP}}.
\]
Combining the two bounds proves \(\eqref{eq:threshold_saturation}\). Finally, if \(h^{\mathrm{FP}}=\log q\), the upper bound \(b_w\le\log q\) is automatic, so the conclusion follows from the lower bound without the stability assumption.
\subsection{Proof of Lemma V.6}
Put \(\pi_w^{(s)}
=
\mathrm{Law}(J_1+s,J_2,\ldots,J_{d_l})\) where \(\boldsymbol{J}\sim\pi_w\). Recall that \(\eta_w(s)=d_{\mathrm{TV}}(\pi_w,\pi_w^{(s)})\). For both ensembles, \(\pi_w\) and \(\pi_w^{(s)}\) are uniform on supports of equal cardinality. Therefore,  
\begin{equation}
\eta_w(s) =\frac12\sum_{\boldsymbol j}
\left|\pi_w(\boldsymbol j)-\pi_w^{(s)}(\boldsymbol j)\right| =1-
\frac{\left|
	\operatorname{supp}\pi_w
	\cap\operatorname{supp}\pi_w^{(s)}
	\right|}
{|\operatorname{supp}\pi_w|}.\label{eq:Q1}
\end{equation}
For the standard ensemble, \(J_1,\ldots,J_{d_l}\) are independent and uniform on \(\{0,\ldots,w-1\}\), so we have
\[
\operatorname{supp}\pi_w
 =\{0,\ldots,w-1\}^{d_l},\qquad
\operatorname{supp}\pi_w^{(s)} =\{s,\ldots,w+s-1\}
\times\{0,\ldots,w-1\}^{d_l-1}. 
\]
If \(1\le s<w\), then \(\operatorname{supp}\pi_w
\cap\operatorname{supp}\pi_w^{(s)}=\{s,\ldots,w-1\}
\times\{0,\ldots,w-1\}^{d_l-1}\); if \(s\ge w\), then \(\operatorname{supp}\pi_w
\cap\operatorname{supp}\pi_w^{(s)}=\varnothing\).\,Thus
\[
\eta_w(s)=
\begin{cases}
	\dfrac{s}{w},&1\le s<w,\\[4pt]
	1,&s\ge w.
\end{cases}
\]
Hence, \(\eta_w=\min_{s\ge1}\eta_w(s)=\frac1w\). For the improved ensemble, \(\boldsymbol J\) is uniform over all ordered \(d_l\)-tuples of distinct elements of \(\{0,\ldots,w-1\}\). Thus,
\[
|\operatorname{supp}\pi_w|
=(w)_{d_l}
=\frac{w!}{(w-d_l)!},
\]
where \((n)_k=n(n-1)\cdots(n-k+1)\) denotes the falling factorial. For \(1\le s<w\), a tuple \(\boldsymbol j=(j_1,\ldots,j_{d_l})\in\{0,\ldots,w-1\}^{d_l}\) belongs to both supports if and only if   (i) \(s\le j_1\le w-1\); (ii) \(j_2,\ldots,j_{d_l}\) are pairwise distinct; and (iii) \(j_a\notin\{j_1,j_1-s\}\) for all \(2\le a\le d_l\). Membership in \(\operatorname{supp}\pi_w\) requires the coordinates of \(\boldsymbol j\) to be distinct, whereas membership in \(\operatorname{supp}\pi_w^{(s)}\) requires the coordinates of \((j_1-s,j_2,\ldots,j_{d_l})\) to be distinct. There are \(w-s\) choices for \(j_1\). For each such choice, \(j_1\ne j_1-s\), and thus the remaining \(d_l-1\) coordinates must be distinct and chosen from the \(w-2\) elements outside \(\{j_1,j_1-s\}\). Therefore,
\[
\left|
\operatorname{supp}\pi_w
\cap\operatorname{supp}\pi_w^{(s)}
\right|
=(w-s)(w-2)_{d_l-1}.
\]
Using \eqref{eq:Q1}, we obtain
\[
\eta_w(s) =1-\frac{(w-s)(w-2)_{d_l-1}}{(w)_{d_l}} =1-\frac{(w-s)(w-d_l)}{w(w-1)},
\qquad 1\le s<w.
\] 
For \(s\ge w\), the supports are disjoint, so \(\eta_w(s)=1\). Since \(w\ge d_l\), we have \(\eta_w(1)
=
1-\frac{w-d_l}{w}
=
\frac{d_l}{w}
\le1\),  and
\[
\eta_w(s)-\eta_w(1)
=
\frac{(s-1)(w-d_l)}{w(w-1)}
\ge0,
\qquad 1\le s<w.
\]
Thus \(\eta_w=\min_{s\ge1}\eta_w(s)=\frac{d_l}{w}\). Finally, for fixed \(d_l\), both ensembles satisfy \(\eta_w\to0\) as \(w\to\infty\). When \(d_l\ge3\), the stability condition at \(h^{\mathrm{FP}}\) holds automatically. Hence, Lemma V.5 gives \(\eqref{eq:threshold_saturation}\). The explicit conditions on \(w\) follow directly by substituting the two expressions for \(\eta_w\) into the condition
\[
\frac12 C_{q,d_l,d_r}\eta_w
<
\Delta E(\mathsf c_h)
\]
in Theorem V.4 i). When \(\Delta E(\mathsf c_h)=+\infty\), the conclusion holds without this restriction.  
 
}


\bibliographystyle{IEEEtran}
\bibliography{IEEEabrv,my_ref}

\begin{thebibliography}{10}
\providecommand{\url}[1]{#1}
\csname url@samestyle\endcsname
\providecommand{\newblock}{\relax}
\providecommand{\bibinfo}[2]{#2}
\providecommand{\BIBentrySTDinterwordspacing}{\spaceskip=0pt\relax}
\providecommand{\BIBentryALTinterwordstretchfactor}{4}
\providecommand{\BIBentryALTinterwordspacing}{\spaceskip=\fontdimen2\font plus
\BIBentryALTinterwordstretchfactor\fontdimen3\font minus
  \fontdimen4\font\relax}
\providecommand{\BIBforeignlanguage}[2]{{%
\expandafter\ifx\csname l@#1\endcsname\relax
\typeout{** WARNING: IEEEtran.bst: No hyphenation pattern has been}%
\typeout{** loaded for the language `#1'. Using the pattern for}%
\typeout{** the default language instead.}%
\else
\language=\csname l@#1\endcsname
\fi
#2}}
\providecommand{\BIBdecl}{\relax}
\BIBdecl

\bibitem{LDPCC1}
A.~J. Felstr{\"o}m and K.~S. Zigangirov, ``{Time-varying periodic convolutional
  codes with low-density parity-check matrices},'' \emph{IEEE Trans. Inf.
  Theory}, vol.~45, no.~6, pp. 2181--2191, Sep. 1999.

\bibitem{LDPCC2}
A.~Sridharan, M.~Lentmaier, D.~J. Costello, Jr., and K.~S. Zigangirov,
  ``{Convergence analysis for a class of {LDPC} convolutional codes on the
  erasure channel},'' in \emph{Proc. Annu. Allerton Conf. Commun., Control,
  Comput.}, Monticello, IL, USA, Oct. 2004, pp. 953--962.

\bibitem{LDPCC3}
M.~Lentmaier, A.~Sridharan, K.~S. Zigangirov, and D.~J. Costello, Jr.,
  ``{Terminated {LDPC} convolutional codes with thresholds close to
  capacity},'' in \emph{Proc. IEEE Int. Symp. Inf. Theory (ISIT)}, Adelaide,
  Australia, Sep. 2005, pp. 1372--1376.

\bibitem{LDPCC4}
M.~Lentmaier and G.~P. Fettweis, ``{On the thresholds of generalized {LDPC}
  convolutional codes based on protographs},'' in \emph{Proc. IEEE Int. Symp.
  Inf. Theory (ISIT)}, Austin, TX, USA, Jun. 2010, pp. 709--713.

\bibitem{LDPCC5}
M.~Lentmaier, A.~Sridharan, K.~S. Zigangirov, and D.~J. Costello, Jr.,
  ``{Iterative decoding threshold analysis for {LDPC} convolutional codes},''
  \emph{IEEE Trans. Inf. Theory}, vol.~56, no.~10, pp. 5274--5289, Oct. 2010.

\bibitem{SCLDPC1}
S.~Kudekar, T.~J. Richardson, and R.~L. Urbanke, ``{Threshold saturation via
  spatial coupling: Why convolutional {LDPC} ensembles perform so well over the
  {BEC}},'' \emph{IEEE Trans. Inf. Theory}, vol.~57, no.~2, pp. 803--834, Feb.
  2011.

\bibitem{SCLDPC}
------, ``{Spatially coupled ensembles universally achieve capacity under
  belief propagation},'' \emph{IEEE Trans. Inf. Theory}, vol.~59, no.~12, pp.
  7761--7813, Dec. 2013.

\bibitem{SCLDPC3}
A.~Yedla, Y.-Y. Jian, P.~S. Nguyen, and H.~D. Pfister, ``{A simple proof of
  {Maxwell} saturation for coupled scalar recursions},'' \emph{IEEE Trans. Inf.
  Theory}, vol.~60, no.~11, pp. 6943--6965, Nov. 2014.

\bibitem{SCLDPC4}
S.~Kumar, A.~J. Young, N.~Macris, and H.~D. Pfister, ``{Threshold saturation
  for spatially-coupled {LDPC} and {LDGM} codes on {BMS} channels},''
  \emph{IEEE Trans. Inf. Theory}, vol.~60, no.~12, pp. 7389--7415, Dec. 2014.

\bibitem{SCLDPC_minimum_distance1}
D.~G.~M. Mitchell, M.~Lentmaier, and D.~J. Costello, Jr., ``{On the minimum
  distance of generalized spatially coupled {LDPC} codes},'' in \emph{Proc.
  IEEE Int. Symp. Inf. Theory (ISIT)}, Istanbul, Turkey, Jul. 2013, pp.
  1874--1878.

\bibitem{SCLDPC_minimum_distance2}
------, ``{Spatially coupled {LDPC} codes constructed from protographs},''
  \emph{IEEE Trans. Inf. Theory}, vol.~61, no.~9, pp. 4866--4889, Sep. 2015.

\bibitem{SCLDPC_minimum_distance3}
D.~G.~M. Mitchell, P.~M. Olmos, M.~Lentmaier, and D.~J. Costello, Jr.,
  ``{Spatially coupled generalized {LDPC} codes: Asymptotic analysis and finite
  length scaling},'' \emph{IEEE Trans. Inf. Theory}, vol.~67, no.~6, pp.
  3708--3723, Jun. 2021.

\bibitem{SCLDPC_minimum_distance4}
D.~G.~M. Mitchell, A.~E. Pusane, and D.~J. Costello, Jr., ``{Minimum distance
  and trapping set analysis of protograph-based {LDPC} convolutional codes},''
  \emph{IEEE Trans. Inf. Theory}, vol.~59, no.~1, pp. 254--281, Jan. 2013.

\bibitem{SC_magzine}
D.~J. Costello, Jr., L.~Dolecek, T.~Fuja, J.~Kliewer, D.~Mitchell, and
  R.~Smarandache, ``{Spatially coupled sparse codes on graphs: Theory and
  practice},'' \emph{IEEE Commun. Mag.}, vol.~52, no.~7, pp. 168--176, Jul.
  2014.

\bibitem{SCLDPC_finite_length1}
P.~M. Olmos and R.~L. Urbanke, ``{A scaling law to predict the finite-length
  performance of spatially-coupled {LDPC} codes},'' \emph{IEEE Trans. Inf.
  Theory}, vol.~61, no.~6, pp. 3164--3184, Jun. 2015.

\bibitem{SCLDPC_finite_length2}
M.~Stinner and P.~M. Olmos, ``{On the waterfall performance of finite-length
  {SC-LDPC} codes constructed from protographs},'' \emph{IEEE J. Sel. Areas
  Commun.}, vol.~34, no.~2, pp. 345--361, Feb. 2016.

\bibitem{SCLDPC_finite_length3}
R.~Sokolovskii, A.~Graell~i Amat, and F.~Br{\"a}nnstr{\"o}m, ``{Finite-length
  scaling of {SC-LDPC} codes with a limited number of decoding iterations},''
  \emph{IEEE Trans. Inf. Theory}, vol.~69, no.~8, pp. 4869--4888, Aug. 2023.

\bibitem{Davey1998}
M.~C. Davey and D.~J.~C. MacKay, ``{Low-density parity check codes over
  {$\mathrm{GF}(q)$}},'' \emph{IEEE Commun. Lett.}, vol.~2, no.~6, pp.
  165--167, Jun. 1998.

\bibitem{NBSC1}
A.~Piemontese, A.~Graell~i Amat, and G.~Colavolpe, ``{Nonbinary
  spatially-coupled {LDPC} codes on the binary erasure channel},'' in
  \emph{Proc. IEEE Int. Conf. Commun. (ICC)}, Budapest, Hungary, Jun. 2013, pp.
  3270--3274.

\bibitem{NBSC2}
L.~Wei, T.~Koike-Akino, D.~G.~M. Mitchell, T.~E. Fuja, and D.~J. Costello, Jr.,
  ``{Threshold analysis of non-binary spatially-coupled {LDPC} codes with
  windowed decoding},'' in \emph{Proc. IEEE Int. Symp. Inf. Theory (ISIT)},
  Honolulu, HI, USA, Jun. 2014, pp. 881--885.

\bibitem{NBSC3}
I.~Andriyanova and A.~Graell~i Amat, ``{Threshold saturation for nonbinary
  {SC-LDPC} codes on the binary erasure channel},'' \emph{IEEE Trans. Inf.
  Theory}, vol.~62, no.~5, pp. 2622--2638, May 2016.

\bibitem{NBSC4}
K.~Huang, D.~G.~M. Mitchell, L.~Wei, X.~Ma, and D.~J. Costello, Jr.,
  ``{Performance comparison of {LDPC} block and spatially coupled codes over
  {$\mathrm{GF}(q)$}},'' \emph{IEEE Trans. Commun.}, vol.~63, no.~3, pp.
  592--604, Mar. 2015.

\bibitem{Zhang2018}
J.~Zhang, B.~Bai, D.~Deng, M.~Zhu, H.~Xu, and M.~Guan, ``{Non-uniform
  spatially-coupled {LDPC} codes over {$\mathrm{GF}(2^m)$}},'' in \emph{Proc.
  IEEE Int. Symp. Inf. Theory (ISIT)}, Vail, CO, USA, Jun. 2018, pp. 816--820.

\bibitem{Hareedy2019}
A.~Hareedy, C.~Lanka, N.~Guo, and L.~Dolecek, ``{A combinatorial methodology
  for optimizing non-binary graph-based codes: Theoretical analysis and
  applications in data storage},'' \emph{IEEE Trans. Inf. Theory}, vol.~65,
  no.~4, pp. 2128--2154, Apr. 2019.

\bibitem{ModernCode}
T.~J. Richardson and R.~Urbanke, \emph{{Modern Coding Theory}}.\hskip 1em plus
  0.5em minus 0.4em\relax Cambridge, U.K.: Cambridge Univ. Press, 2008.

\bibitem{Huang2015Distance}
K.~Huang, D.~G.~M. Mitchell, X.~Ma, and D.~J. Costello, Jr., ``{Asymptotic
  distance properties of protograph-based spatially coupled {LDPC} codes over
  {$\mathrm{GF}(q)$}},'' in \emph{Proc. IEEE Inf. Theory Workshop (ITW)}, 2015,
  pp. 1--5.

\bibitem{OT}
C.~Villani, \emph{{Optimal Transport: Old and New}}, ser. Grundlehren der
  mathematischen Wissenschaften.\hskip 1em plus 0.5em minus 0.4em\relax Berlin,
  Germany: Springer, 2009, vol. 338.

\bibitem{Gallager}
R.~G. Gallager, \emph{{Low-Density Parity-Check Codes}}.\hskip 1em plus 0.5em
  minus 0.4em\relax Cambridge, MA, USA: MIT Press, 1963.

\bibitem{GoodCodes}
D.~J.~C. MacKay, ``{Good error-correcting codes based on very sparse
  matrices},'' \emph{IEEE Trans. Inf. Theory}, vol.~45, no.~2, pp. 399--431,
  Mar. 1999.

\bibitem{Luby}
M.~Luby, M.~Mitzenmacher, A.~Shokrollahi, and D.~A. Spielman, ``{Improved
  low-density parity-check codes using irregular graphs},'' \emph{IEEE Trans.
  Inf. Theory}, vol.~47, no.~2, pp. 585--598, Feb. 2001.

\bibitem{LDPC1}
T.~Richardson and R.~Urbanke, ``{The capacity of low-density parity-check codes
  under message-passing decoding},'' \emph{IEEE Trans. Inf. Theory}, vol.~47,
  no.~2, pp. 599--618, Feb. 2001.

\bibitem{margulis1982explicit}
G.~A. Margulis, ``{Explicit constructions of graphs without short cycles and
  low density codes},'' \emph{Combinatorica}, vol.~2, no.~1, pp. 71--78, 1982.

\bibitem{vasic2004combinatorial}
B.~Vasic and O.~Milenkovic, ``{Combinatorial constructions of low-density
  parity-check codes for iterative decoding},'' \emph{IEEE Trans. Inf. Theory},
  vol.~50, no.~6, pp. 1156--1176, Jun. 2004.

\bibitem{Weight_Distribution1}
A.~Bennatan and D.~Burshtein, ``{On the application of {LDPC} codes to
  arbitrary discrete-memoryless channels},'' \emph{IEEE Trans. Inf. Theory},
  vol.~50, no.~3, pp. 417--438, Mar. 2004.

\bibitem{Weight_Distribution2}
G.~Como and F.~Fagnani, ``{Average spectra and minimum distances of low-density
  parity-check codes over {Abelian} groups},'' \emph{SIAM J. Discrete Math.},
  vol.~23, no.~1, pp. 19--53, 2008.

\bibitem{Weight_Finite_Field}
S.~Yang, T.~Honold, Y.~Chen, Z.~Zhang, and P.~Qiu, ``{Weight distributions of
  regular low-density parity-check codes over finite fields},'' \emph{IEEE
  Trans. Inf. Theory}, vol.~57, no.~11, pp. 7507--7521, Nov. 2011.

\bibitem{Di2002}
C.~Di, D.~Proietti, I.~E. Telatar, T.~J. Richardson, and R.~L. Urbanke,
  ``{Finite-length analysis of low-density parity-check codes on the binary
  erasure channel},'' \emph{IEEE Trans. Inf. Theory}, vol.~48, no.~6, pp.
  1570--1579, Jun. 2002.

\bibitem{Orlitsky2005}
A.~Orlitsky, K.~Viswanathan, and J.~Zhang, ``{Stopping set distribution of
  {LDPC} code ensembles},'' \emph{IEEE Trans. Inf. Theory}, vol.~51, no.~3, pp.
  929--953, Mar. 2005.

\bibitem{Bound_binary}
G.~Miller and D.~Burshtein, ``{Bounds on the maximum-likelihood decoding error
  probability of low-density-parity-check codes},'' \emph{IEEE Trans. Inf.
  Theory}, vol.~47, no.~7, pp. 2696--2710, Nov. 2001.

\bibitem{Sridharan2007}
A.~Sridharan, D.~Truhachev, M.~Lentmaier, D.~J. Costello, Jr., and K.~S.
  Zigangirov, ``{Distance bounds for an ensemble of {LDPC} convolutional
  codes},'' \emph{IEEE Trans. Inf. Theory}, vol.~53, no.~12, pp. 4537--4555,
  Dec. 2007.

\bibitem{Truhachev2010}
D.~Truhachev, K.~S. Zigangirov, and D.~J. Costello, Jr., ``{Distance bounds for
  periodically time-varying and tail-biting {LDPC} convolutional codes},''
  \emph{IEEE Trans. Inf. Theory}, vol.~56, no.~9, pp. 4301--4308, Sep. 2010.

\bibitem{Mitchell2011FreeDistance}
D.~G.~M. Mitchell, A.~E. Pusane, M.~Lentmaier, and D.~J. Costello, Jr.,
  ``{Exact free distance and trapping set growth rates for {LDPC} convolutional
  codes},'' in \emph{Proc. IEEE Int. Symp. Inf. Theory (ISIT)}, 2011, pp.
  1096--1100.

\bibitem{Gallager_information}
R.~G. Gallager, \emph{{Information Theory and Reliable Communication}}.\hskip
  1em plus 0.5em minus 0.4em\relax New York, NY, USA: Wiley, 1968.

\bibitem{QPEC}
R.~Cohen and Y.~Cassuto, ``{Iterative decoding of {LDPC} codes over the
  {$q$}-ary partial erasure channel},'' \emph{IEEE Trans. Inf. Theory},
  vol.~62, no.~5, pp. 2658--2672, May 2016.

\bibitem{Coset1}
A.~Kav{\v c}i{\'c}, X.~Ma, and M.~Mitzenmacher, ``{Binary intersymbol
  interference channels: {Gallager} codes, density evolution and code
  performance bounds},'' \emph{IEEE Trans. Inf. Theory}, vol.~49, no.~7, pp.
  1636--1652, Jul. 2003.

\bibitem{Coset2}
J.~Hou, P.~H. Siegel, L.~B. Milstein, and H.~D. Pfister,
  ``{Capacity-approaching bandwidth-efficient coded modulation schemes based on
  low density parity-check codes},'' \emph{IEEE Trans. Inf. Theory}, vol.~49,
  no.~9, pp. 2141--2155, Sep. 2003.

\bibitem{Coset3}
A.~Bennatan and D.~Burshtein, ``{Design and analysis of nonbinary {LDPC} codes
  for arbitrary discrete-memoryless channels},'' \emph{IEEE Trans. Inf.
  Theory}, vol.~52, no.~2, pp. 549--583, Feb. 2006.

\bibitem{Strassen}
V.~Strassen, ``{The existence of probability measures with given marginals},''
  \emph{Ann. Math. Statist.}, vol.~36, no.~2, pp. 423--439, Apr. 1965.

\bibitem{BauschkeCombettes2017}
H.~H. Bauschke and P.~L. Combettes, \emph{{Convex Analysis and Monotone
  Operator Theory in Hilbert Spaces}}, 2nd~ed., ser. CMS Books in
  Mathematics.\hskip 1em plus 0.5em minus 0.4em\relax Springer, 2017.

\bibitem{GEXIT}
C.~M{\'e}asson, A.~Montanari, T.~J. Richardson, and R.~Urbanke, ``{The
  generalized area theorem and some of its consequences},'' \emph{IEEE Trans.
  Inf. Theory}, vol.~55, no.~11, pp. 4793--4821, Nov. 2009.

\end{thebibliography}
\end{document}